%% file: main.tex
\RequirePackage{lineno}
\documentclass[pdftex, twocolumn, epjc3]{svjour3}
\pdfoutput=1

\usepackage{placeins}
\usepackage{multirow}
\usepackage[utf8]{inputenc}
\usepackage{color}
\input{common/preamble.tex}

\usepackage{cite}

\def\bracketbar{\hbox{\kern-9pt\raise1pt%
    \hbox{{\tiny(}{\lower1.5pt\hbox{\bf--}}{\tiny)}}}}

\journalname{Eur. Phys. J. C}
\hypersetup{draft}
\begin{document}

\title{Long-baseline neutrino oscillation physics potential of the DUNE experiment}
\date{\today}
\authorrunning{DUNE Collaboration}
\input{sections/author_list}

\onecolumn
\maketitle
\twocolumn
\sloppy
\begin{abstract}
The sensitivity of the Deep Underground Neutrino Experiment (DUNE)  to neutrino oscillation is determined, based on a full simulation, reconstruction, and event selection of the far detector and a full simulation and parameterized analysis of the near detector. Detailed uncertainties due to the flux prediction, neutrino interaction model, and detector effects are included. DUNE will resolve the neutrino mass ordering to a precision of 5$\sigma$, for all \deltacp values, after 2 years of running with the nominal detector design and beam configuration. It has the potential to observe charge-parity violation in the neutrino sector to a precision of 3$\sigma$ (5$\sigma$) after an exposure of 5 (10) years, for 50\% of all \deltacp values. It will also make precise measurements of other parameters governing long-baseline neutrino oscillation, and after an exposure of 15 years will achieve a similar sensitivity to $\sin^{2} 2\theta_{13}$ to current reactor experiments.
\end{abstract}

\input{sections/overview}
\input{sections/flux}
\input{sections/nuint}
\input{sections/nd}
\input{sections/fd}
\input{sections/rate}
\input{sections/syst}

\input{sections/methods}
\input{sections/sens}
\input{sections/conclusion}
\input{sections/acknowledgements}
\bibliographystyle{utphys}
\bibliography{common/tdr-citedb}

\end{document}

%% file: sections/author_list.tex
\author {The DUNE Collaboration: B.~Abi\thanksref{Oxford}
	 \and R.~Acciarri\thanksref{Fermi}
	 \and M.~A.~Acero\thanksref{Atlantico}
	 \and G.~Adamov\thanksref{Georgian}
	 \and D.~Adams\thanksref{Brookhaven}
	 \and M.~Adinolfi\thanksref{Bristol}
	 \and Z.~Ahmad\thanksref{VariableEnergy}
	 \and J.~Ahmed\thanksref{Warwick}
	 \and T.~Alion\thanksref{Sussex}
	 \and S.~Alonso Monsalve\thanksref{CERN}
% ...
}

\thankstext{corr1}{E-Mail: callum.wilkinson@lhep.unibe.ch}
\thankstext{corr2}{E-Mail: marshall@lbl.gov}
\thankstext{corr3}{E-Mail: etw@bnl.gov}

\institute {
% ...
\pagebreak[0] Universidad del Atl{\'a}ntico, Atl{\'a}ntico, Colombia\label{Atlantico}
% ...
\and\pagebreak[0] University of Bristol, Bristol BS8 1TL, United Kingdom\label{Bristol}
	 \and\pagebreak[0] Brookhaven National Laboratory, Upton, NY 11973, USA\label{Brookhaven}
% ...
\and\pagebreak[0] CERN, The European Organization for Nuclear Research, 1211 Meyrin, Switzerland\label{CERN}
% ...
\and\pagebreak[0] Fermi National Accelerator Laboratory, Batavia, IL 60510, USA\label{Fermi}
% ...
\and\pagebreak[0] Georgian Technical University, Tbilisi, Georgia\label{Georgian}
% ...
\and\pagebreak[0] University of Oxford, Oxford, OX1 3RH, United Kingdom\label{Oxford}
% ...
\and\pagebreak[0] University of Sussex, Brighton, BN1 9RH, United Kingdom\label{Sussex}
% ...
\and\pagebreak[0] Variable Energy Cyclotron Centre, 700 064 West Bengal, India\label{VariableEnergy}
% ...
\and\pagebreak[0] University of Warwick, Coventry CV4 7AL, United Kingdom\label{Warwick}
% ...
}

%% file: sections/overview.tex
\section{Introduction}
\label{sec:intro}

The \dword{dune} is a next-generation, long-baseline neutrino oscillation experiment which will carry out a detailed study of neutrino mixing utilizing high-intensity \numu and \anumu beams measured over a long baseline.
\dword{dune} is designed to make significant contributions to the completion of the standard three-flavor picture by measuring all the parameters governing $\nu_1$--$\nu_3$ and $\nu_2$--$\nu_3$ mixing in a single experiment. Its main scientific goals are the definitive determination of the neutrino mass ordering, the definitive observation of \dword{cpv} for more than 50\% of possible true values of the charge-parity violating phase, \deltacp,  
and precise measurement of oscillation parameters, particularly \deltacp, \sinstt{13}, and the octant of $\theta_{23}$.
These measurements will help guide theory in understanding if there are new symmetries in the neutrino sector and whether there is a relationship between the generational structure of quarks and leptons~\cite{Qian:2015waa}. Observation of \dword{cpv} in neutrinos would be an important step in understanding the origin of the baryon asymmetry of the universe~\cite{Fukugita:1986hr, Davidson:2008bu}.

The \dword{dune} experiment will observe neutrinos from a high-power neutrino beam peaked at $\sim$2.5 GeV but with a broad range of neutrino energies, a \dword{nd} located at Fermi National Accelerator Laboratory, in Batavia, Illinois, USA, and a large \dword{lartpc} \dword{fd} located at the 4850 ft level of Sanford Underground Research Facility (SURF), in Lead, South Dakota, USA, 1285~km from the neutrino production point. The neutrino beam provided by \dword{lbnf}~\cite{Abi:2020wmh} is produced using protons from Fermilab's Main Injector, which are guided onto a graphite target, and a traditional horn-focusing system to select and focus particles produced in the target~\cite{Abi:2020evt}. The polarity of the focusing magnets can be reversed to produce a beam dominated by either muon neutrinos or muon antineutrinos. A highly capable \dword{nd} will constrain many systematic uncertainties for the oscillation analysis. The 40-kt (fiducial) \dword{fd} is composed of four 10 kt (fiducial) LArTPC modules~\cite{Acciarri:2016crz,Acciarri:2015uup,Acciarri:2016ooe}.
The deep underground location of the \dword{fd} reduces cosmogenic and atmospheric sources of background, which also provides sensitivity to nucleon decay and low-energy neutrino detection, for example, the possible observation of neutrinos from a core-collapse supernova~\cite{Abi:2020evt}.

The entire complement of neutrino oscillation experiments to date has measured five of the neutrino mixing parameters~\cite{Esteban:2018azc,deSalas:2017kay,Capozzi:2017yic}: the three mixing angles $\theta_{12}$, $\theta_{23}$, and $\theta_{13}$, and the two squared-mass differences $\Delta m^{2}_{21}$ and $|\Delta m^{2}_{31}|$, where $\Delta m^2_{ij} = m^2_{i} - m^{2}_{j}$ is the difference between the squares of the neutrino mass states in eV$^{2}$.
The neutrino mass ordering (i.e., the sign of $\Delta m^{2}_{31}$) is unknown, though recent results show a weak preference for the normal ordering~\cite{Abe:2018wpn,PhysRevD.97.072001,PhysRevLett.123.151803}.
The value of \deltacp is not well known, though neutrino oscillation data are beginning to provide some information on its value~\cite{Abe:2018wpn,Abe:2019vii}.

\begin{figure}[htbp]
  \centering
  \includegraphics[width=0.98\linewidth]{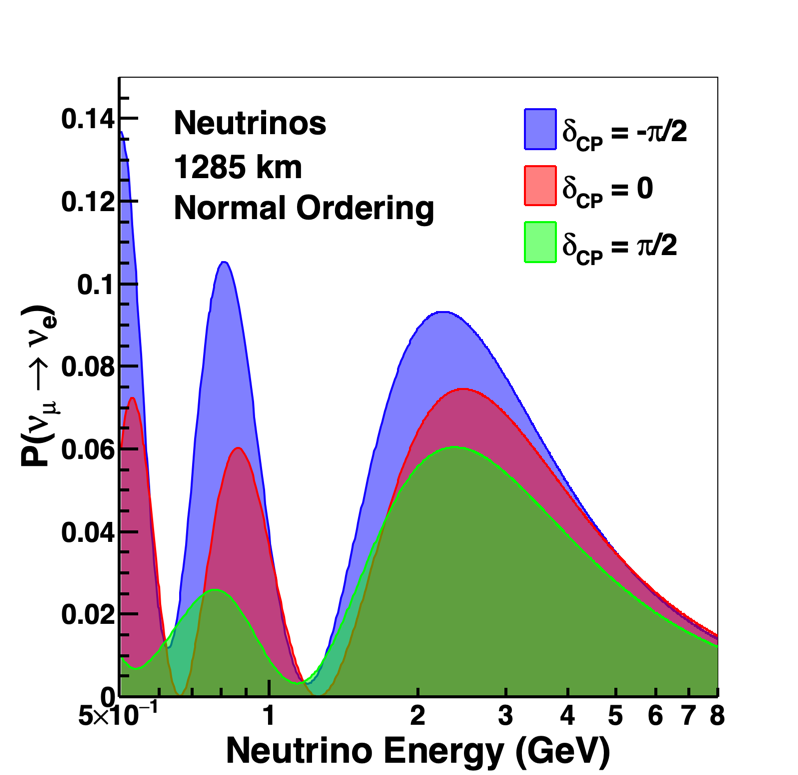}\\
  \includegraphics[width=0.98\linewidth]{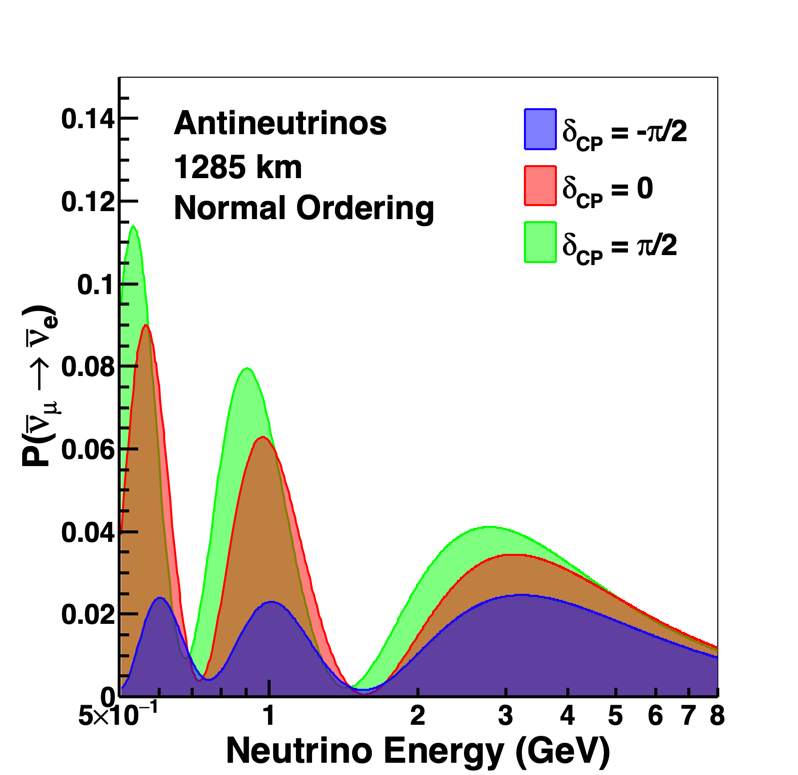}
  \caption[Appearance probabilities for \nue and \anue at \SI{1285}{\km}]{The appearance probability at a baseline of \SI{1285}\km{},
  as a function of neutrino energy, for \deltacp = $-\pi/2$ (blue), 
  0 (red), and $\pi/2$ (green), for neutrinos (top) and antineutrinos
  (bottom), for normal ordering.}
  \label{fig:oscprob}
\end{figure}

The oscillation probability of \numu $\rightarrow$ \nue through matter in the standard three-flavor model and a constant density approximation is, to first order~\cite{Nunokawa:2007qh}:
\begin{equation}
  \begin{aligned}
    P(\;\nu^{\bracketbar}_\mu \rightarrow \nu^{\bracketbar}_e) & \simeq \sin^2 \theta_{23} \sin^2 2 \theta_{13} 
    \frac{ \sin^2(\Delta_{31} - aL)}{(\Delta_{31}-aL)^2} \Delta_{31}^2 \\
    & + \sin 2 \theta_{23} \sin 2 \theta_{13} \sin 2 \theta_{12}\frac{ \sin(\Delta_{31} - aL)}{(\Delta_{31}-aL)} \Delta_{31} \\
    &\times \frac{\sin(aL)}{(aL)} \Delta_{21} \cos (\Delta_{31} \pm \mdeltacp) & \\
    & + \cos^2 \theta_{23} \sin^2 2 \theta_{12} \frac {\sin^2(aL)}{(aL)^2} \Delta_{21}^2,
  \end{aligned}
  \label{eqn:appprob}
\end{equation}
where
\begin{equation*}
  a = \pm \frac{G_{\mathrm{F}}N_e}{\sqrt{2}} \approx \pm\frac{1}{3500~\mathrm{km}}\left(\frac{\rho}{3.0~\mathrm{g/cm}^{3}}\right),
\end{equation*}
$G_{\mathrm{F}}$ is the Fermi constant, $N_e$ is the number density of electrons in the Earth's crust, $\Delta_{ij} = 1.267 \Delta m^2_{ij} L/E_\nu$, $L$ is the baseline in km, and $E_\nu$ is the neutrino energy in GeV. 
Both \deltacp and $a$ terms are positive for
$\nu_\mu \to \nu_e$ and negative for $\bar{\nu}_\mu \to \bar{\nu}_e$ oscillations; i.e.,
a neutrino-antineutrino asymmetry is introduced both by \dword{cpv} (\deltacp)
and the matter effect ($a$). The origin of the matter effect asymmetry 
is simply the presence of electrons and absence of positrons in the Earth~\cite{Wolfenstein:1977ue,Mikheev:1986gs}.
The (anti-)electron neutrino appearance probability
is shown in 
Figure~\ref{fig:oscprob} at the \dword{dune} baseline of \SI{1285}\km{} as a function of neutrino 
energy for several values of \deltacp.

\dword{dune} has a number of features that give it unique physics reach, complementary to other existing and planned experiments~\cite{Ayres:2007tu,Abe:2011ks,Abe:2018uyc}. Its broad-band beam makes it sensitive to the shape of the oscillation spectrum for a range of neutrino energies. \dword{dune}'s relatively high energy neutrino beam enhances the size of the matter effect and will allow \dword{dune} to measure \deltacp and the mass ordering simultaneously. The unique \dword{lartpc} detector technology will enhance the resolution on \dword{dune}'s measurement of the value of \deltacp, and along with the increased neutrino energy, gives \dword{dune} a different set of systematic uncertainties to other experiments, making \dword{dune} complementary with them.

This paper describes studies that quantify DUNE's expected sensitivity to long-baseline neutrino oscillation, using the accelerator neutrino beam. Note that atmospheric neutrino samples would provide additional sensitivity to some of the same physics, but are not included in this work. The flux simulation and associated uncertainties are described in Section~\ref{sec:flux}. Section~\ref{sec:nuint} describes the neutrino interaction model and systematic variations. The near and far detector simulation, reconstruction, and event selections are described in Sections~\ref{sec:nd} and \ref{sec:fd}, respectively, with a nominal set of event rate predictions given in Section~\ref{sec:rate}. Detector uncertainties are described in Section~\ref{sec:syst}. The methods used to extract oscillation sensitivities are described in Section~\ref{sec:methods}. The primary sensitivity results are presented in Section~\ref{sec:sens}. We present our conclusions in Section~\ref{sec:conclude}.

%% file: sections/flux.tex
\section{Neutrino Beam Flux and Uncertainties}
\label{sec:flux}

The expected neutrino flux is generated using G4LBNF~\cite{Aliaga:2016oaz,Abi:2020evt}, a \dword{geant4}-based~\cite{Agostinelli:2002hh} simulation of the \dword{lbnf} neutrino beam. The simulation uses a detailed description of the \dword{lbnf} optimized beam design~\cite{Abi:2020evt}, which includes a target and horns designed to maximize sensitivity to \dword{cpv} given the physical constraints on the beamline design.   

\begin{figure}
\begin{center}
\includegraphics[width=0.98\linewidth]{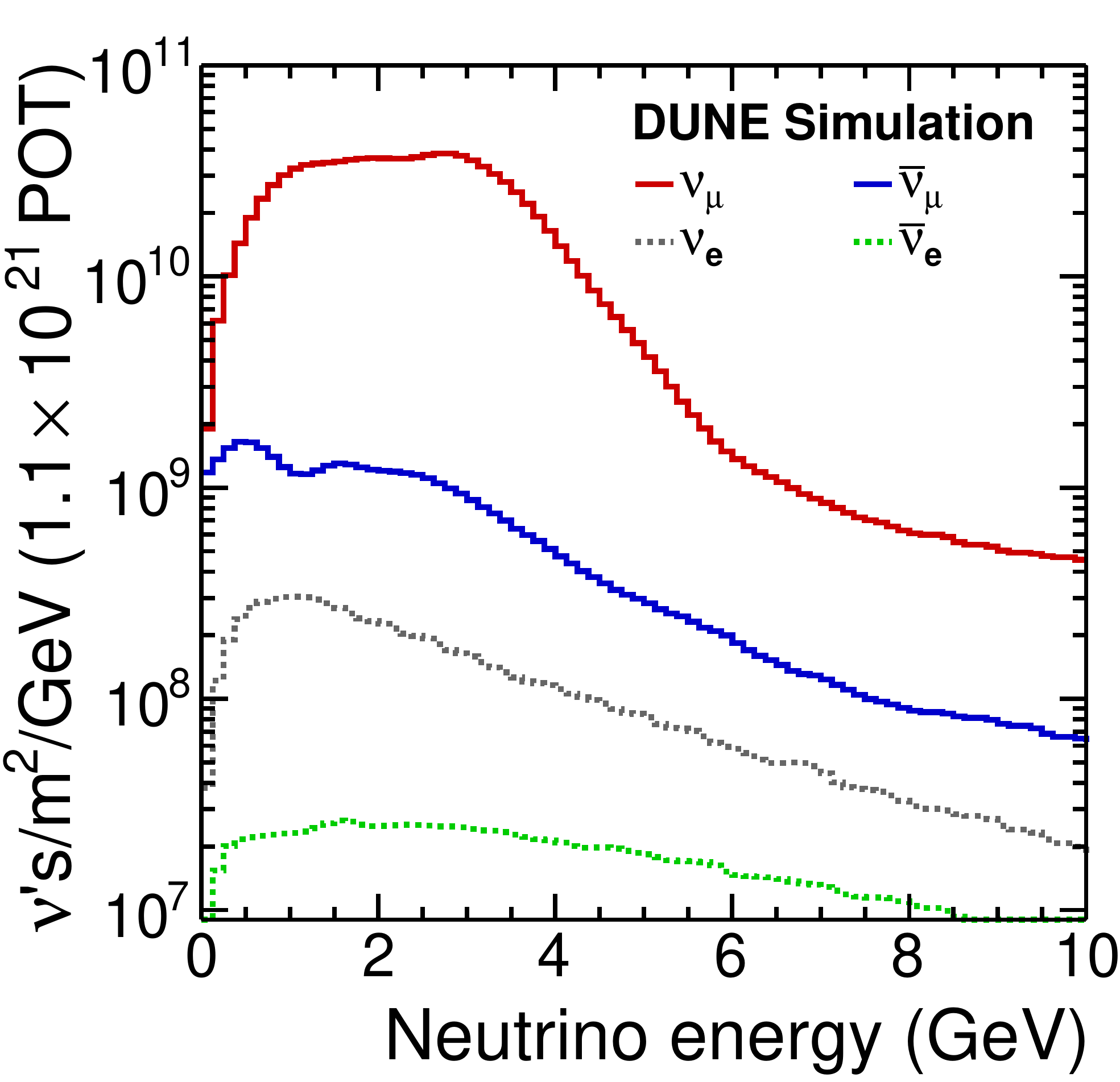}
\includegraphics[width=0.98\linewidth]{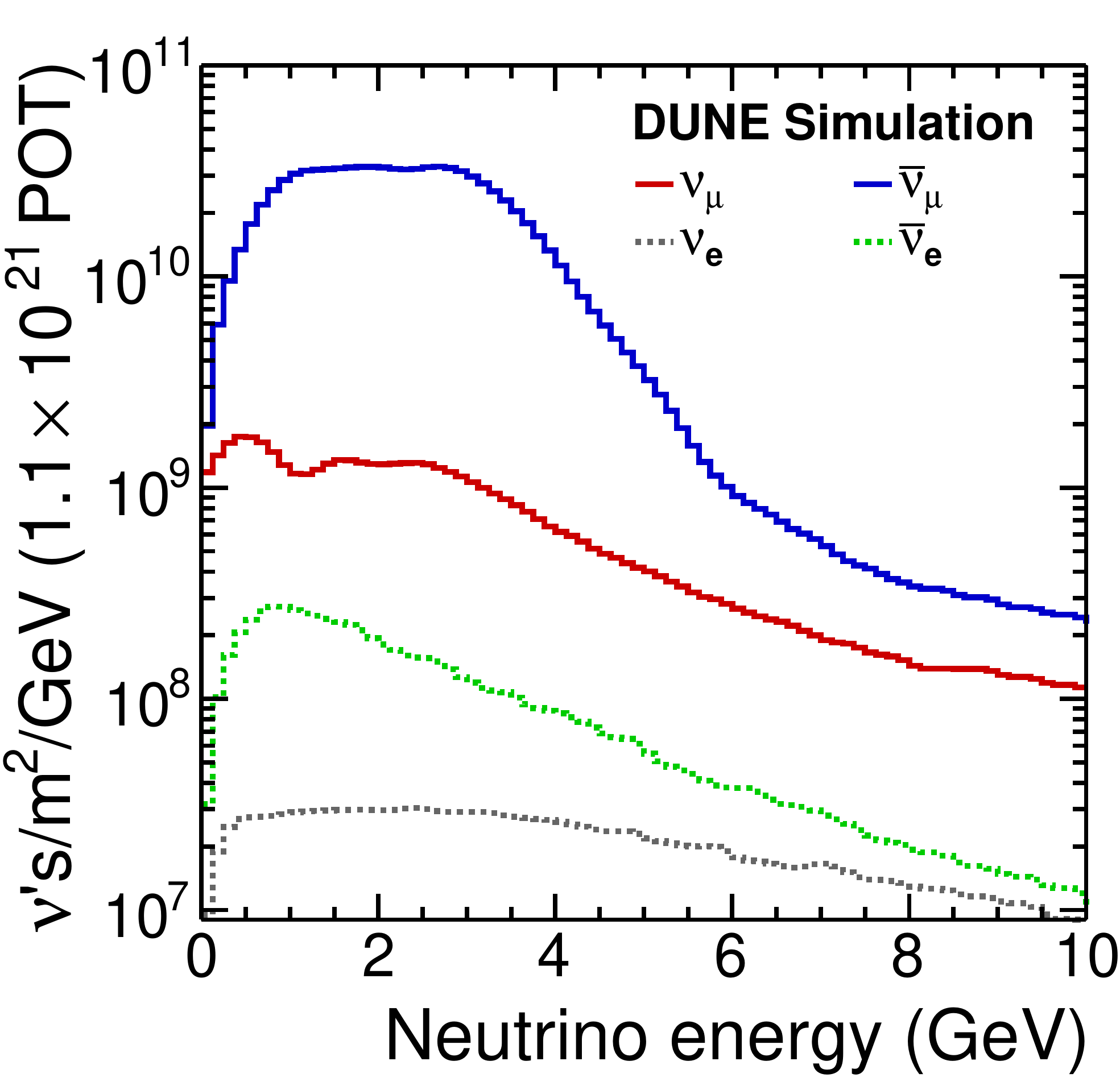}
\caption{Neutrino fluxes at the \dword{fd} for neutrino-enhanced, FHC, beam running (top) and antineutrino, RHC, beam running (bottom).}
\label{fig:flux_flavor}
\end{center}
\end{figure}

\begin{figure}
\includegraphics[width=0.98\linewidth]{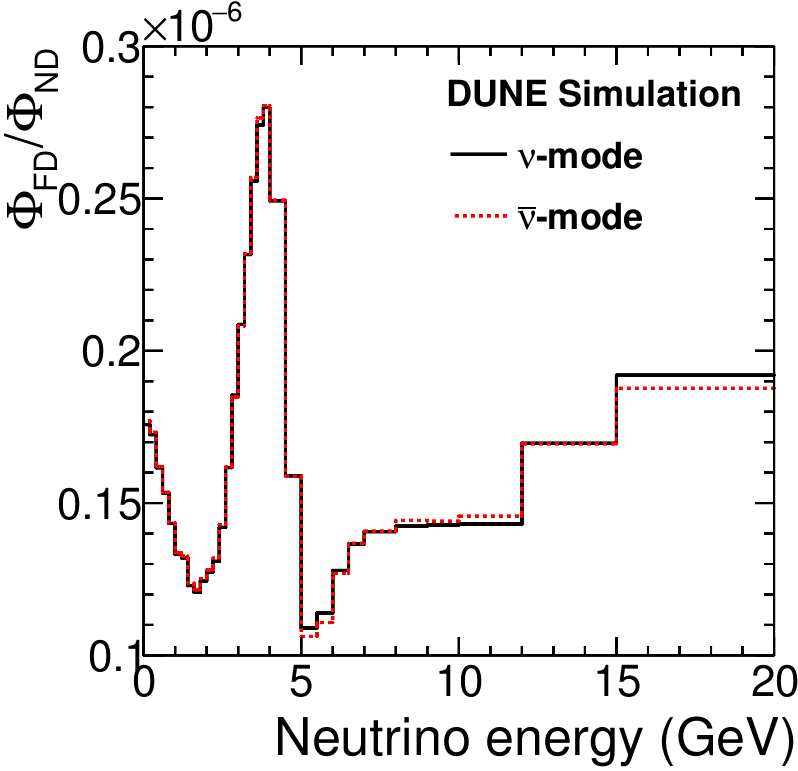}
\includegraphics[width=0.98\linewidth]{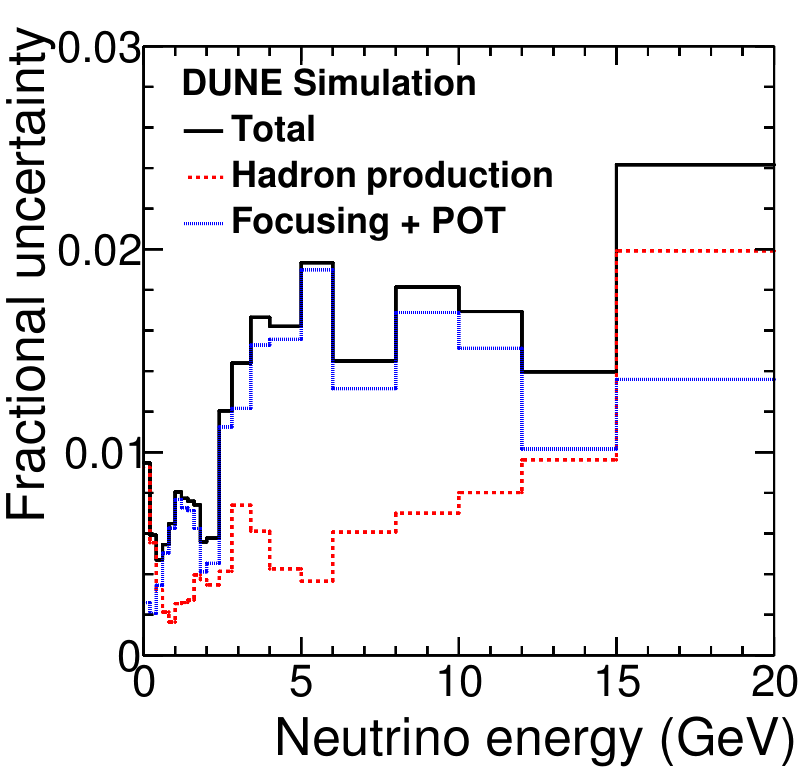}
\caption{Ratio of \dword{nd} and \dword{fd} fluxes show for the muon neutrino component of the \dword{fhc} flux and the muon antineutrino component of the \dword{rhc} flux (top) and uncertainties on the \dword{fhc} muon neutrino ratio (bottom).}
\label{fig:flux_nearfar}
\end{figure}

Neutrino fluxes for neutrino-enhanced, forward horn current (FHC), and antineutrino-enhanced, reverse horn current (RHC), configurations of \dword{lbnf} are shown in Figure~\ref{fig:flux_flavor}.  Uncertainties on the neutrino fluxes arise primarily from uncertainties in hadrons produced off the target and uncertainties in the design parameters of the beamline, such as horn currents and horn and target positioning (commonly called ``focusing uncertainties'')~\cite{Abi:2020evt}. Given current measurements of hadron production and \dword{lbnf} estimates of alignment tolerances, flux uncertainties are approximately 8\% at the first oscillation maximum and 12\% at the second.  These uncertainties are highly correlated across energy bins and neutrino flavors.
The unoscillated fluxes at the \dword{nd} and \dword{fd} are similar, but not identical. The relationship is well understood, and flux uncertainties mostly cancel for the ratio of fluxes between the two detectors. Uncertainties on the ratio are dominated by focusing uncertainties and are $\sim$1\% or smaller except at the falling edge of the focusing peak ($\sim$4 GeV), where they rise to 2\%. The rise is due to the presence of many particles which are not strongly focused by the horns in this energy region, which are particularly sensitive to focusing and alignment uncertainties. The near-to-far flux ratio and uncertainties on this ratio are shown in Fig.~\ref{fig:flux_nearfar}.

\begin{figure}[htbp]
    \includegraphics[width=0.98\linewidth]{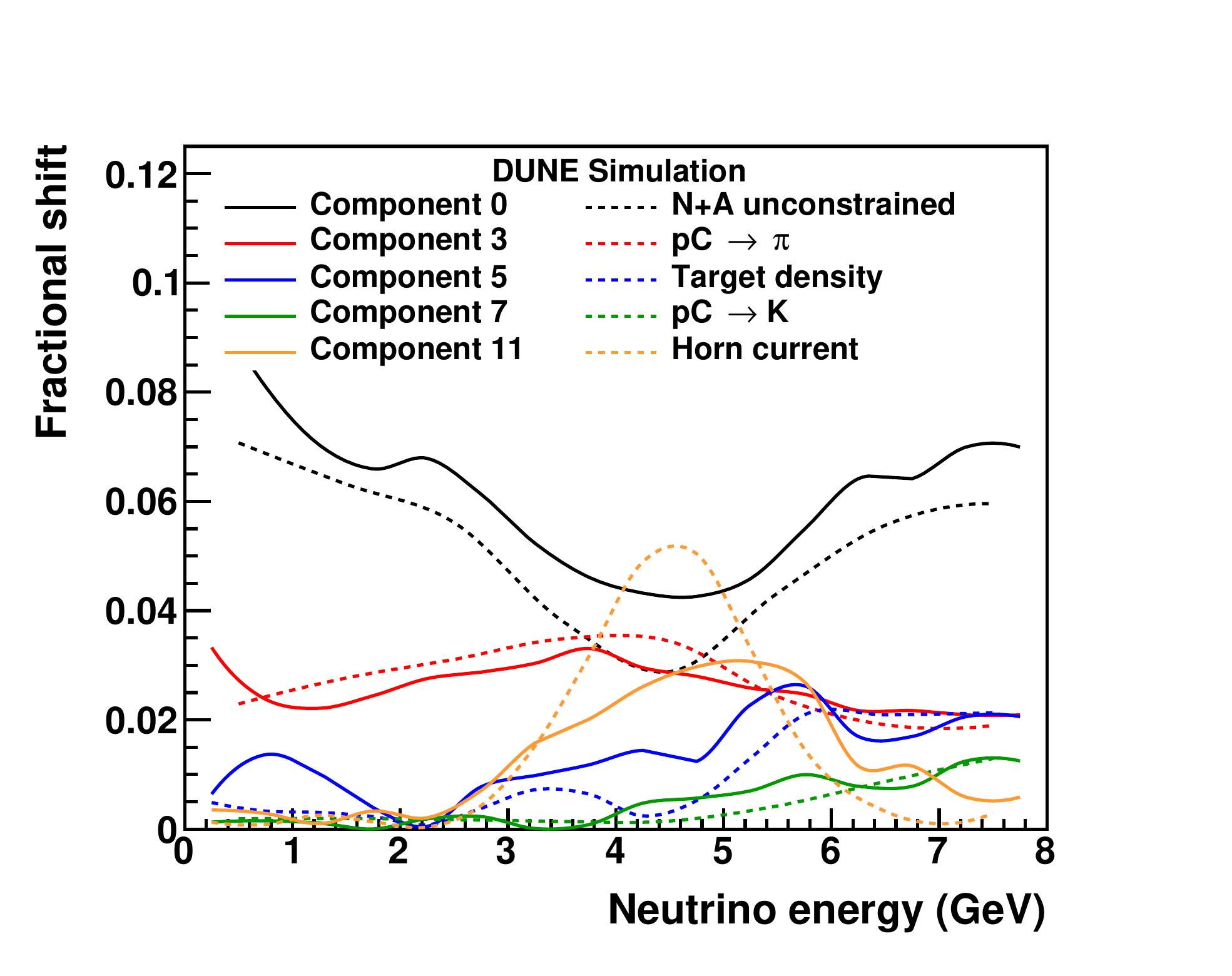}
    \caption{Select flux principal components are compared to specific underlying uncertainties from the hadron production and beam focusing models. Note that while these are shown as positive shifts, the absolute sign is arbitrary.}
    \label{fig:fluxPCA}
\end{figure}

Beam-focusing and hadron-production uncertainties on the flux prediction are evaluated by reproducing the full beamline simulation many times with variations of the input model according to those uncertainties. The resultant uncertainty on the neutrino flux prediction is described through a covariance matrix, where each bin corresponds to an energy range of a particular beam mode and neutrino species, separated by flux at the \dword{nd} and \dword{fd}. The output covariance matrix has $208 \times 208$ bins, despite having only $\sim$30 input uncertainties. To reduce the number of parameters used in the fit, the covariance matrix is diagonalized, and each principal component is treated as an uncorrelated nuisance parameter. The 208 principal components are ordered by the magnitude of their corresponding eigenvalues, which is the variance along the principal component (eigenvector) direction, and only the first $\sim$30 are large enough that they need to be included. This was validated by including more flux parameters and checking that there was no significant change to the sensitivity for a small number of test cases. By the 10th principal component, the eigenvalue is 1\% of the largest eigenvalue. As may be expected, the largest uncertainties correspond to the largest principal components as shown in Figure~\ref{fig:fluxPCA}. The largest principal component (component 0) matches the hadron production uncertainty on nucleon-nucleus interactions in a phase space region not covered by data. Components 3 and 7 correspond to the data-constrained uncertainty on proton interactions in the target producing pions and kaons, respectively. Components 5 and 11 correspond to two of the largest focusing uncertainties, the density of the target and the horn current, respectively. Other components not shown either do not fit a single uncertain parameter or may represent two or more degenerate systematics or ones that produce anti-correlations in neighboring energy bins.

Future hadron production measurements are expected to improve the quality of, and the resulting constraints on, these flux uncertainty estimates.  Approximately 40\% of the interactions that produce neutrinos in the \dword{lbnf} beam simulation have no direct data constraints. Large uncertainties are assumed for these interactions. The largest unconstrained sources of uncertainty are proton quasielastic interactions and pion and kaon rescattering in beamline materials. The proposed EMPHATIC experiment~\cite{emphatic} at Fermilab will be able to constrain quasielastic and low-energy interactions that dominate the lowest neutrino energy bins. The NA61 experiment at CERN has taken data that will constrain many higher energy interactions, and also plans to measure hadrons produced on a replica \dword{lbnf} target, which would provide tight constraints on all interactions occurring in the target. A similar program at NA61 has reduced flux uncertainties for the T2K experiment from $\sim$10\% to $\sim$5\%~\cite{Vladisavljevic:2018prd}. Another proposed experiment, the \dword{lbnf} spectrometer~\cite{DOCDB4604}, would measure hadrons after both production and focusing in the horns to further constrain the hadron production uncertainties, and could also be used to experimentally assess the impact of shifted alignment parameters on the focused hadrons (rather than relying solely on simulation).

%% file: sections/nuint.tex
\section{Neutrino interaction model and uncertainties}\label{sec:nuint}
A framework for considering the impact of neutrino interaction model uncertainties on the oscillation analysis has been developed. The default interaction model is implemented in v2.12.10 of the \dword{genie} generator~\cite{Andreopoulos:2009rq,Andreopoulos:2015wxa}. Variations in the cross sections are implemented in various ways: using \dword{genie} reweighting parameters (sometimes referred to as ``\dword{genie} knobs''); with {\em ad hoc} weights of events that are designed to parameterize uncertainties or cross-section corrections currently not implemented within \dword{genie}; or through discrete alternative model comparisons. The latter are achieved through alternative generators, alternative \dword{genie} configurations, or custom weightings, which made extensive use of the NUISANCE package~\cite{Stowell:2016jfr} in their development.

The interaction model components and uncertainties can be divided into seven groups: (1) initial state, (2) hard scattering and nuclear modifications to the quasielastic, or one-particle one-hole ($1p1h$) process, (3) multinucleon, or two-particle two-hole ($2p2h$), hard scattering processes, (4) hard scattering in pion production processes, (5) higher invariant mass ($W$) and \dword{nc} processes, (6) \dword{fsi}, (7) neutrino flavor dependent differences. Uncertainties are intended to reflect current theoretical freedom, deficiencies in implementation, and/or current experimental knowledge.

The default nuclear model in \dword{genie} describing the initial state of nucleons in the nucleus is the Bodek-Ritchie global Fermi gas model~\cite{BodekRitchie}. There are significant deficiencies that are known in global Fermi gas models: these include a lack of consistent incorporation of the high-momentum tails in the nucleon momentum distribution that result from correlations among nucleons; the lack of correlation between location within the nucleus and momentum of the nucleon; and an incorrect relationship between momentum and energy of the off-shell, bound nucleon within the nucleus. They have also been shown to agree poorly with neutrino-nucleus scattering data~\cite{Wilkinson:2016wmz}. \dword{genie} modifies the nucleon momentum distribution empirically to account for short-range correlation effects, which populates the high-momentum tail above the Fermi cutoff, but the other deficiencies persist. Alternative initial state models, such as spectral functions~\cite{Benhar:1994hw,Nieves:2004wx}, the mean field model of GiBUU~\cite{Gallmeister:2016dnq}, or continuum random phase approximation (CRPA) calculations~\cite{Pandey:2014tza} may provide better descriptions of the nuclear initial state~\cite{Sobczyk:2017mts}, but are not considered further here.

The primary uncertainties considered in $1p1h$ interactions ($\nu_{l}+n \rightarrow l^{-}+p$, $\bar{\nu}_{l}+p \rightarrow l^{+}+n$) are the axial form factor of the nucleon and the nuclear screening---from the so-called \dword{rpa} calculations---of low momentum transfer reactions. The Valencia group's~\cite{nieves1,nieves2} description of \dword{rpa} comes from summation of $W^\pm$ self-energy terms. In practice, this modifies the $1p1h$ (quasielastic) cross section in a non-trivial way, with associated uncertainties presented in Ref.~\cite{nieves_uncert}, which were evaluated as a function of $Q^2$. Here we use T2K's 2017/8 parameterization of the Valencia RPA effect~\cite{Abe:2018wpn}. The shape of the correction and error is parameterized with a third-order Bernstein polynomial up to $Q^2=1.2\text{ GeV}^2$ where the form transitions to a decaying exponential. The BeRPA (Bernstein RPA) function has three parameters controlling the behavior at increasing $Q^2$ (A, B and D), a fourth parameter (E) that controls the high-$Q^2$ tail, and a fifth (U), which changes the position at which the behaviour changes from polynomial to exponential. The BeRPA parameterization modifies the central value of the model prediction, as decribed in Table~\ref{table:NuXSecKnobs_Central}. BeRPA parameters E and U are not varied in the analysis described here, the parameters A and B have a prefit uncertainty of 20\%, and D has a prefit unertainty of 15\%. The axial form factor parameterization we use, a dipole, is known to be inadequate~\cite{Meyer:2016oeg}. However, the convolution of BeRPA uncertainties with the limited axial form factor uncertainties do provide more freedom as a function of $Q^2$, and the two effects combined likely provide adequate freedom for the $Q^2$ shape in quasielastic events. BBBA05 vector form factors are used~\cite{bbba05}.

The $2p2h$ contribution to the cross section comes from the Valencia model~\cite{nieves1,nieves2}, the implementation in \dword{genie} is described in Ref.~\cite{Schwehr:2016pvn}. However, \minerva~\cite{Rodrigues:2015hik} and \nova~\cite{NOvA:2018gge} have shown that this model underpredicts observed event rates on carbon. The extra strength from the ``\minerva tune'' to $2p2h$ is applied as a two-dimensional Gaussian in $(q_0,q_3)$ space, where $q_0$ is the energy transfer from the leptonic system, and $q_3$ is the magnitude of the three momentum transfer) to fit reconstructed \minerva CC-inclusive data~\cite{Rodrigues:2015hik}. Reasonable predictions of \minerva's data are found by attributing the missing strength to any of $2p2h$ from $np$ initial state pairs, $2p2h$ from $nn$ initial state pairs, or $1p1h$ (quasielastic) processes.  The default tune uses an enhancement of the $np$ and $nn$ initial strengths in the ratio predicted by the Valencia model, and alternative systematic variation tunes (``MnvTune'' 1-3) attribute the missing strength to the individual interaction processes above. We add uncertainties for the energy dependence of this missing strength based on the \minerva results~\cite{Rodrigues:2015hik}, and assume a generic form for the energy dependence of the cross section using the ``A'' and ``B'' terms taken from Ref.~\cite{llewelyn-smith}. These uncertainties are labeled $E_{2p2h}$ and are separated for neutrinos and antineutrinos.
We add uncertainties on scaling the $2p2h$ prediction from carbon to argon on electron-scattering measurements of short-range correlated (SRC) pairs taken on multiple targets~\cite{Colle:2015ena}, separately for neutrinos (ArC$2p2h$ $\nu$) and antineutrinos (ArC$2p2h$ $\bar{\nu}$).

\dword{genie} uses a modified version of the Rein-Sehgal (R-S) model for pion production~\cite{Rein:1980wg}, including only the 16 resonances recommended by the Particle Data group~\cite{Tanabashi:2018oca}, and excluding interferences between resonances. The cross section is cut off at invariant masses, $W \geq 1.7$ GeV (2 GeV in the original R-S model). No in-medium modifications to the resonances are included, and by default they decay isotropically in their rest frame, although there is a parameter denoted here as ``$\theta_{\pi}$ from $\Delta$-decay'', for changing the angular distribution of pions produced through $\Delta$ resonance decays to match the experimentally observed distributions used in the original R-S paper~\cite{Rein:1980wg}. Resonance decays to $\eta$ and $\gamma$ (plus a nucleon) are included from Ref.~\cite{Tanabashi:2018oca}. We use a tuning of the \dword{genie} model to reanalyzed neutrino--deuterium bubble chamber data~\cite{Wilkinson:2014yfa,Rodrigues:2016xjj} as our base model, as noted in Table~\ref{table:NuXSecKnobs_Central}. We note that an improved Rein-Sehgal-like resonance model has been developed~\cite{minoo}, and has been implemented in Monte Carlo generators, although is not used as the default model in the present work.

The \dword{dis} model implemented in \dword{genie} uses the Bodek-Yang parametrization~\cite{Bodek:2002ps}, using GRV98 parton distribution functions~\cite{Gluck:1998xa}. Hadronization is described by the AKGY model~\cite{Yang:2009zx}, which uses the KNO scaling model~\cite{Koba:1972ng} for invariant masses $W \leq 2.3$ GeV and PYTHIA6~\cite{Sjostrand:2006za} for invariant masses $W \geq 3$ GeV, with a smooth transition between the two for intermediate invariant masses. A number of variable parameters affecting \dword{dis} processes are included in \dword{genie}, as listed in Table~\ref{table:NuXSecKnobs_Central}, and described in Ref.~\cite{Bodek:2002ps}. In \dword{genie}, the \dword{dis} model is extrapolated to all values of invariant mass, and replaces the non-resonant background to pion production in the R-S model.

The \nova experiment~\cite{nova_2018} developed uncertainties beyond those provided by \dword{genie} to describe their single pion to \dword{dis} transition region data. We follow their findings, and implement separate, uncorrelated uncertainties for all perturbations of 1, 2, and $\geq 3$ pion final states, CC/NC, neutrinos/antineutrinos, and interactions on protons/neutrons, with the exception of CC neutrino 1-pion production, where interactions on protons and neutrons are merged, following \cite{Rodrigues:2016xjj}, which modifies the central value of the model prediction, as listed in Table~\ref{table:NuXSecKnobs_Central}. This leads to 23 distinct uncertainty channels with a label to denote the process it affects: NR [$\nu$,$\bar{\nu}$] [CC,NC] [n,p] [1$\pi$,2$\pi$,3$\pi$]. Each channel has an uncertainty of 50\% for $W \leq 3$ GeV, and an uncertainty which drops linearly above $W = 3$ GeV until it reaches a flat value of 5\% at $W = 5$ GeV, where external measurements better constrain this process.

\dword{genie} includes a large number of final state uncertainties on its final state cascade model~\cite{Dytman:2011zz,Dytman:2015taa,intranuke_2009}, which are summarized in Table~\ref{table:HadTranspKnobs}. A recent comparison of the underlying interaction probabilities used by GENIE is compared with other available simulation packages in Ref.~\cite{PinzonGuerra:2018rju}.

The cross sections include terms proportional to the lepton mass, which are significant contributors at low energies where quasielastic processes dominate.  Some of the form factors in these terms have significant uncertainties in the nuclear environment.  Ref.~\cite{Day:2012gb} ascribes the largest possible effect to the presence of poorly constrained second-class current vector form factors in the nuclear environment, and proposes a variation in the cross section ratio of $\sigma_\mu/\sigma_e$ of $\pm 0.01/{\rm\textstyle Max}(0.2~{\rm\textstyle GeV},E_\nu)$ for neutrinos and $\mp 0.018/{\rm\textstyle Max}(0.2~ {\rm\textstyle GeV},E_\nu)$ for antineutrinos.  Note the anticorrelation of the effect in neutrinos and antineutrinos. This parameter is labeled $\nu_{e}$/$\bar{\nu}_{e}$ norm.

An additional normalization uncertainty (\dword{nc} norm.) of 20\% is applied to all \dword{nc} events at the \dword{nd} in this analysis to investigate whether the small contamination of \dword{nc} events which passed the simple selection cuts had an effect on the analysis. Although a similar systematic could have been included (uncorrelated) at the \dword{fd}, it was not in this analysis.

Finally, some electron-neutrino interactions occur at four-momentum transfers where a corresponding muon-neutrino interaction is kinematically forbidden, therefore the nuclear response has not been constrained by muon-neutrino cross-section measurements.  This region at lower neutrino energies has a significant overlap with the Bodek-Ritchie tail of the nucleon momentum distribution in the Fermi gas model~\cite{BodekRitchie}. There are significant uncertainties in this region, both from the form of the tail itself and from the lack of knowledge about the effect of RPA and $2p2h$ in this region.
Here, a 100\% uncertainty is applied in the phase space present for $\nu_e$ but absent for $\nu_\mu$ (labeled $\nu_{e}$ phase space (PS)).

The complete set of interaction model uncertainties includes \dword{genie} implemented uncertainties 
(Tables~\ref{table:NuXSecKnobs} and \ref{table:HadTranspKnobs}), 
and new uncertainties developed for this effort (Table~\ref{tab:nuintsystlist}) which represent uncertainties beyond those implemented in the \dword{genie} generator.  

\begin{table}[ptb]
\centering
\global\long\def\arraystretch{1.75}
\scalebox{0.9}{
\begin{tabular}{ll}
\hline
Description & 1$\sigma$  \tabularnewline
\hline\hline
\textbf{Quasielastic}&\tabularnewline
$M_{\mathrm{A}}^{\mathrm{QE}}$, Axial mass for CCQE & ${}^{+0.25}_{-0.15}$~GeV \tabularnewline
QE FF, CCQE vector form factor shape & N/A \tabularnewline
$p_{\mathrm{F}}$ Fermi surface momentum for Pauli blocking &   $\pm$30\% \tabularnewline
\hline
\textbf{Low $\mathbf{W}$}&\tabularnewline
$M_{\mathrm{A}}^{\mathrm{RES}}$, Axial mass for CC resonance & $\pm$0.05~GeV \tabularnewline
$M_{\mathrm{V}}^{\mathrm{RES}}$ Vector mass for CC resonance & $\pm$10\% \tabularnewline
$\Delta$-decay ang., $\theta_{\pi}$ from $\Delta$ decay (isotropic $\rightarrow$ R-S) & N/A \tabularnewline
\hline
\textbf{High $\mathbf{W}$} (BY model)&\tabularnewline
$A_{\mathrm{HT}}$, higher-twist in scaling variable $\xi_{w}$  & $\pm$25\% \tabularnewline
$B_{\mathrm{HT}}$, higher-twist in scaling variable $\xi_{w}$  & $\pm$25\% \tabularnewline
$C_{\mathrm{V1u}}$, valence GRV98 PDF correction & $\pm$30\% \tabularnewline
$C_{\mathrm{V2u}}$, valence GRV98 PDF correction & $\pm$40\% \tabularnewline
\hline
\textbf{Other neutral current}&\tabularnewline
$M_{\mathrm{A}}^{\mathrm{NCRES}}$, Axial mass for \dword{nc} resonance & $\pm$10\% \tabularnewline
$M_{\mathrm{V}}^{\mathrm{NCRES}}$, Vector mass for \dword{nc} resonance & $\pm$5\% \tabularnewline
\hline
\end{tabular}
}
\\[2pt]
\caption[Neutrino interaction cross-section systematic parameters considered in GENIE]
{Neutrino interaction cross-section systematic parameters considered in \dword{genie}. \dword{genie} default central values and uncertainties are used for all parameters except the CC resonance axial mass. The central values are the \dword{genie} nominals, and the 1$\sigma$ uncertainty is as given. Missing \dword{genie} parameters were omitted where uncertainties developed for this analysis significantly overlap with the supplied \dword{genie} freedom, the response calculation was too slow, or the variations were deemed unphysical.
}
\label{table:NuXSecKnobs}
\end{table}

\begin{table}[ptb]
\centering

\global\long\def\arraystretch{1.75}
\begin{tabular}{ll}
\hline
Description  & 1$\sigma$  \tabularnewline
\hline\hline
N. CEX, Nucleon charge exchange probability  & $\pm$50\%  \tabularnewline
N. EL, Nucleon elastic reaction probability  & $\pm$30\%  \tabularnewline
N. INEL, Nucleon inelastic reaction probability  & $\pm$40\%  \tabularnewline
N. ABS, Nucleon absorption probability  & $\pm$20\%  \tabularnewline
N. PROD, Nucleon $\pi$-production probability  & $\pm$20\%  \tabularnewline
$\pi$ CEX, $\pi$ charge exchange probability  & $\pm$50\%  \tabularnewline
$\pi$ EL, $\pi$ elastic reaction probability  & $\pm$10\%  \tabularnewline
$\pi$ INEL, $\pi$ inelastic reaction probability  & $\pm$40\%  \tabularnewline
$\pi$ ABS, $\pi$ absorption probability  & $\pm$20\%  \tabularnewline
$\pi$ PROD, $\pi$ $\pi$-production probability  & $\pm$20\%  \tabularnewline
\hline
\end{tabular}\\[2pt] \caption[Intra-nuclear hadron transport systematic parameters implemented in GENIE]
{The intra-nuclear hadron transport systematic parameters implemented in \dword{genie} with associated uncertainties considered in
this work. Note that the `mean free path' parameters are omitted for both N-N and $\pi$-N interactions as they produced unphysical variations in observable analysis variables. Table adapted from Ref~\cite{Andreopoulos:2015wxa}.
}
\label{table:HadTranspKnobs}
\end{table}

Tunes which are applied to the default model, using the dials described, which represent known deficiencies in \dword{genie}'s description of neutrino data, are listed in Table~\ref{table:NuXSecKnobs_Central}.
\begin{table}[ptb]
\centering
\global\long\def\arraystretch{1.75}
\scalebox{0.9}{
\begin{tabular}{ll}
\hline
Description  & Value  \tabularnewline
\hline\hline
\textbf{Quasielastic} & \tabularnewline
BeRPA & \tabularnewline
$A$ controls low $Q^2$ & $A: 0.59$ \tabularnewline
$B$ controls low-mid $Q^2$ & $B: 1.05$ \tabularnewline
$D$ controls mid $Q^2$ & $D: 1.13$ \tabularnewline
$E$ controls high $Q^2$ fall-off & $E: 0.88$ \tabularnewline
$U$ controls transition from polynomial to exponential & $U: 1.20$ \tabularnewline
\hline
\textbf{$2p2h$}&\tabularnewline
$q0,q3$ dependent correction to $2p2h$ events&\\
\hline
\textbf{Low $\mathbf{W}$ single pion production} & \tabularnewline
Axial mass for CC resonance in \dword{genie}& $0.94$ \tabularnewline
Normalization of CC1$\pi$ non-resonant interaction & $0.43$  \tabularnewline
\hline
\end{tabular}
}
\\[2pt]
\caption[Neutrino interaction cross-section systematic parameters that receive a central-value tune]{Neutrino interaction cross-section systematic parameters that receive a central-value tune and modify the nominal event rate predictions.}
\label{table:NuXSecKnobs_Central}
\end{table}

The way model parameters are treated in the analysis is described by three categories:
\begin{itemize}
\item Category 1: expected to be constrained with on-axis data; uncertainties are implemented in the same way for \dword{nd} and \dword{fd}.
\item Category 2: implemented in the same way for \dword{nd} and \dword{fd}, but on-axis \dword{nd} data alone is not sufficient to constrain these parameters. They may be constrained by additional \dword{nd} samples in future analyses, such as off-axis measurements.
\item Category 3: implemented only in the \dword{fd}.  Examples are parameters which only affect $\nu_e$ and $\overline{\nu}_e$ rates which are small and difficult to precisely isolate from background at the \dword{nd}.
\end{itemize}
All \dword{genie} uncertainties (original or modified), given in Tables~\ref{table:NuXSecKnobs} and~\ref{table:HadTranspKnobs}, are all treated as Category 1.
Table~\ref{tab:nuintsystlist}, which describes the uncertainties beyond those available within \dword{genie}, includes a column identifying which of these categories describes the treatment of each additional uncertainty. 

\begin{table}[htbp]
  \begin{center}
\scalebox{0.9}{
\begin{tabular}{lcc}\hline
Uncertainty & Mode & Category  \\  \hline \hline
BeRPA [$A$,$B$,$D$] & $1p1h$/QE & 1  \\
ArC$2p2h$ [$\nu$,$\bar{\nu}$] & $2p2h$ & 1 \\
$E_{2p2h}$ [A,B] [$\nu$,$\bar{\nu}$] & $2p2h$ & 2 \\
NR [$\nu$,$\bar{\nu}$] [CC,NC] [n,p] [1$\pi$,2$\pi$,3$\pi$] & Non-res. pion & 1 \\
$\nu_e$ PS & $\nu_e$,$\overline{\nu}_e$ inclusive &  3 \\
$\nu_e$/$\overline{\nu}_e$ norm & $\nu_e$,$\overline{\nu}_e$ inclusive & 3 \\
NC norm. & NC & 2*\\
\hline
\end{tabular}
} 

\caption[List of extra interaction model uncertainties in addition to those provided by GENIE]{List of extra interaction model uncertainties in addition to those provided by GENIE, and the category to which they belong in the analysis. Note that in this analysis, the NC norm. systematic is not applied at the \dword{fd}, as described in the text.}
\label{tab:nuintsystlist}
\end{center}
\end{table}

%% file: sections/nd.tex
\section{The Near Detector Simulation and Reconstruction}\label{sec:nd}
The \dword{nd} hall will be located at \dword{fnal}, 574 m from where the protons hit the beam target, and 60 m underground. The baseline design for the \dword{dune} \dword{nd} system consists of a \dword{lartpc} with a downstream magnetized \dword{mpd}, and an on-axis beam monitor. Additionally, it is planned for the \dword{lartpc} and \dword{mpd} to be movable perpendicular to the beam axis, to take measurements at a number of off-axis angles. The use of off-axis angles is complementary to the on-axis analysis described in this work through the DUNE-PRISM concept, originally developed in the context of the J-PARC neutrino beamline in Ref.~\cite{Bhadra:2014oma}. We note that there are many possible \dword{nd} samples which are not included in the current analysis, but which may either help improve the sensitivity in future, or will help control uncertainties to the level assumed here. These include: neutrino--electron scattering studies, which can independently constrain the flux normalization to $\sim$2\%~\cite{dune_nue}; additional flux constraints from the low-$\nu$ method, which exploits the fact that the low energy transfer (low-$\nu$) cross section is roughly constant with neutrino energy~\cite{Quintas:1992yv,Yang:2000ju,Tzanov:2005kr,Adamson:2009ju,DeVan:2016rkm,Ren:2017xov}; and using interactions on the \dword{gar} in the \dword{mpd}. There is also the potential to include events where the muon does not pass through the \dword{mpd}, e.g. using multiple Coulomb scattering to estimate the muon momentum~\cite{Abratenko:2017nki}.

The \dword{lartpc} is a modular detector based on the ArgonCube design~\cite{argoncube_loi}, with fully-\threed pixelated readout~\cite{Dwyer:2018phu} and optical segmentation~\cite{arclight}. These features greatly reduce reconstruction ambiguities that hamper monolithic, projective-readout \dwords{tpc}, and enable the \dword{nd} to function in the high-intensity environment of the \dword{dune} \dword{nd} site. Each module is itself a \dword{lartpc} with two anode planes and a shared central cathode. The active dimensions of each module are $1 \times 3 \times 1$~m ($x \times y \times z$), where the $z$ direction is along the neutrino beam axis, and the $y$ direction points upward. Charge drifts in the $\pm x$ direction, with a maximum drift distance of 50 cm for ionization electrons. The full \dword{lar} detector consists of an array of modules in a single cryostat. The minimum active size required for full containment of hadronic showers in the \dword{lbnf} beam is $3 \times 4 \times 5$~m. High-angle muons can also be contained by extending the width to 7 m. For this analysis, 35 modules are arranged in an array 5 modules deep in the $z$ direction and 7 modules across in $x$ so that the total active dimensions are $7 \times 3 \times 5$~m. The total active \dword{lar} volume is $105$~m$^{3}$, corresponding to a mass of 147 tons.

The \dword{mpd} used in the analysis consists of a high-pressure \dword{gartpc} in a cylindrical pressure vessel at 10 bar, surrounded by a granular, high-performance electromagnetic calorimeter, which sits immediately downstream of the \dword{lar} cryostat. The pressure vessel is 5 m in diameter and 5 m long. The \dword{tpc} is divided into two drift regions by a central cathode, and filled with a 90\%:10\% Ar:CH$_{4}$ gas mixture, such that 97\% of neutrino interactions will occur on the Ar target. The \dword{gartpc} is described in detail in Ref.~\cite{Abi:2020evt}. The \dword{ecal} is composed of a series of absorber layers followed by arrays of scintillator and is described in Ref. ~\cite{Emberger:2018pgr}. The entire \dword{mpd} sits inside a magnetic field, which allows the \dword{mpd} to precisely measure the momentum and discriminate the sign of particles passing through it.

Neutrino interactions are simulated in the active volume of the \dword{lartpc}. The propagation of neutrino interaction products through the \dword{lartpc} and \dword{mpd} detector volumes is simulated using a \dword{geant4}-based model~\cite{Agostinelli:2002hh}. Pattern recognition and reconstruction software has not yet been developed for the \dword{nd}. Instead, we perform a parameterized reconstruction based on true energy deposits in active detector volumes as simulated by \dword{geant4}.

\begin{figure}
 \includegraphics[width=0.98\linewidth]{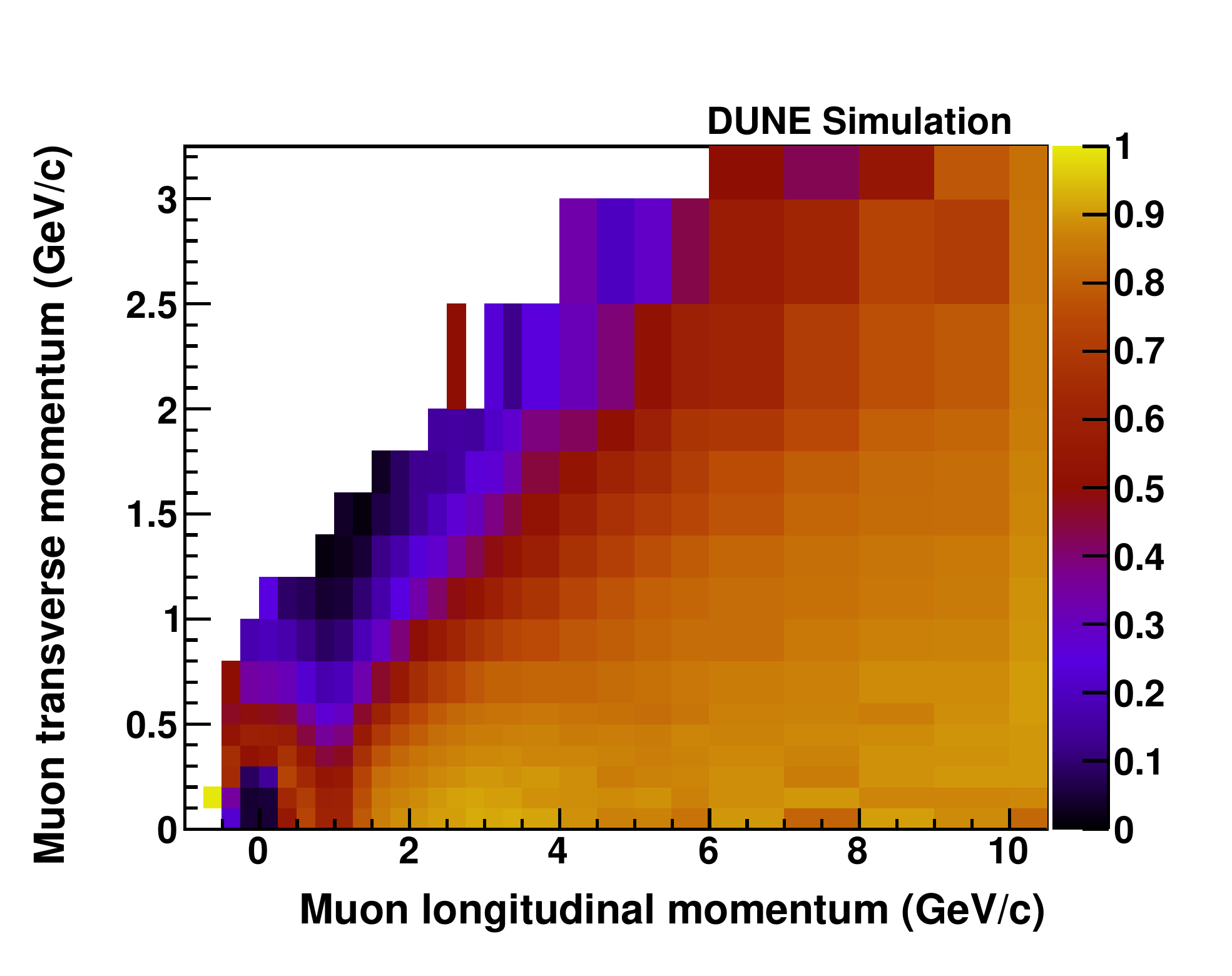}\\
 \includegraphics[width=0.98\linewidth]{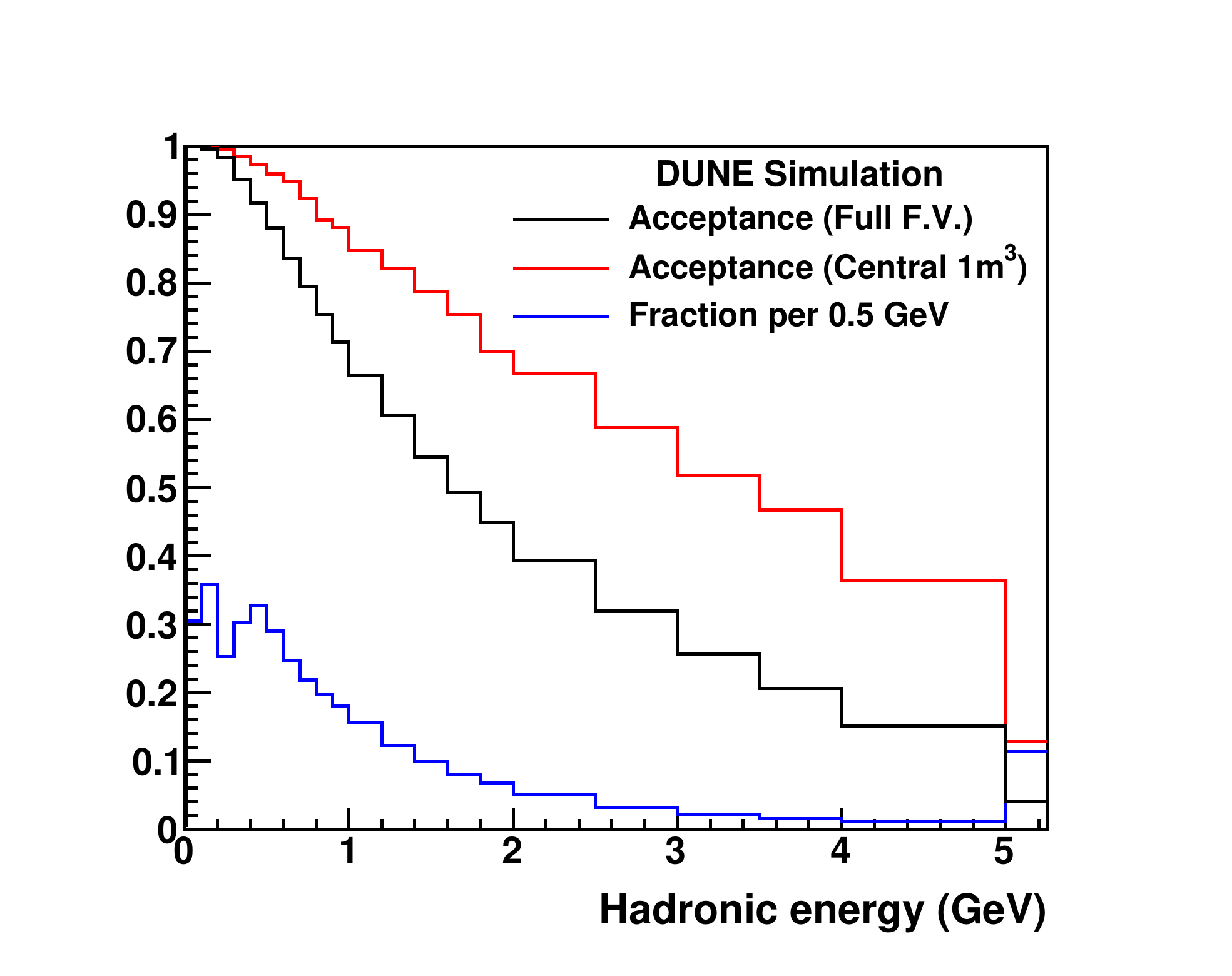}
 \caption{Top: LAr+MPD acceptance for $\nu_{\mu}$ \dword{cc} events as a function of muon transverse and longitudinal momentum. Bottom: Acceptance as a function of hadronic energy; the black line is for the full Fiducial Volume (FV) while the red line is for a $1 \times 1 \times 1$~m$^{3}$ volume in the center, where the acceptance is higher due to the better hadron containment. The blue curve shows the expected distribution of true hadronic energy in the \dword{dune} \dword{nd} flux normalized to unity; 56\% of events have hadronic energy below 1 GeV where the acceptance is high.}
 \label{fig:NDacceptance}
\end{figure}
The analysis described here uses events originating in the \dword{lar} component, within a \dword{fv} that excludes 50 cm from the sides and upstream edge, and 150 cm from the downstream edge of the active region, for a total of $6 \times 2 \times 3$~m$^{2}$. Muons with kinetic energy greater than $\sim$1 GeV typically exit the \dword{lar}. An energetic forward-going muon will pass through the \dword{ecal} and into the gaseous \dword{tpc}, where its momentum and charge are reconstructed by curvature. For these events, it is possible to differentiate between $\mu^{+}$ and $\mu^{-}$ event by event. Muons that stop in the \dword{lar} or \dword{ecal} are reconstructed by range. Events with wide-angle muons that exit the \dword{lar} and do not match to the \dword{gartpc} are rejected, as the muon momentum is not reconstructed. The asymmetric transverse dimensions of the \dword{lar} volume make it possible to reconstruct wide-angle muons with some efficiency.

The charge of muons stopping in the \dword{lar} volume cannot be determined event by event. However, the wrong-sign flux is predominantly concentrated in the high-energy tail, where leptons are more likely to be forward and energetic. In \dword{fhc} beam running, the wrong-sign background in the focusing peak is negligibly small, and $\mu^{-}$ is assumed for all stopping muon tracks. In \dword{rhc} beam running, the wrong-sign background is larger in the peak region. Furthermore, high-angle leptons are generally at higher inelasticity, which enhances the wrong-sign contamination in the contained muon subsample. To mitigate this, a Michel electron is required at the end of the track. The wrong-sign $\mu^{-}$ captures on Ar with 75\% probability, effectively suppressing the relative $\mu^{-}$ component by a factor of four.

True muons and charged pions are evaluated as potential muon candidates. The track length is determined by following the true particle trajectory until it undergoes a hard scatter or ranges out. The particle is classified as a muon if its track length is at least 1 m, and the mean energy deposit per centimeter of track length is less than 3 MeV. The mean energy cut rejects tracks with detectable hadronic interactions. The minimum length requirement imposes an effective threshold on true muons of about 200 MeV kinetic energy, but greatly suppresses potential NC backgrounds with low-energy, non-interacting charged pions. Charged-current events are required to have exactly one muon, and if the charge is reconstructed, it must be of the appropriate charge. 

As in the \dword{fd} reconstruction described in Section~\ref{sec:fd}, hadronic energy in the \dword{nd} is reconstructed by summing all charge deposits in the \dword{lar} active volume that are not associated with the muon. To reject events where the hadronic energy is poorly reconstructed due to particles exiting the detector, a veto region is defined as the outer 30 cm of the active volume on all sides. Events with more than 30 MeV total energy deposit in the veto region are excluded from the analysis. This leads to an acceptance that decreases as a function of hadronic energy, as shown in the bottom panel of Figure~\ref{fig:NDacceptance}. Neutron energy is typically not observed, resulting in poor reconstruction of events with energetic neutrons, as well as in events where neutrons are produced in secondary interactions inside the detector. The reconstructed neutrino energy is the sum of the reconstructed hadronic energy and the reconstructed muon energy.

\begin{figure*}
  \centering
  \includegraphics[width=0.8\linewidth]{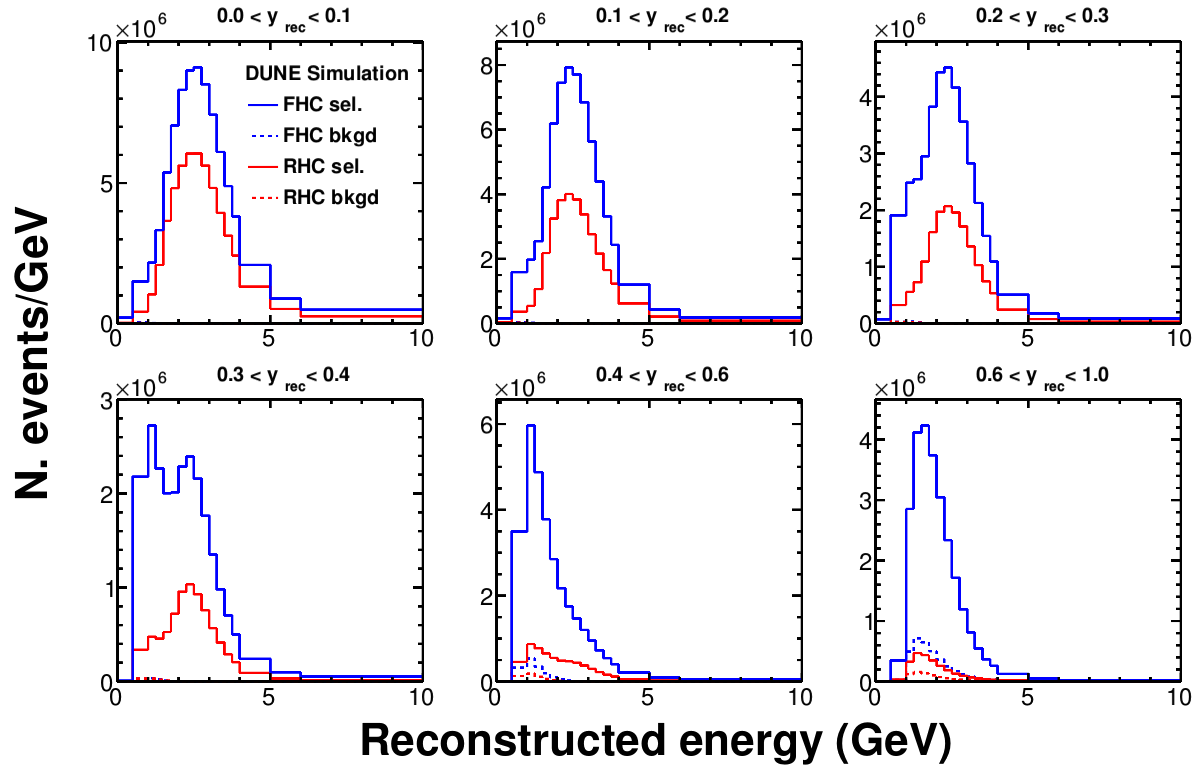}
  \caption{\dword{nd} samples in both \dword{fhc} (blue) and \dword{rhc} (red), shown in the reconstructed neutrino energy and reconstructed inelasticity binning used in the analysis, shown for a 7 year staged exposure, with an equal split between \dword{fhc} and \dword{rhc}. Backgrounds are also shown (dashed lines), which are dominated by \dword{nc} events, although there is some contribution from wrong-sign \numu background events in \dword{rhc}.}
 \label{fig:nd_samples}
\end{figure*}

The oscillation analysis presented here includes samples of $\nu_{\mu}$ and $\bar{\nu}_{\mu}$ charged-current interactions originating in the \dword{lar} portion of the ND, as shown in Figure~\ref{fig:nd_samples}. These samples are binned in two dimensions as a function of reconstructed neutrino energy and inelasticity, $y_{\mathrm{rec}} = 1 - E^{\mathrm{rec}}_{\mu}/E^{\mathrm{rec}}_{\nu}$, where $E^{\mathrm{rec}}_{\mu}$ and $E^{\mathrm{rec}}_{\nu}$ are the reconstructed muon and neutrino energies, respectively. Backgrounds to $\nu^{\bracketbar}_{\mu}$ \dword{cc} arise from \dword{nc} $\pi^{\pm}$ production where the pion leaves a long track and does not shower. Muons below about 400 MeV kinetic energy have a significant background from charged pions, so these \dword{cc} events are excluded from the selected sample. Wrong-sign contamination in the beam is an additional background, particularly at low reconstructed neutrino energies in RHC.

%% file: sections/fd.tex
\section{The Far Detector Simulation and Reconstruction}\label{sec:fd}

The 40-kt \dword{dune} \dword{fd} consists of four separate \dword{lartpc} \dwords{detmodule}, each with a \dword{fv} of at least \nominalmodsize, installed $\sim$1.5 km underground at the \mbox{\dword{surf}}~\cite{Abi:2018dnh}. \dword{dune} is committed to deploying both single-phase~\cite{Abi:2018alz} and dual-phase~\cite{Abi:2018rgm} \dword{lartpc} technologies, and is investigating advanced detector designs for the fourth \dword{detmodule}. As such, the exact order of construction and number of modules of each design is unknown. In this work, the \dword{fd} reconstruction performance is assessed assuming a single-phase design for all four modules, which does not fully exploit the benefits of different technologies with independent systematics in the sensitivity studies. A full simulation chain has been developed, from the generation of neutrino events in a \dword{geant4} model of the \dword{fd} geometry, to efficiencies and reconstructed neutrino energy estimators of all samples used in the sensitivity analysis.

The total active \dword{lar} volume of each single-phase \dword{dune} \dword{fd} \dword{detmodule} is 13.9 m long, 12.0 m high and 13.3 m wide, with the 13.3 m width in the drift direction subdivided into four independent drift regions, with two shared cathodes. Full details of the single-phase \dword{detmodule} design can be found in Ref.~\cite{Abi:2020loh}. The total active volume of each module is $\sim$13 kt, the \dword{fv} of at least \nominalmodsize is defined by studies of neutrino energy resolution, using the neutrino energy estimators described below. At the anode, there are two wrapped-wire readout induction planes, which are offset by $\pm$35.7$^{\circ}$ to the vertical, and a vertical collection plane.

Neutrino interactions of all flavors are simulated such that weights can be applied to produce samples for any set of oscillation parameters. The interaction model described in Section~\ref{sec:nuint} was used to model the neutrino-argon interactions in the volume of the cryostat, and the final-state particles are propagated in the detector through \dword{geant4}. The electronics response to the ionization electrons and scintillation light is simulated to produce digitized signals in the wire planes and \dwords{pd} respectively.

Raw detector signals are processed using algorithms to remove the impact of the \dword{lartpc} electric field and electronics response from the measured signal, to identify hits, and to cluster hits that may be grouped together due to proximity in time and space. Clusters from different wire planes are matched to form high-level objects such as tracks and showers. These high level objects are used as inputs to the neutrino energy reconstruction algorithm.

The energy of the incoming neutrino in \dword{cc} events is estimated by adding the lepton and hadronic energies reconstructed using the Pandora toolkit~\cite{Marshall:2015rfa,Acciarri:2017hat}. If the event is selected as $\nu_{\mu}$ \dword{cc}, the neutrino energy is estimated as the sum of the energy of the longest reconstructed track and the hadronic energy. The energy of the longest reconstructed track is estimated from its range if the track is contained in the detector. If the longest track exits the detector, its energy is estimated from multiple Coulomb scattering. The hadronic energy is estimated from the charge of reconstructed hits that are not in the longest track, and corrections are applied to each hit charge for recombination and the electron lifetime. An additional correction is made to the hadronic energy to account for missing energy due to neutral particles and final-state interactions.

If the event is selected as $\nu_{e}$ \dword{cc}, the energy of the neutrino is estimated as the sum of the energy of the reconstructed electromagnetic (EM) shower with the highest energy and the hadronic energy. The former is estimated from the charges of the reconstructed hits in the shower, and the latter from the hits not in the shower; the recombination and electron lifetime corrections are applied to the charge of each hit. The same hadronic shower energy calibration is used for both $\nu$ and $\bar{\nu}$ based on a sample of $\nu$ and $\bar{\nu}$ events. 

In the energy range of 0.5--4 GeV that is relevant for oscillation measurements, the observed neutrino energy resolution is $\sim$15--20\%, depending on lepton flavor and reconstruction method. The muon energy resolution is 4\% for contained tracks and 18\% for exiting tracks. The electron energy resolution is approximately $4\% \oplus 9\%/\sqrt{E}$, with some shower leakage that gives rise to a non-Gaussian tail that is anticorrelated with the hadronic energy measurement. The hadronic energy resolution is 34\%, which could be further improved by identifying individual hadrons, adding masses of charged pions, and applying particle-specific recombination corrections. It may also be possible to identify final-state neutrons by looking for neutron-nucleus scatters, and use event kinematics to further inform the energy estimate. These improvements are under investigation and are not included in this analysis.

Event classification is carried out through image recognition techniques using a convolutional neural network, named \dword{cvn}. Detailed descriptions of the \dword{cvn} architecture can be found in Ref.~\cite{cvn_paper}. The primary goal of the \dword{cvn} is to efficiently and accurately produce event selections of the following interactions: $\nu_{\mu}$ \dword{cc} and $\nu_{e}$ \dword{cc} in \dword{fhc}, and $\bar{\nu}_\mu$ \dword{cc} and $\bar{\nu}_e$ \dword{cc} in \dword{rhc}.

\begin{figure}[htbp] 
  \centering
  \includegraphics[width=0.8\linewidth]{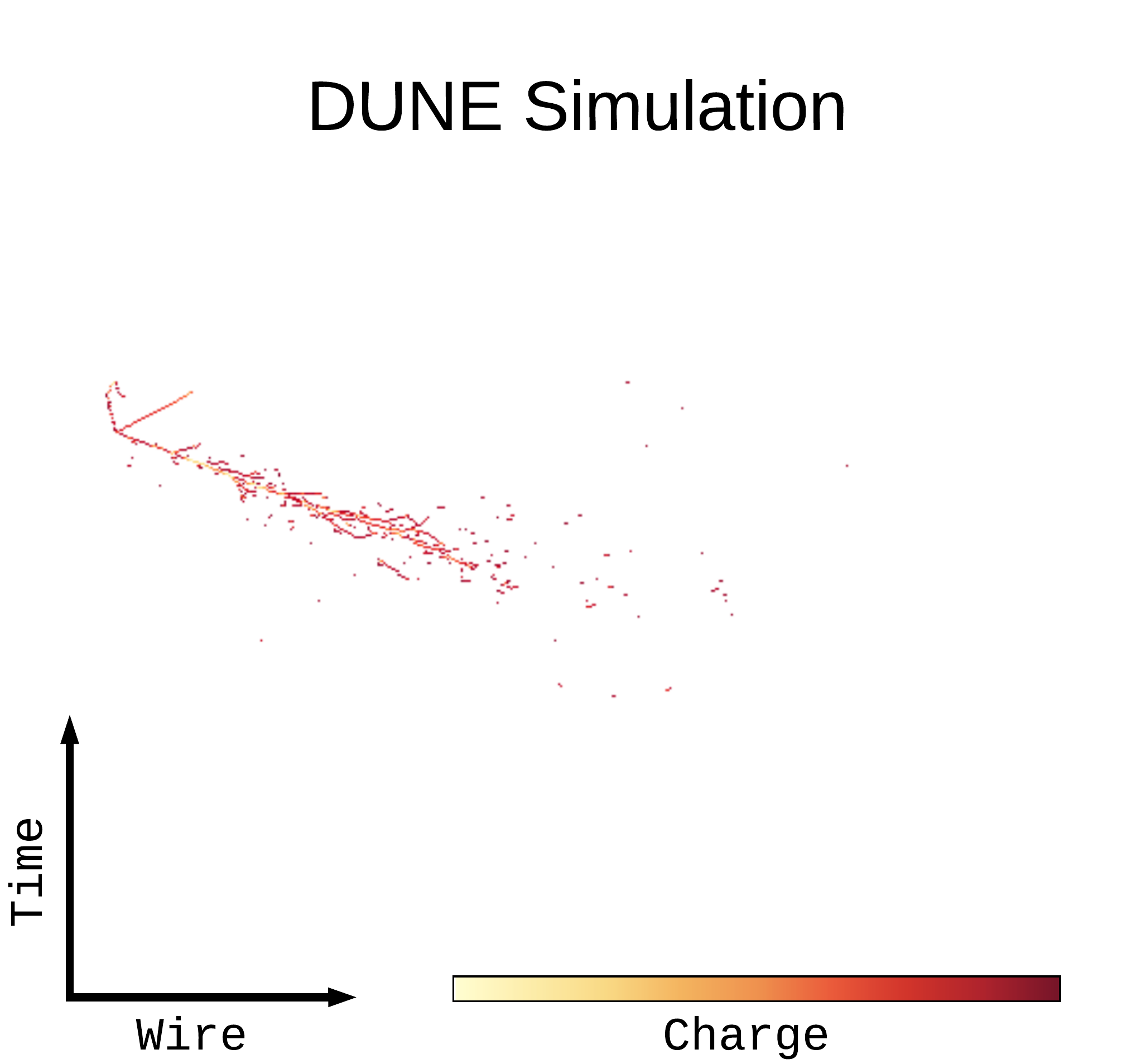}
  \caption[A simulated \SI{2.2}{GeV} \nue CC interaction viewed by collection wires in the SP \lartpc]{A simulated \SI{2.2}{GeV} $\nu_{e}$ \dword{cc} interaction shown in the collection view of the \dword{dune} \dwords{lartpc}. The horizontal axis shows the wire number of the readout plane and the vertical axis shows time. The colorscale shows the charge of the energy deposits on the wires. The interaction looks similar in the other two views. Reproduced from Ref.~\cite{cvn_paper}.}
  \label{fig:views}
\end{figure}
In order to build the training input to the \dword{dune} \dword{cvn} three images of the neutrino interactions are produced, one for each of the three readout views of the \dword{lartpc}, using the reconstructed hits on individual wire planes. Each pixel contains information about the integrated charge in that region. An example of a simulated 2.2\,GeV $\nu_{e}$ \dword{cc} interaction is shown in a single view in Figure~\ref{fig:views} demonstrating the fine-grained detail available from the \dword{lartpc} technology.

The \dword{cvn} is trained using approximately three million simulated neutrino interactions. A statistically independent sample is used to generate the physics measurement sensitivities. The training sample is chosen to ensure similar numbers of training examples from the different neutrino flavors. Validation is performed to ensure that similar classification performance is obtained for the training and test samples to ensure that the \dword{cvn} is not overtrained.

\begin{figure}[htbp]
  \centering
  \includegraphics[width=0.98\linewidth]{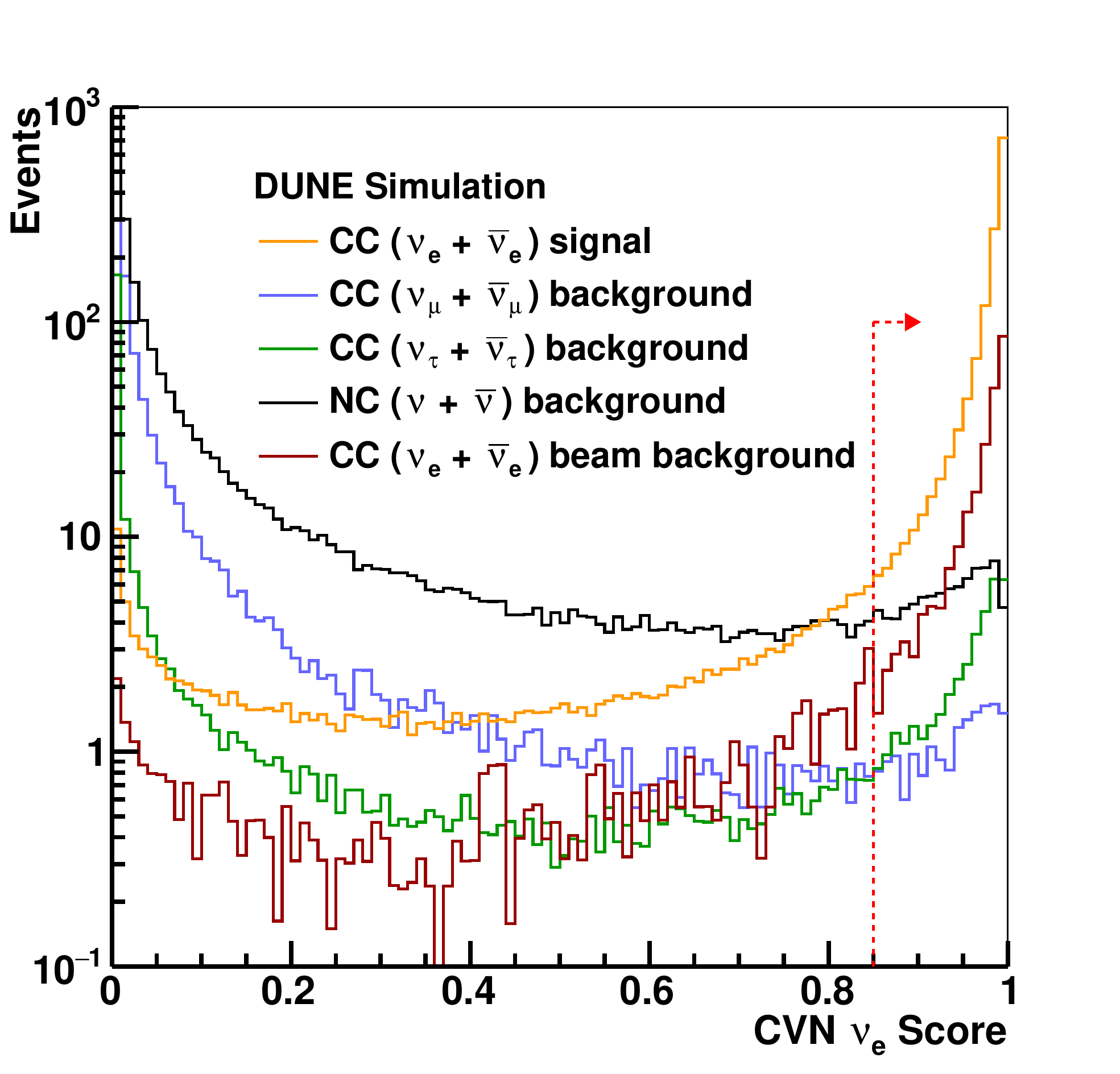}\\
  \includegraphics[width=0.98\linewidth]{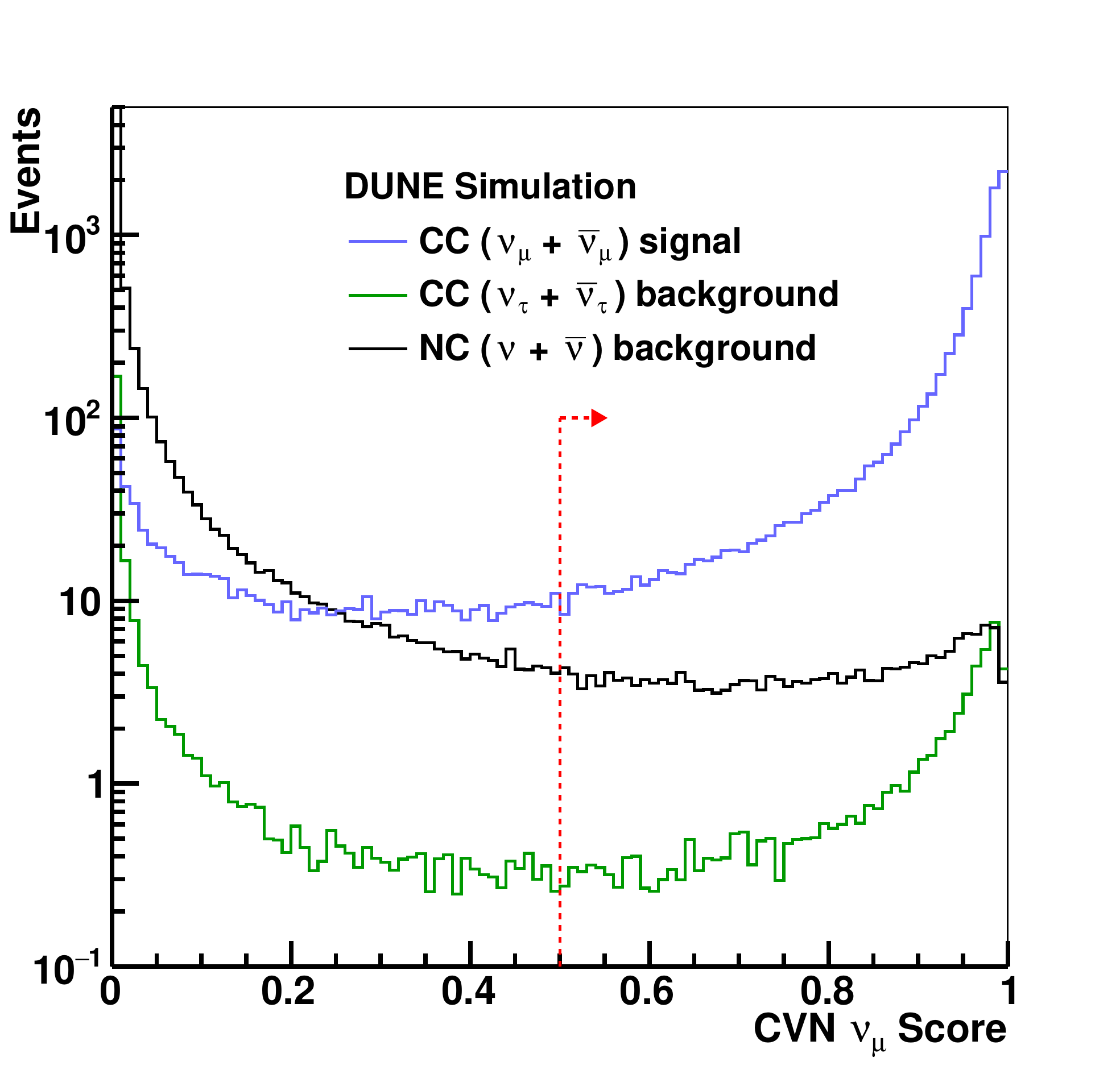} 
  \caption[The distribution of CVN \nue CC and \numu CC scores for \dword{fhc}]{The distribution of \dword{cvn} $\nu_e$ \dword{cc} (top) and $\nu_\mu$ \dword{cc} scores (bottom) for \dword{fhc} shown with a log scale. Reproduced from Ref.~\cite{cvn_paper}.}
  \label{fig:cvnprob}
\end{figure}
For the analysis presented here, we use the \dword{cvn} score for each interaction to belong to one of the following classes: $\nu_{\mu}$ \dword{cc}, $\nu_{e}$ \dword{cc}, $\nu_{\tau}$ \dword{cc} and \dword{nc}. The $\nu_{e}$ \dword{cc} score distribution, $P(\nu_e \textrm{ \dword{cc}})$, and the $\nu_\mu$ \dword{cc} score distribution, $P(\nu_\mu \textrm{ \dword{cc}})$, are shown in Figure~\ref{fig:cvnprob}. Excellent separation between the signal and background interactions is seen in both cases. The event selection requirement for an interaction to be included in the  $\nu_e$ \dword{cc} ($\nu_\mu$ \dword{cc}) is $P(\nu_e \textrm{ \dword{cc}}) > 0.85$ ($P(\nu_\mu \textrm{ \dword{cc}}) > 0.5$), optimized to produce the best sensitivity to \dword{cp} violation. Since all of the flavor classification scores must sum to unity, the interactions selected in the two event selections are completely independent. The same selection criteria are used for both \dword{fhc} and \dword{rhc} beam running.

\begin{figure*}[htbp]
  \centering
  \includegraphics[width=0.8\linewidth, trim={0cm 0cm 0cm 2cm}, clip]{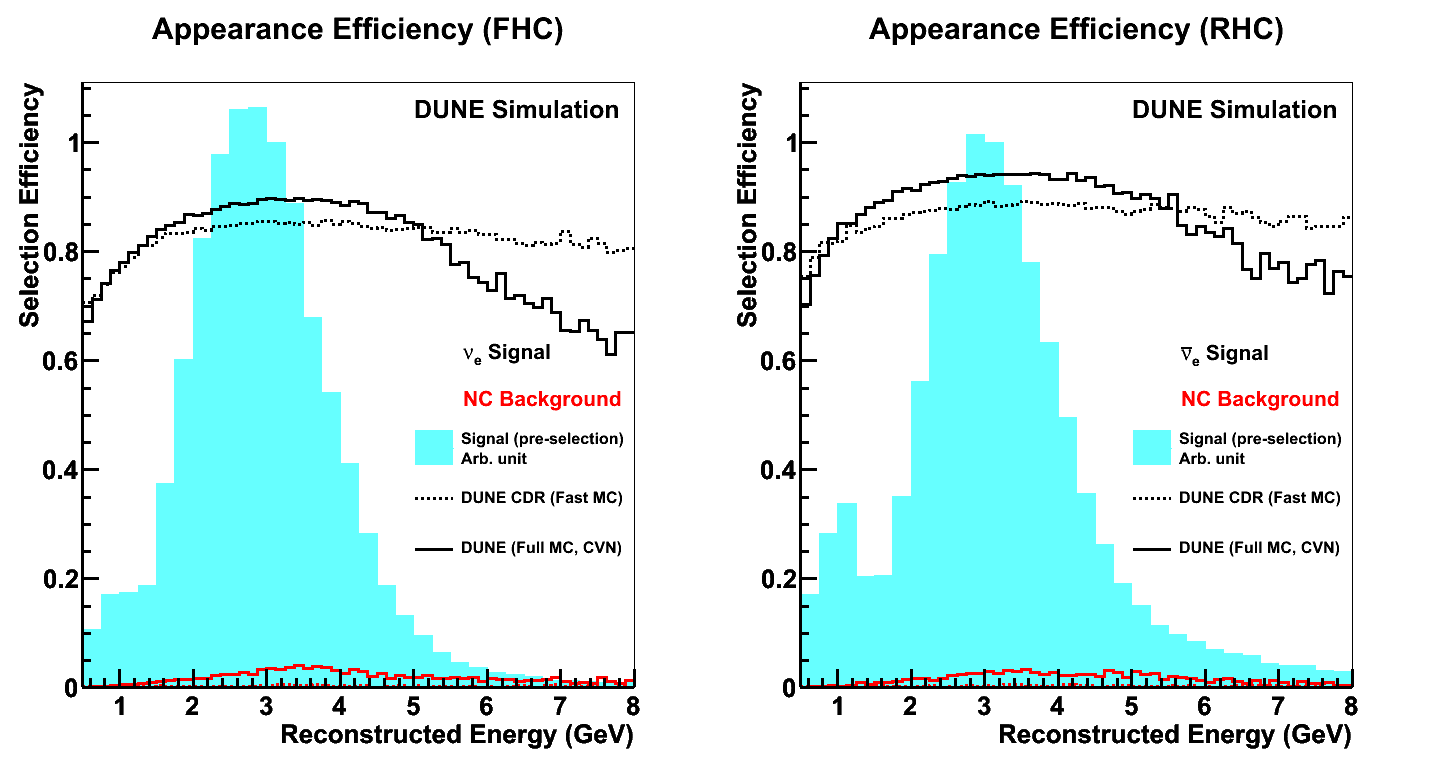}
  \includegraphics[width=0.8\linewidth, trim={0cm 0cm 0cm 2cm}, clip]{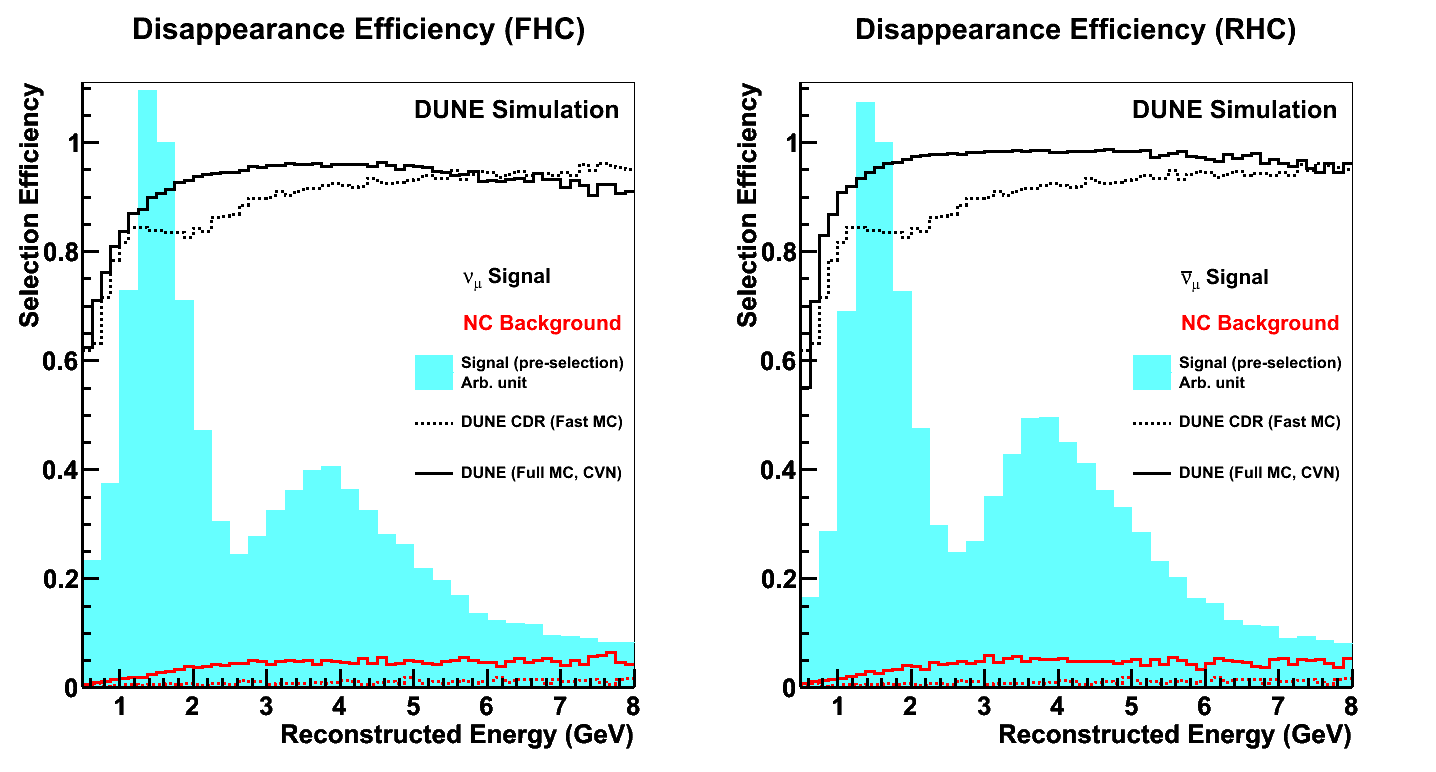} 
  \caption{Top: the $\nu_e$ \dword{cc} selection efficiency for \dword{fhc} (left) and \dword{rhc} (right) simulation with the criterion $P(\nu_e \textrm{ \dword{cc}}) > 0.85$. Bottom: the \numu \dword{cc} selection efficiency for \dword{fhc} (left) and \dword{rhc} (right) simulation with the criterion $P(\numu \textrm{ \dword{cc}}) > 0.5$. The results from \dword{dune}'s Conceptual Design Report (CDR) are shown for comparison~\cite{Acciarri:2015uup}. The solid (dashed) lines show results from the \dword{cvn} (\dword{cdr}) for signal $\nu_e$ \dword{cc} and $\bar{\nu}_e$ \dword{cc} events in black and \dword{nc} background interaction in red. The blue region shows the oscillated flux (A.U.) to illustrate the most important regions of the energy distribution. Reproduced from Ref.~\cite{cvn_paper}.}
    \label{fig:nueeff}
\end{figure*}

Figure~\ref{fig:nueeff} shows the efficiency as a function of reconstructed energy (under the electron neutrino hypothesis) for the $\nu_e$ event selection, and the corresponding selection efficiency for the $\nu_\mu$ event selection. The \nue and \numu efficiencies in both \dword{fhc} and \dword{rhc} beam modes all exceed 90\% in the neutrino flux peak.

The ability of the \dword{cvn} to identify neutrino flavor is dependent on its ability to resolve and identify the charged lepton. Backgrounds originate from the mis-identification of charged pions for $\nu_{\mu}$ disappearance, and photons for $\nu_{e}$ appearance. The probability for these backgrounds to be introduced varies with the momentum and isolation of the energy depositions from the pions and photons. The efficiency was also observed to drop as a function of track/shower angle (with respect to the incoming neutrino beam direction) when energy depositions aligned with wire planes. The shapes of the efficiency functions in lepton momentum, lepton angle, and hadronic energy fraction (inelasticity) are all observed to be consistent with results from previous studies, including hand scans of \dword{lartpc} simulations. The \dword{cvn} is susceptible to bias if there are features in the data that are not present in the simulation, so before its use on data, it will be important to comprehensively demonstrate that the selection is not sensitive to the choice of reference models. A discussion of the bias studies performed so far, and those planned in future, can be found in Ref.~\cite{cvn_paper}.

%% file: sections/rate.tex
\section{Expected Far Detector Event Rate and Oscillation Parameters}
\label{sec:rate}

In this work, \dword{fd} event rates are calculated assuming the following nominal deployment plan, which is based on a technically limited schedule:
\begin{itemize}
    \item Start of beam run: two \dword{fd} module
    volumes for total fiducial mass of 20 kt, 1.2 MW beam
    \item After one year: add one \dword{fd} module  volume for total fiducial mass of 30 kt
    \item After three years: add one \dword{fd} module  volume for total fiducial mass of \fdfiducialmass
    \item After six years: upgrade to 2.4 MW beam
\end{itemize}
Table~\ref{tab:exposures} shows the conversion between number of years under the nominal staging plan, and  kt-MW-years, which are used to indicate the exposure in this analysis. For all studies shown in this work, a 50\%/50\% ratio of FHC to RHC data was assumed, based on studies which showed a roughly equal mix of running produced a nearly optimal \deltacp and mass ordering sensitivity. The exact details of the run plan are not included in the staging plan.

\begin{table}[htbp]
  \centering
  \begin{tabular}{cc}
    \hline
    Years & kt-MW-years \\
    \hline\hline
    7 & 336 \\
    10 & 624 \\
    15 & 1104 \\
    \hline
  \end{tabular}
  \caption{Conversion between number of years in the nominal staging plan, and kt-MW-years, the two quantities used to indicate exposure in this analysis.}
  \label{tab:exposures}
\end{table}

Event rates are calculated with the assumption of 1.1 $\times 10^{21}$ \dword{pot} per year, which assumes a combined uptime and efficiency of the \dword{fnal} accelerator complex and the \dword{lbnf} beamline of 57\%~\cite{Abi:2020evt}.

\begin{figure}[htbp]
 \includegraphics[width=0.98\linewidth]{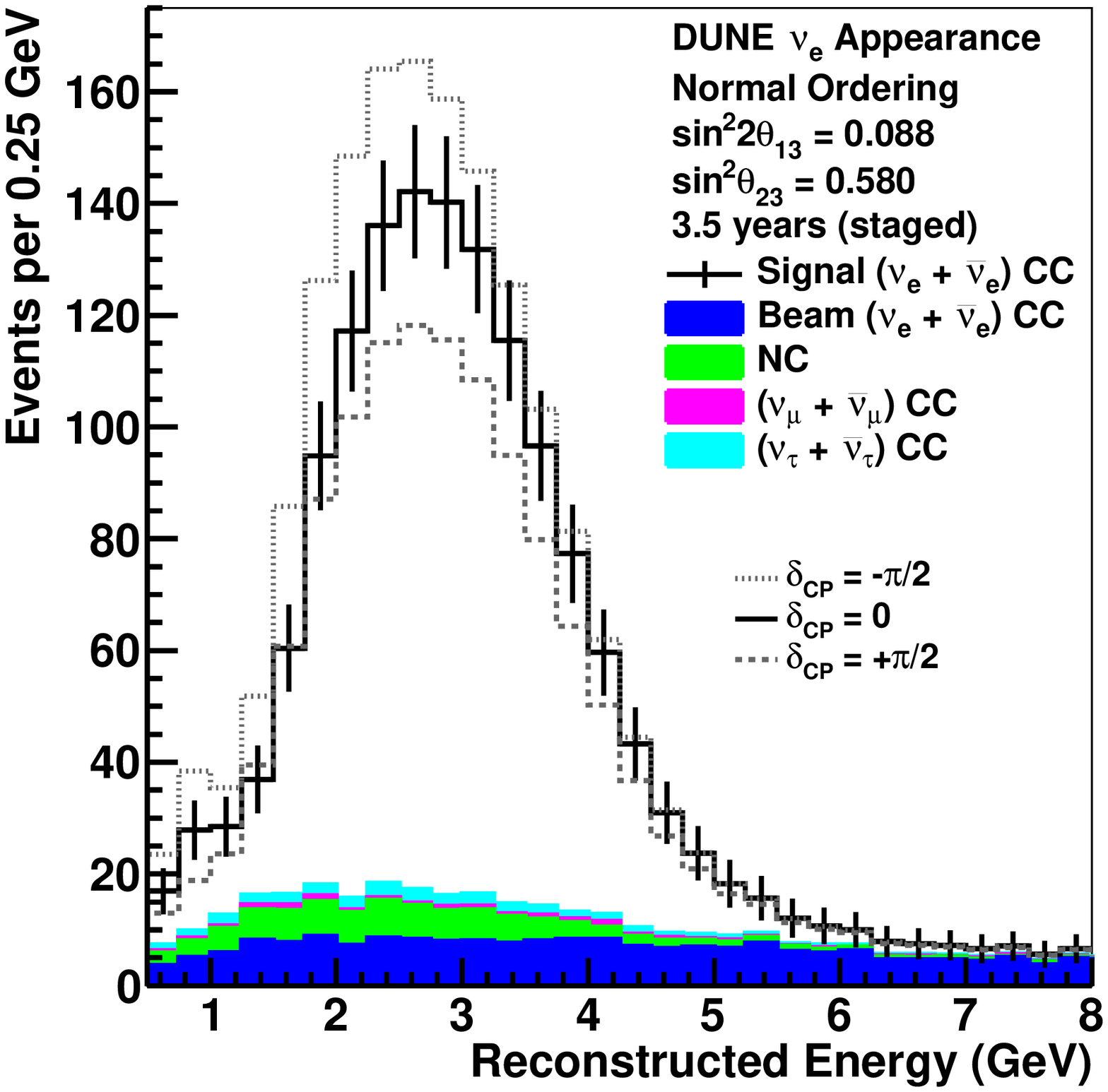}\\
 \includegraphics[width=0.98\linewidth]{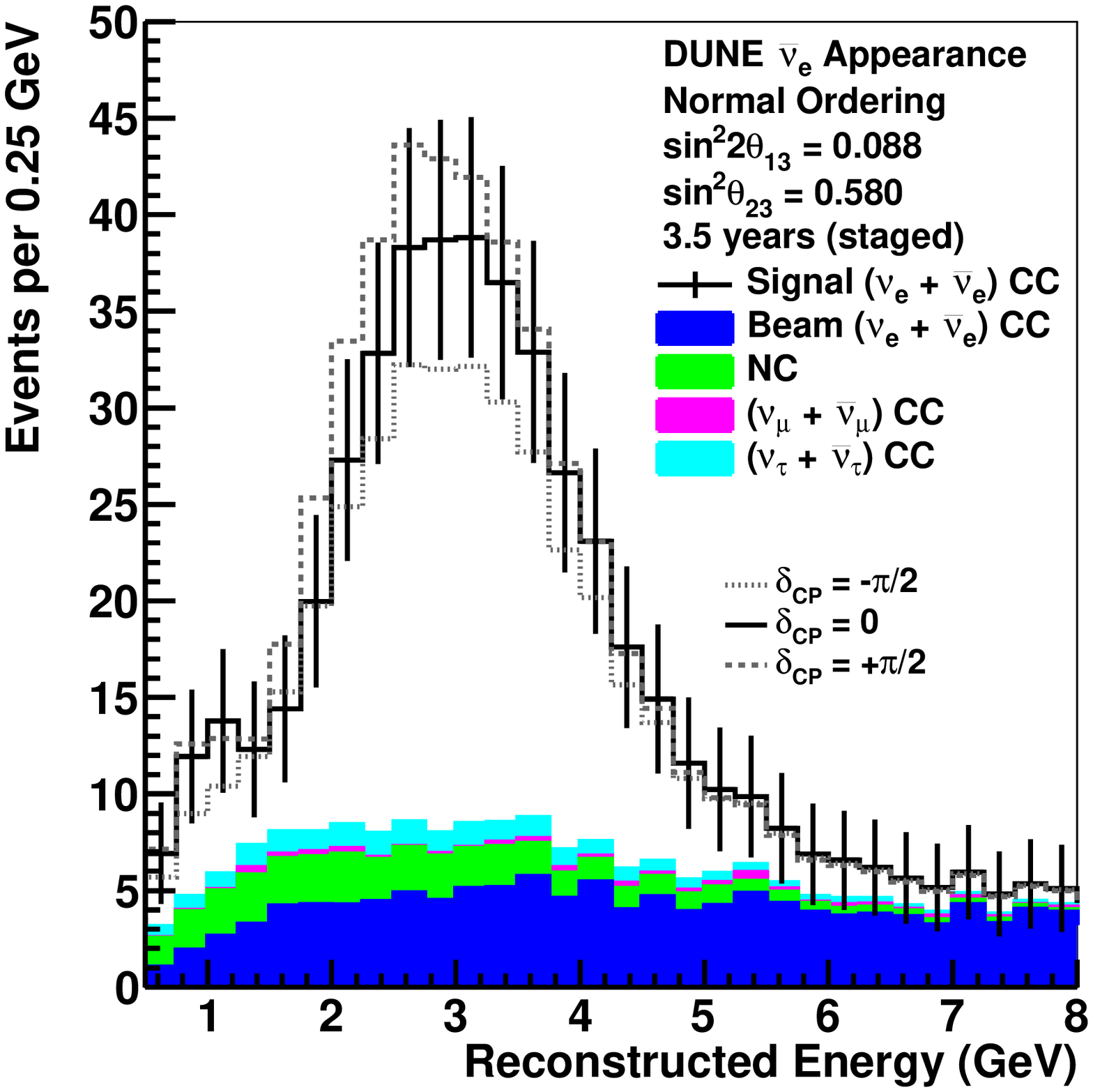}
 \caption{\nue and \anue appearance spectra: reconstructed energy distribution of selected \nue \dword{cc}-like events assuming 3.5 years (staged) running in the neutrino-beam mode (top) and antineutrino-beam mode (bottom), for a total of seven years (staged) exposure. Statistical uncertainties are shown on the datapoints. The plots assume normal mass ordering and include curves for $\mdeltacp = -\pi/2, 0$, and $\pi/2$.}
 \label{fig:appspectra}
\end{figure}

\begin{figure}[htbp]
\includegraphics[width=0.98\linewidth]{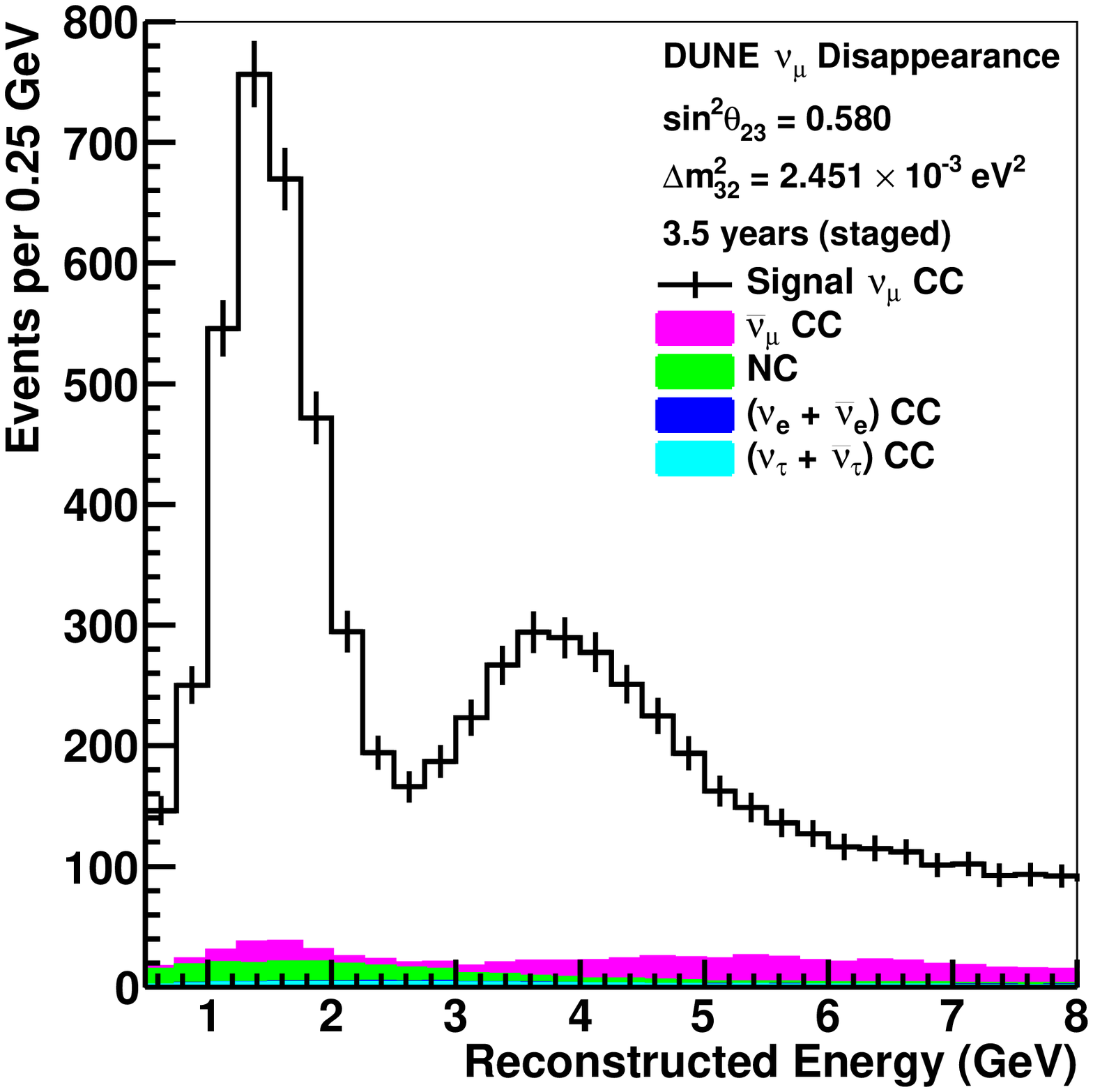}\\
\includegraphics[width=0.98\linewidth]{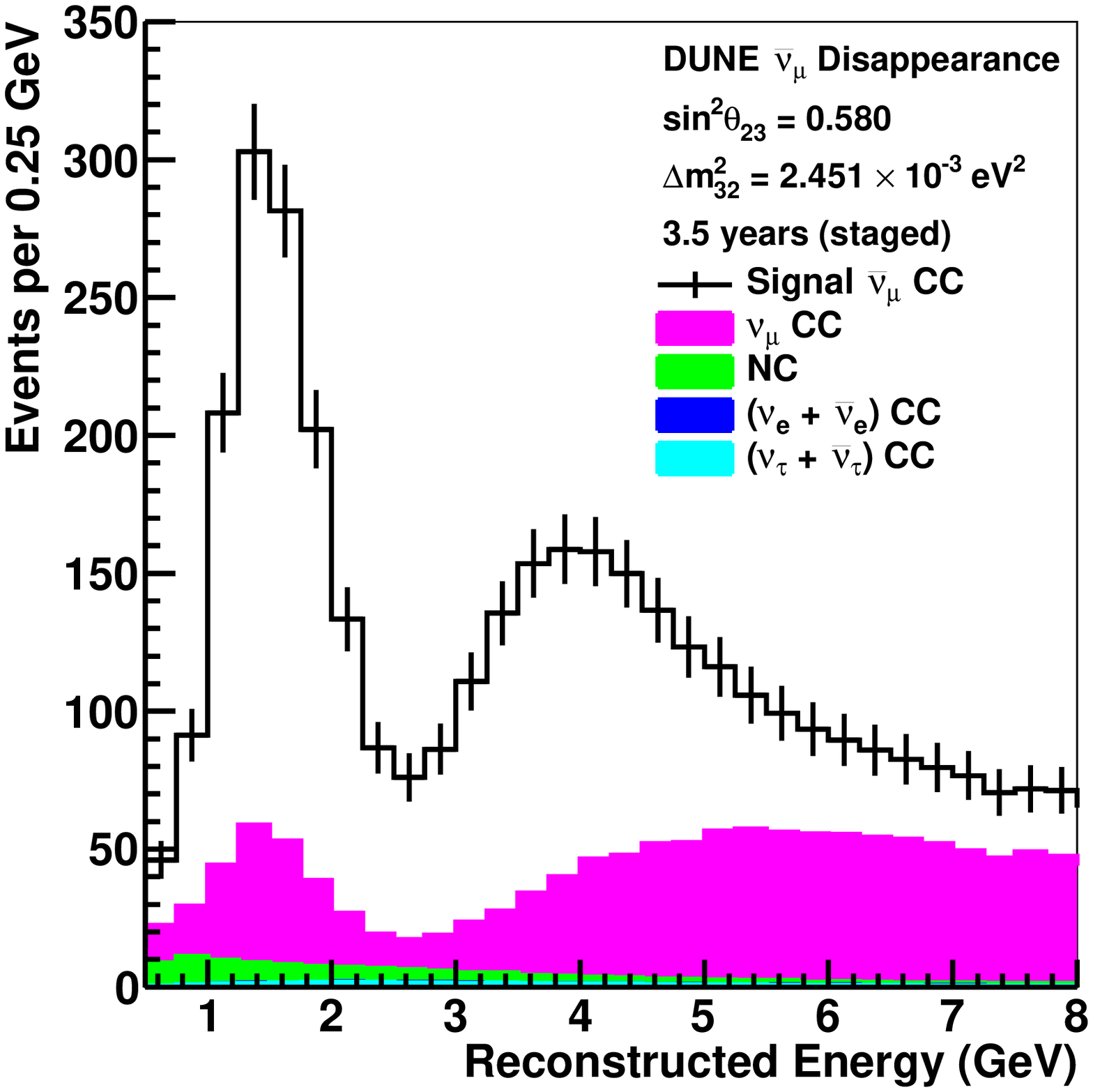}
\caption{\numu and \anumu disappearance spectra: reconstructed energy distribution of selected $\nu_{\mu}$ \dword{cc}-like events assuming 3.5 years (staged) running in the neutrino-beam mode (top) and antineutrino-beam mode (bottom), for a total of seven years (staged) exposure. Statistical uncertainties are shown on the datapoints. The plots assume normal mass ordering.}
\label{fig:disspectra}
\end{figure}

Figures~\ref{fig:appspectra} and~\ref{fig:disspectra} show the expected rate of selected events for \nue appearance and \numu disappearance, respectively, including expected flux, cross section, and oscillation probabilities, as a function of reconstructed neutrino energy at a baseline of
\num{1285}~km. The spectra are shown for a \num{3.5}~year (staged) exposure each for \dword{fhc} and \dword{rhc} beam modes, for a total run time of seven 
years. The rates shown are scaled to obtain different exposures. Tables~\ref{tab:apprates} and~\ref{tab:disrates} give the integrated rate for the \nue 
appearance and \numu disappearance spectra, respectively. Note that the total rates are integrated over the range of reconstructed neutrino energies used in the analysis, 0.5--10 GeV. The nominal neutrino oscillation parameters used in Figures~\ref{fig:appspectra} and~\ref{fig:disspectra} and the uncertainty on those parameters (used later in the analysis) are taken from the \dword{nufit}~\cite{Esteban:2018azc,nufitweb} global fit to neutrino data, and their values are given in Table~\ref{tab:oscpar_nufit}. See also
Refs.~\cite{deSalas:2017kay} and \cite{Capozzi:2017yic} for other recent global fits.

As can be seen in Figure~\ref{fig:appspectra}, the background to \nue appearance is composed of: (1) \dword{cc} interactions of \nue and \anue intrinsic to the beam; (2) misidentified \dword{nc} interactions;  (3) misidentified \numu and \anumu \dword{cc} interactions; and (4) $\nu_\tau$ and $\bar{\nu}_\tau$ \dword{cc} interactions in which the $\tau$'s decay leptonically into electrons/positrons. \dword{nc} and $\nu_\tau$ backgrounds emanate from interactions of higher-energy neutrinos that feed down to lower reconstructed neutrino energies due to missing energy in unreconstructed final-state neutrinos. The selected NC and \dword{cc} \numu generally include an asymmetric decay of a relatively high energy $\pi^0$ coupled with a prompt photon conversion. As can be seen in Figure~\ref{fig:disspectra}, the backgrounds to the \numu disappearance are due to wrong-sign \numu interactions, which cannot easily be distinguished in the unmagnetized \dword{dune} \dword{fd}, and NC interactions, where a pion has been misidentified as the primary muon. As expected, the \numu background in RHC is much larger than the \anumu background in FHC.

\begin{table}[htbp]
  \centering
  \begin{tabular}{lcccc}
    \hline
    Sample & \multicolumn{4}{c}{Expected Events} \\
    & \multicolumn{2}{c}{$\mdeltacp = 0\;$} & \multicolumn{2}{c}{$\mdeltacp =-\frac{\pi}{2}$} \\
    & NO & IO & NO & IO \\ \hline\hline
    \textbf{$\nu$ mode} & & & & \\
    Oscillated \nue & 1155 & 526 & 1395 & 707 \\
    Oscillated \anue & 19 & 33 & 14 & 28 \\
    \hline
    Total oscillated & 1174 & 559 & 1409 & 735 \\
    \hline 
    Beam $\nu_{e}+\bar{\nu}_{e}$ \dword{cc} background & 228 & 235 & 228 & 235 \\
    \dword{nc} background & 84 & 84 & 84 & 84 \\
    $\nu_{\tau}+\bar{\nu}_{\tau}$ \dword{cc} background & 36 & 36 & 35 & 36 \\
    $\nu_{\mu}+\bar{\nu}_{\mu}$ \dword{cc} background & 15 & 15 & 15 & 15 \\
    \hline
    Total background & 363 & 370 & 362 & 370 \\
    \hline\hline
    \textbf{$\bar{\nu}$ mode} & & & & \\
    Oscillated \nue & 81 & 39 & 95 & 53 \\
    Oscillated \anue & 236 & 492 & 164 & 396 \\
    \hline
    Total oscillated & 317 & 531 & 259 & 449 \\
    \hline 
    Beam $\nu_{e}+\bar{\nu}_{e}$ \dword{cc} background & 145 & 144 & 145 & 144 \\
    \dword{nc} background & 40 & 40 & 40 & 40 \\
    $\nu_{\tau}+\bar{\nu}_{\tau}$ \dword{cc} background & 22 & 22 & 22 & 22 \\
    $\nu_{\mu}+\bar{\nu}_{\mu}$ \dword{cc} background & 6 & 6 & 6 & 6 \\
    \hline 
    Total background & 216 & 215 & 216 & 215 \\
    \hline
  \end{tabular}
 \caption{\nue and \anue appearance rates: integrated rate of selected $\nu_e$ \dword{cc}-like events between 0.5 and 10.0~GeV assuming a \num{3.5}-year (staged) exposure in the neutrino-beam mode and antineutrino-beam mode.  The rates are shown for both \dword{no} and \dword{io}, and signal events are shown for both $\mdeltacp = 0$ and $\mdeltacp = -\pi/2$.}
 \label{tab:apprates}
\end{table}

\begin{table}[htbp]
  \centering
  \begin{tabular}{lcc}
    \hline
    Sample & \multicolumn{2}{c}{Expected Events} \\
    & NO & IO \\
    \hline\hline
    \textbf{$\nu$ mode} & & \\
    \numu Signal & 7235 & 7368 \\
    \hline 
    \anumu \dword{cc} background & 542 & 542 \\
    \dword{nc} background & 213 & 213 \\
    $\nu_{\tau}+\bar{\nu}_{\tau}$ \dword{cc} background & 53 & 54 \\
    $\nu_e+\bar{\nu}_e$ \dword{cc} background & 9 & 5 \\
    \hline\hline
    \textbf{$\bar{\nu}$ mode}  & & \\
    \anumu Signal & 2656 & 2633 \\
    \hline 
    \numu \dword{cc} background & 1590 & 1600 \\
    \dword{nc} background & 109 & 109 \\
    $\nu_{\tau}+\bar{\nu}_{\tau}$ \dword{cc} background & 31 & 31 \\
    $\nu_e+\bar{\nu}_e$ \dword{cc} background & 2 & 2 \\
    \hline
  \end{tabular}
  \caption{\numu and \anumu disappearance rates: integrated rate of selected $\nu_{\mu}$ \dword{cc}-like events between 0.5 and 10.0~GeV assuming a \num{3.5}-year (staged) exposure in the neutrino-beam mode and antineutrino-beam mode. The rates are shown for both \dword{no} and \dword{io}, with $\mdeltacp = 0$.}
 \label{tab:disrates}
\end{table}

\begin{table}[htbp]
    \centering
    \begin{tabular}{lcc}
      \hline
 Parameter &    Central value & Relative uncertainty \\
\hline\hline
$\theta_{12}$ & 0.5903 & 2.3\% \\ \hline
$\theta_{23}$ (NO) & 0.866  & 4.1\% \\ 
$\theta_{23}$ (IO) & 0.869  & 4.0\% \\ \hline
$\theta_{13}$ (NO) & 0.150  & 1.5\% \\ 
$\theta_{13}$ (IO) & 0.151  & 1.5\% \\ \hline
$\Delta m^2_{21}$ & 7.39$\times10^{-5}$~eV$^2$ & 2.8\% \\ \hline
$\Delta m^2_{32}$ (NO) & 2.451$\times10^{-3}$~eV$^2$ &  1.3\% \\
$\Delta m^2_{32}$ (IO) & -2.512$\times10^{-3}$~eV$^2$ &  1.3\% \\
\hline
$\rho$ & 2.848 g cm$^{-3}$ & 2\% \\
\hline
    \end{tabular}
    \caption[Parameter values and uncertainties from a global fit to neutrino oscillation data]{Central value and relative uncertainty of neutrino oscillation parameters from a global fit~\cite{Esteban:2018azc,nufitweb} to neutrino oscillation data. The matter density is taken from Ref.~\cite{Roe:2017zdw}. Because the probability distributions are somewhat non-Gaussian (particularly for $\theta_{23}$), the relative uncertainty is computed using 1/6 of the 3$\sigma$ allowed range from the fit, rather than 1/2 of the 1$\sigma$ range. For $\theta_{23}$, $\theta_{13}$, and $\Delta m^2_{31}$, the best-fit values and uncertainties depend on whether normal mass ordering (NO) or inverted mass ordering (IO) is assumed.}
    \label{tab:oscpar_nufit}
\end{table}

%% file: sections/syst.tex
\section{Detector Uncertainties}
\label{sec:syst}

Detector effects impact the event selection efficiency as well as the reconstruction of quantities used in the oscillation fit, such as neutrino energy. The main sources of detector systematic uncertainties are limitations of the expected calibration and modeling of particles in the detector.

The \dword{nd} \dword{lartpc} uses similar technology to the \dword{fd}, but important differences lead to uncertainties that do not fully correlate between the two detectors. First, the readout technology is different, as the \dword{nd} \dword{lartpc} uses pixels as well as a different, modular photon detector. Therefore, the charge response will be different between near and far detectors due to differences in electronics readout, noise, and local effects like alignment.  Second, the high-intensity environment of the \dword{nd} complicates associating detached energy deposits to events, a problem which is not present in the \dword{fd}. Third, the calibration strategies will be different. For example, the \dword{nd} has a high-statistics calibration sample of through-going, momentum-analyzed muons from neutrino interactions in the upstream rock, which is not available with high statistics for the \dword{fd}.
Finally, the reconstruction efficiency will be inherently different due to the relatively small size of the \dword{nd}. Containment of charged hadrons will be significantly worse at the \dword{nd}, especially for events with energetic hadronic showers or with vertices near the edges of the \dword{fv}.

An uncertainty on the overall energy scale is included in the analysis presented here, as well as particle response uncertainties that are separate and uncorrelated between four species: muons, charged hadrons, neutrons, and electromagnetic showers. In the \dword{nd}, muons reconstructed by range in \dword{lar} and by curvature in the \dword{mpd} are treated separately. The energy scale and particle response uncertainties are allowed to vary with energy; each term is described by three free parameters:

\begin{equation}
\label{eq:escale_unc}    
E^{\prime}_{rec} = E_{rec} \times (p_{0} + p_{1}\sqrt{E_{rec}} + \frac{p_{2}}{\sqrt{E_{rec}}})
\end{equation}

\noindent
where $E_{rec}$ is the nominal reconstructed energy, $E^{\prime}_{rec}$ is the shifted energy, and $p_{0}$, $p_{1}$, and $p_{2}$ are free fit parameters that are allowed to vary within \textit{a priori} constraints. Note that the parameters produce a shift to the kinematic variables in an event, as opposed to simply assigning a weight to each simulated event. The energy scale and resolution parameters are conservatively treated as uncorrelated between the \dword{nd} and \dword{fd}. With a better understanding of the relationship between \dword{nd} and \dword{fd} calibration and reconstruction techniques, it may be possible to correlate some portion of the energy response. The full list of assumed energy scale uncertainties is given as Table~\ref{tab:EscaleSysts}. In addition to the uncertainties on the energy scale, uncertainties on energy resolutions are also included. These are treated as fully uncorrelated between the near and far detectors and are taken to be 2\% for muons, charged hadrons, and EM showers and 40\% for neutrons.

\begin{table}[htb]
  \centering
  \begin{tabular}{lccc}
    \hline
    Particle type & \multicolumn{3}{c}{Allowed variation} \\ 
    & $p_{0}$ & $p_{1}$ & $p_{2}$ \\ \hline\hline
    all (except muons) & 2\%   & 1\%   & 2\%   \\
    $\mu$ (range)      & 2\%   & 2\%   & 2\%   \\
    $\mu$ (curvature)  & 1\%   & 1\%   & 1\%   \\
    p, $\pi^{\pm}$     & 5\%   & 5\%   & 5\%   \\
    e, $\gamma$, $\pi^{0}$ & 2.5\%   & 2.5\%   & 2.5\%   \\
    n                  & 20\%  & 30\%  & 30\%  \\
    \hline
    \end{tabular}
    \caption{Uncertainties applied to the energy response of various particles. $p_{0}$, $p_{1}$, and $p_{2}$ correspond to the constant, square root, and inverse square root terms in the energy response parameterization given in Equation~\ref{eq:escale_unc}. All are treated as uncorrelated between the \dword{nd} and \dword{fd}.}
    \label{tab:EscaleSysts}
\end{table} 

The scale of these assumed uncertainties is motivated by what has been achieved in recent experiments, including calorimetric based approaches (\dword{nova}, \dword{minerva}) and \dwords{lartpc} (\dword{lariat}, \dword{microboone}, \dword{argoneut}). The \dword{dune} performance is expected to significantly exceed the performance of these current surface-based experiments. \dword{nova}~\cite{NOvA:2018gge} has achieved $<1$\% (5\%) uncertainties on the energy scale of muons (protons). Uncertainties associated to the pion and proton re-interactions in the detector medium are expected to be controlled from \dword{protodune} and \dword{lariat} data, as well as the combined analysis of low density (gaseous) and high density (\dword{lar}) \dwords{nd}. Uncertainties in the \efield also contribute to the energy scale uncertainty~\cite{Adams:2019qrr}, and calibration is needed (with cosmics at \dword{nd}, laser system at \dword{fd}) to constrain the overall energy scale. The recombination model will continue to be validated by the suite of \dword{lar} experiments and is not expected to be an issue for nominal field provided minimal \efield distortions. Uncertainties in the electronics response are controlled with a dedicated charge injection system and validated with intrinsic sources, Michel electrons and \Ar39.

The response of the detector to neutrons is a source of active study and will couple strongly to detector technology. The validation of neutron interactions in \dword{lar} will continue to be characterized by dedicated measurements (e.g., CAPTAIN~\cite{Berns:2013usa,Bhandari:2019rat}) and the \dword{lar} program (e.g., \dword{argoneut}~\cite{Acciarri:2018myr}).  However, the association of the identification of a neutron scatter or capture to the neutron's true energy has not been demonstrated, and significant reconstruction issues exist, so a large uncertainty (20\%) is assigned comparable to the observations made by \dword{minerva}~\cite{Elkins:2019vmy} assuming they are attributed entirely to the detector model. Selection of photon candidates from $\pi^0$ is also a significant reconstruction challenge, but a recent measurement from \dword{microboone} indicates this is possible and the reconstructed $\pi^0$ invariant mass has an uncertainty of 5\%, although with some bias~\cite{Adams:2018sgn}.

The $p_{1}$ and $p_{2}$ terms in Equation~\ref{eq:escale_unc} allow the energy response to vary as a function of energy. The energy dependence is conservatively assumed to be of the same order as the absolute scale uncertainties given by the $p_{0}$ terms.

In addition to impacting energy reconstruction, the \efield model also affects the definition of the \dword{fd} fiducial volume, which is sensitive to electron drift. An additional 1\% uncertainty is assumed on the total fiducial mass, which is conservatively treated as uncorrelated between the $\nu_{\mu}$ and $\nu_{e}$ samples due to the potential distortion caused by large electromagnetic showers in the electron sample. These uncertainties affect only the overall normalization, and are called \texttt{FV numu FD} and \texttt{FV nue FD} in Figure~\ref{fig:postfit_unc_ndfd}.

The \dword{nd} and \dword{fd} have different acceptance to \dword{cc} events due to the very different detector sizes. The \dword{fd} is sufficiently large that acceptance is not expected to vary significantly as a function of event kinematics. However, the \dword{nd} selection requires that hadronic showers be well contained in \dword{lar} to ensure a good energy resolution, resulting in a loss of acceptance for events with energetic hadronic showers. The \dword{nd} also has regions of muon phase space with lower acceptance due to tracks exiting the side of the \dword{tpc} but failing to match to the \dword{mpd}, which are currently not used in the sensitivity analysis.

Uncertainties are evaluated on the muon and hadron acceptance of the \dword{nd}. The detector acceptance for muons and hadrons is shown in Figure~\ref{fig:NDacceptance}. Inefficiency at very low lepton energy is due to events being misreconstructed as neutral current. For high energy, forward muons, the inefficiency is only due to events near the edge of the \dword{fv} where the muon happens to miss the \dword{mpd}. At high transverse momentum, muons begin to exit the side of the \dword{lar} active volume, except when they happen to go along the 7~m axis. The acceptance is sensitive to the modeling of muons in the detector. An uncertainty is estimated based on the change in the acceptance as a function of muon kinematics.

Inefficiency at high hadronic energy is due to the veto on more than 30 MeV deposited in the outer 30 cm of the \dword{lar} active volume. Rejected events are typically poorly reconstructed due to low containment, and the acceptance is expected to decrease at high hadronic energy. Similar to the muon reconstruction, this acceptance is sensitive to detector modeling, and an uncertainty is evaluated based on the change in the acceptance as a function of true hadronic energy.

%% file: sections/methods.tex
\section{Sensitivity Methods}
\label{sec:methods}

Previous DUNE sensitivity predictions have used the GLoBES framework~\cite{Acciarri:2015uup,Alion:2016uaj,Bass:2014vta}. In this work, fits are performed using the \dword{cafana}~\cite{CAFAna} analysis framework, developed originally for the \dword{nova} experiment. Systematics are implemented using one-dimensional response functions for each analysis bin, and oscillation weights are calculated exactly, in fine (50 MeV) bins of true neutrino energy. For a given set of inputs, flux, oscillation parameters, cross sections, detector energy response matrices, and detector efficiency, an expected event rate can be produced. Minimization is performed using the {\sc minuit}~\cite{James:1994vla} package.

Oscillation sensitivities are obtained by simultaneously fitting the \numutonumu, $\bar{\nu}_\mu \rightarrow \bar{\nu}_\mu$ (Figure~\ref{fig:disspectra}), \numutonue, and $\bar{\nu}_\mu \rightarrow \bar{\nu}_e$ (Figure~\ref{fig:appspectra}) \dword{fd} spectra along with the $\nu_{\mu}$ FHC and $\bar{\nu}_{\mu}$ RHC samples from the \dword{nd} (Figure~\ref{fig:nd_samples}). In the studies, all oscillation parameters shown in Table~\ref{tab:oscpar_nufit} are allowed to vary. Gaussian penalty terms (taken from Table~\ref{tab:oscpar_nufit}) are applied to $\theta_{12}$ and \dm{12} and the matter density, $\rho$, of the Earth along the DUNE baseline~\cite{Roe:2017zdw}. Unless otherwise stated, studies presented here include a Gaussian penalty term on $\theta_{13}$ (also taken from Table~\ref{tab:oscpar_nufit}), which is precisely measured by experiments sensitive to reactor antineutrino disappearance~\cite{Abe:2014bwa,Adey:2018zwh,Bak:2018ydk}. The remaining parameters, \sinst{23}, $\Delta m^{2}_{32}$, and \deltacp are allowed to vary freely, with no penalty term. Note that the penalty terms are treated as uncorrelated with each other, or other parameters, which is a simplification. In particular, the reactor experiments that drive the constraint on $\theta_{13}$ in the \dword{nufit} analysis are also sensitive to \dm{32}, so the constraint on $\theta_{13}$ should be correlated with \dm{32}. We do not expect this to have a significant impact on the fits, and this effect only matters for those results with the $\theta_{13}$ Gaussian penalty term included.

Flux, cross section, and \dword{fd} detector parameters are allowed to vary in the fit, but constrained by a penalty term proportional to the pre-fit uncertainty. \dword{nd} detector parameters are not allowed to vary in the fit, but their effect is included via a covariance matrix based on the shape difference between \dword{nd} prediction and the ``data'' (which comes from the simulation in this sensitivity study). The covariance matrix is constructed with a throwing technique. For each ``throw'', all \dword{nd} energy scale, resolution, and acceptance parameters are simultaneously thrown according to their respective uncertainties, and the modified prediction is produced by varying the relevant quantities away from the nominal prediction according to the thrown parameter values. The bin-to-bin covariance is determined by comparing the resulting spectra with the nominal prediction, in the same binning as is used in the oscillation sensitivity analysis. This choice protects against overconstraining that could occur given the limitations of the parameterized \dword{nd} reconstruction described in Section~\ref{sec:nd} taken together with the high statistical power at the \dword{nd}, but is also a simplification.

The compatibility of a particular oscillation hypothesis with both \dword{nd} and \dword{fd} data is evaluated using a negative log-likelihood ratio, which converges to a $\chi^{2}$ at high-statistics~\cite{Tanabashi:2018oca}:
\begin{equation}
\begin{aligned}
  \chi^2(\vec{\vartheta}, \vec{x}) &= -2\log\mathcal{L}(\vec{\vartheta}, \vec{x}) \\
  &= 2\sum_i^{N_{\rm bins}}\left[ M_i(\vec{\vartheta}, \vec{x})-D_i+D_i\ln\left({D_i\over M_i(\vec{\vartheta}, \vec{x})}\right) \right] \\
  &+ \sum_{j}^{N_{\mathrm{systs}}}\left[ \frac{\Delta x_{j}}{\sigma_{j}} \right]^{2} \\
  &+ \sum^{N^{\mathrm{ND}}_{\mathrm{bins}}}_{k}\sum^{N^{\mathrm{ND}}_{\mathrm{bins}}}_{l} \left(M_k(\vec{x})-D_k \right) V^{-1}_{kl}\left(M_l(\vec{x})-D_l \right),
\end{aligned}
\label{eq:chisq}
\end{equation}
where $\vec{\vartheta}$ and $\vec{x}$ are the vector of oscillation parameter and nuisance parameter values respectively; $M_i(\vec{\vartheta}, \vec{x})$ and $D_{i}$ are the \dword{mc} expectation and fake data in the $i$th reconstructed bin (summed over all selected samples), with the oscillation parameters neglected for the \dword{nd}; $\Delta x_{j}$ and $\sigma_{j}$ are the difference between the nominal and current value, and the prior uncertainty on the $j$th nuisance parameter with uncertainties evaluated and described in Sections~\ref{sec:flux},~\ref{sec:nuint} and~\ref{sec:syst}; and $V_{kl}$ is the covariance matrix between \dword{nd} bins described previously. In order to avoid falling into a false minimum, all fits are repeated for four different \deltacp values (-$\pi$, -$\pi$/2, 0, $\pi$/2), both mass orderings, and in both octants, and the lowest $\chi^{2}$ value is taken as the minimum.

\begin{table}
  \centering
  \begin{tabular}{lcc}
    \hline
    Parameter & Prior & Range\\ \hline\hline
    $\sin^{2}\theta_{23}$ & Uniform & [0.4; 0.6] \\
    $|\Delta m^{2}_{32}|$ ($\times 10^{-3}$ eV$^{2}$) & Uniform & $|[2.3;2.7]|$ \\
    \deltacp ($\pi$) & Uniform & [-1;1] \\
    $\theta_{13}$ & Gaussian & \dword{nufit} \\
    & Uniform & [0.13; 0.2] \\
    \hline
  \end{tabular}
  \caption{Treatment of the oscillation parameters for the simulated data set studies. Note that for some studies $\theta_{13}$ has a Gaussian penalty term applied based on the \dword{nufit} value, and for others it is thrown uniformly within a range determined from the \dword{nufit} 3$\sigma$ allowed range.}
  \label{table:OA_throw}
\end{table}
Two approaches are used for the sensitivity studies presented in this work. First, Asimov studies~\cite{Cowan:2010js} are carried out in which the fake (Asimov) dataset is the same as the nominal \dword{mc}. In these, the true value of all systematic uncertainties and oscillation parameters except those of interest (which are fixed at a test point) remain unchanged, and can vary in the fit, but are constrained by their pre-fit uncertainty. Second, studies are performed where many statistical and systematic throws are made according to their pre-fit Gaussian uncertainties, and fits of all parameters are carried out for each throw. A distribution of post-fit values is built up for the parameter of interest. In these, the expected resolution for oscillation parameters is determined from the spread in best-fit values obtained from an ensemble of throws that vary according to both the statistical and systematic uncertainties.  For each throw, the true value of each nuisance parameter is chosen randomly from a distribution determined by the {\it a priori} uncertainty on the parameter. For some studies, oscillation parameters are also randomly chosen as described in Table~\ref{table:OA_throw}. Poisson fluctuations are then applied to all analysis bins, based on the mean event count for each bin after the systematic adjustments have been applied. For each throw in the ensemble, the test statistic is minimized, and the best-fit value of all parameters is determined. The median throw and central 68\% of throws derived from these ensembles are shown.

Sensitivity calculations for \dword{cpv}, neutrino mass ordering, and octant are performed, in addition to studies of oscillation parameter resolution in one and two dimensions. In these cases, the experimental sensitivity is quantified using a likelihood ratio as the test statistic:
\begin{equation}
  \Delta\chi^2 = \chi^2_{\mathrm{B}} - \chi^{2}_{\mathrm{A}},
  \label{eq:dchisq}
\end{equation}
where $\chi^2_{\mathrm{B}}$ and $\chi^2_{\mathrm{A}}$ are both obtained from Equation~\ref{eq:chisq}, using a coherent systematic and statistical throw. The size of $\Delta\chi^2$ is a measure of how well the data can exclude model B in favor of model A, given the uncertainty in the model. For example, the sensitivity for excluding the \dword{io} in favor of the \dword{no} would be given as $\chi^2_{\mathrm{IO}} - \chi^{2}_{\mathrm{NO}}$. Note that the $\Delta\chi^2$ for the mass ordering may be negative, depending on how the test is set up. The sensitivity for discovering \dword{cpv} is the preference for the \dword{cp} violating hypothesis over the \dword{cp} conserving hypothesis, $\chi^2_{0, \pi} - \chi^{2}_{\mathrm{CPV}}$.

\begin{figure}
  \includegraphics[width=0.98\linewidth]{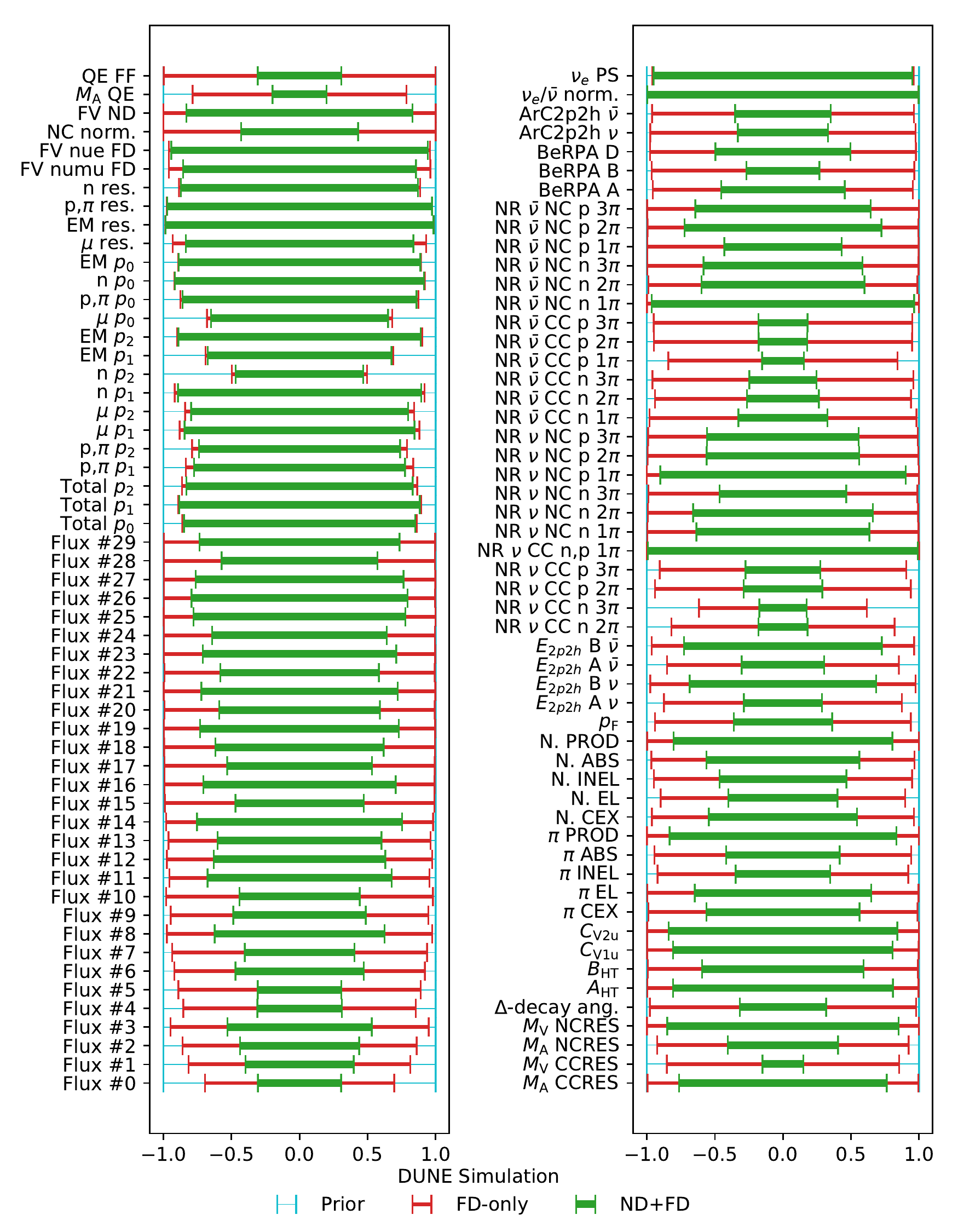}
  \caption{The ratio of post-fit to pre-fit uncertainties for various systematic parameters for a 15-year staged exposure. The red band shows the constraint from the \dword{fd} only in 15 years, while the green shows the \dword{nd}+\dword{fd} constraints. Flux parameters are named ``Flux \#$i$'' representing the $i$th principal flux component, cross-section parameter names are given in Section~\ref{sec:nuint}, and detector systematics are described in Section~\ref{sec:syst}, where the $p_{0}$, $p_1$ and $p_2$ parameters are described in Table~\ref{tab:EscaleSysts}.}
  \label{fig:postfit_unc_ndfd}
\end{figure}
Post-Fit uncertainties on systematic parameters are shown for Asimov fits at the \dword{nufit} best-fit point to both the ND+FD samples, and the FD-only samples in Figure~\ref{fig:postfit_unc_ndfd}, as a fraction of the pre-fit systematic uncertainties described in Sections~\ref{sec:flux}, \ref{sec:nuint}, and \ref{sec:syst}. The FD alone can only weakly constrain the flux and cross-section parameters, which are much more strongly constrained when the \dword{nd} is included. The \dword{nd} is, however, unable to strongly constrain the \dword{fd} detector systematics as they are treated as uncorrelated, and due to the treatment of \dword{nd} detector systematics in a covariance matrix in Equation~\ref{eq:chisq}. Adding the \dword{nd} does slightly increase the constraint on detector parameters as it breaks degeneracies with other parameters. Several important cross-section uncertainties are also not constrained by the \dword{nd}. In particular, an uncertainty on the ratio of \numu to \nue cross sections is totally unconstrained, which is not surprising given the lack of \dword{nd} \nue samples in the current analysis. The most significant flux terms are constrained at the level of ~20\% of their \textit{a priori} values.  Less significant principal components have little impact on the observed distributions at either detector, and receive weaker constraints.

\begin{figure}[htbp]
\includegraphics[width=0.98\linewidth]{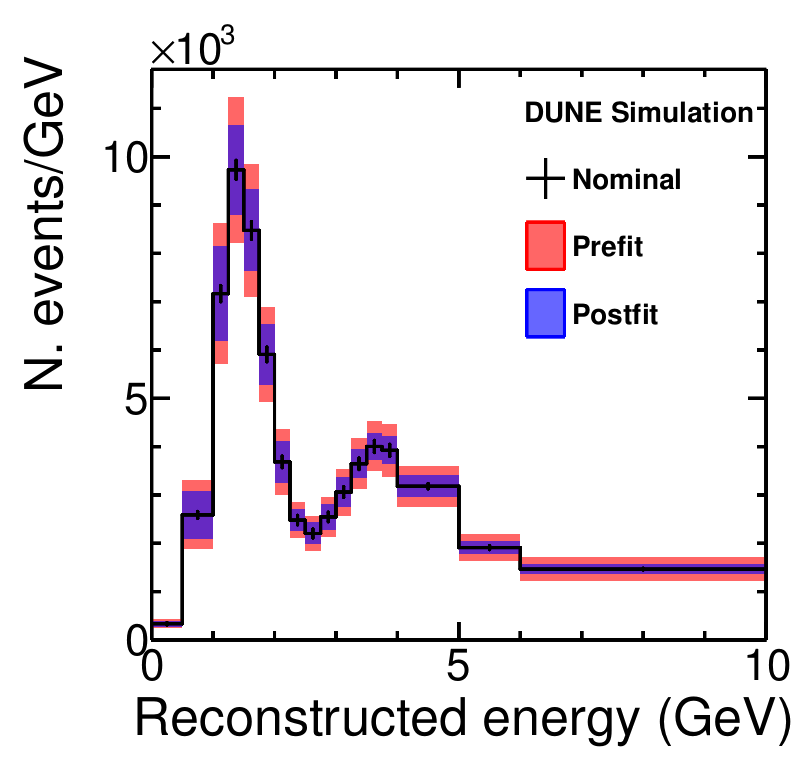}
\includegraphics[width=0.98\linewidth]{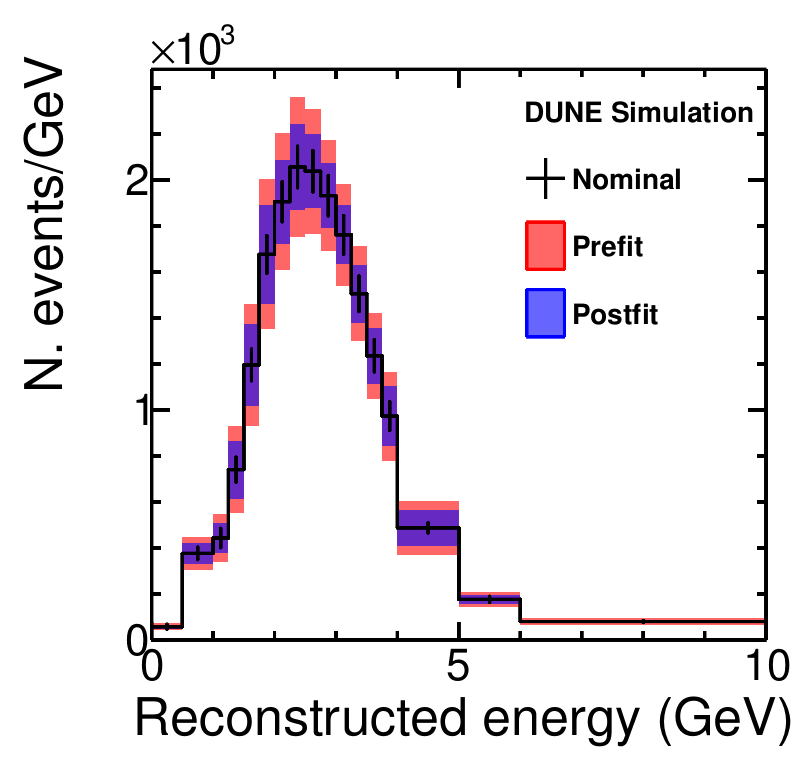}
\caption{\numu (top) and \nue (bottom) \dword{fd} FHC spectra for a 15 year staged exposure with oscillation parameters set to the \dword{nufit} best-fit point, shown as a function of reconstructed neutrino energy. The statistical uncertainty on the total rate is shown on the data points, and the pre- and post-fit systematic uncertainties are shown as shaded bands. The post-fit uncertainty includes the effect of the \dword{nd} samples in the fit, and corresponds to the parameter constraints shown in Figure~\ref{fig:postfit_unc_ndfd}.}
\label{fig:postfit_spectra}
\end{figure}
Figure~\ref{fig:postfit_spectra} shows the pre- and post-fit systematic uncertainties on the \dword{fd} FHC samples for Asimov fits at the \dword{nufit} best-fit point including both \dword{nd} and \dword{fd} samples with a 15 year exposure. It shows how the parameter constraints seen in Figure~\ref{fig:postfit_unc_ndfd} translate to a constraint on the event rate. Similar results are seen for the RHC samples. The large reduction in the systematic uncertainties is largely due to the \dword{nd} constraint on the systematic uncertainties apparent from Figure~\ref{fig:postfit_unc_ndfd}.

%% file: sections/sens.tex
\section{Sensitivities}
\label{sec:sens}

In this section, various sensitivity results are presented. For the sake of simplicity, unless otherwise stated, only true normal ordering is shown. Possible variations of sensitivity are presented in two ways. Results produced using Asimovs are shown as lines, and differences between two Asimov scenarios are shown with a colored band. Note that the band in the Asimov case is purely to guide the eye, and does not denote a confidence interval. For results produced using many throws of oscillation parameters, systematic and statistical uncertainties, $\sim$300,000 throws were used to calculate the sensitivity for each scenario. The median sensitivity is shown with a solid line, and a transparent filled area indicates the region containing the central 68\% of throws, which can be interpreted as the 1$\sigma$ uncertainty on the sensitivity.

\begin{figure}[htbp]
  \centering
  \includegraphics[width=0.98\linewidth, trim={0cm 0cm 0cm 2.3cm}, clip]{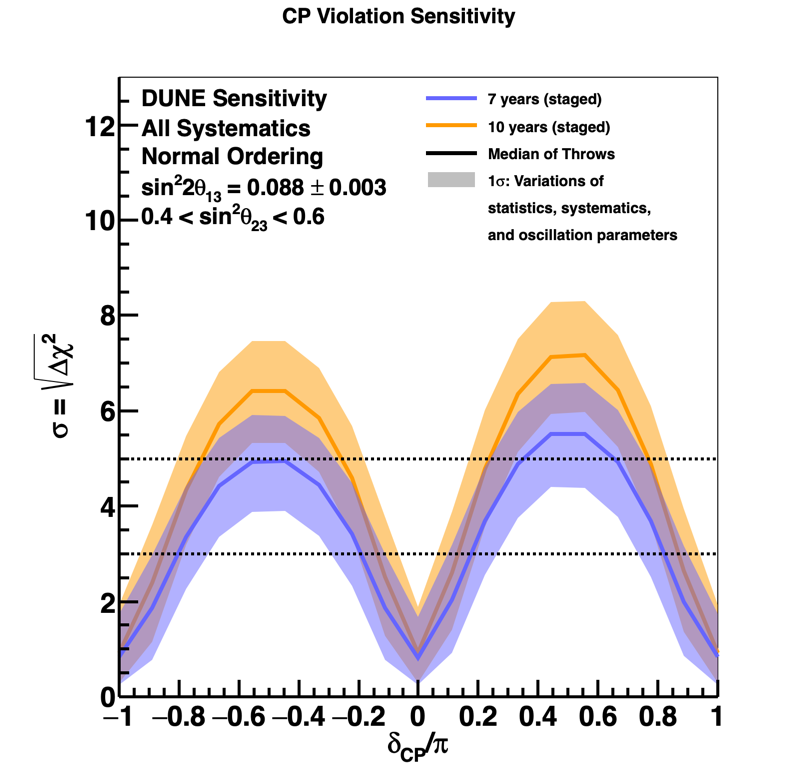}\\
  \includegraphics[width=0.98\linewidth, trim={0cm 0cm 0cm 2.3cm}, clip]{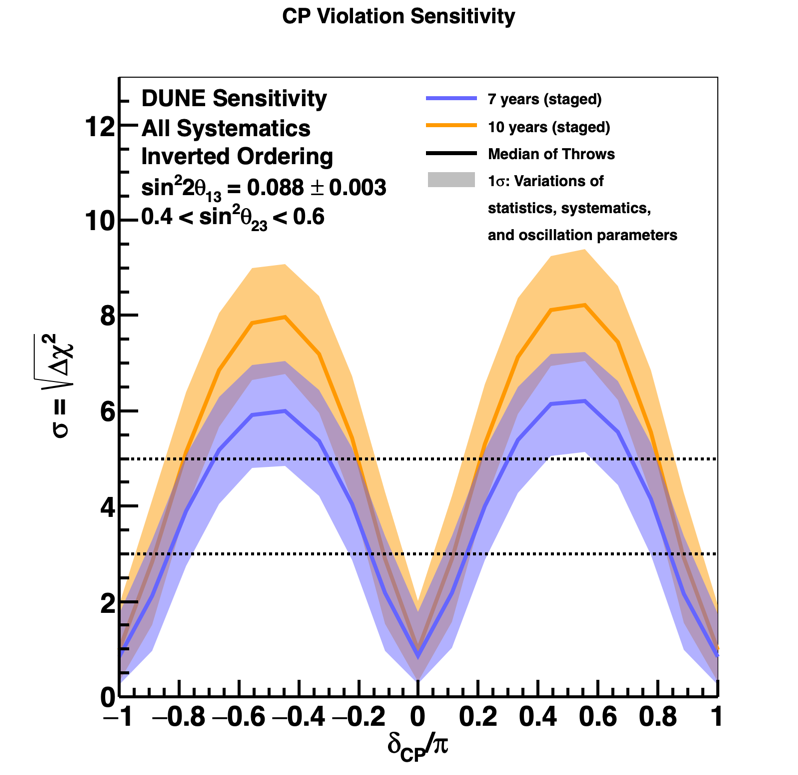}
  \caption[Significance of the DUNE determination of CP-violation as a function of \deltacp in both \dword{no} and \dword{io}]{Significance of the DUNE determination of CP-violation ($\deltacp \neq [0,\pm\pi]$) as a function of the true value of \deltacp, for seven (blue) and ten (orange) years of exposure, in both normal (top) and inverted (bottom) ordering. The width of the transparent bands cover 68\% of fits in which random throws are used to simulate statistical variations and select true values of the oscillation and systematic uncertainty parameters, constrained by pre-fit uncertainties. The solid lines show the median sensitivity.}
  \label{fig:cpv_nominal}
\end{figure}
Figure~\ref{fig:cpv_nominal} shows the significance with which \dword{cpv} ($\deltacp \neq [0, \pm\pi]$) can be observed in both \dword{no} and \dword{io} as a
function of the true value of \deltacp for exposures corresponding to seven and ten years of data, using the staging scenario described in Section~\ref{sec:rate}, and using the toy throwing method described in Section~\ref{sec:methods} to investigate their effect on the sensitivity.
This sensitivity has a characteristic double peak
structure because the significance of a \dword{cpv} measurement
necessarily decreases around CP-conserving values of \deltacp.
The median \dword{cpv} sensitivity reaches 5$\sigma$ for a small range of values after an exposure of seven years in \dword{no}, and a broad range of values after a ten year exposure. In \dword{io}, \dword{dune} has slightly stronger sensitivity to \dword{cpv}, and reaches 5$\sigma$ for a broad range of values after a seven year exposure.
Note that with statistical and systematic throws, the median sensitivity never reaches exactly zero.

\begin{figure}[htbp]
    \centering
    \includegraphics[width=0.98\linewidth, trim={0cm 0cm 0cm 2.3cm}, clip]{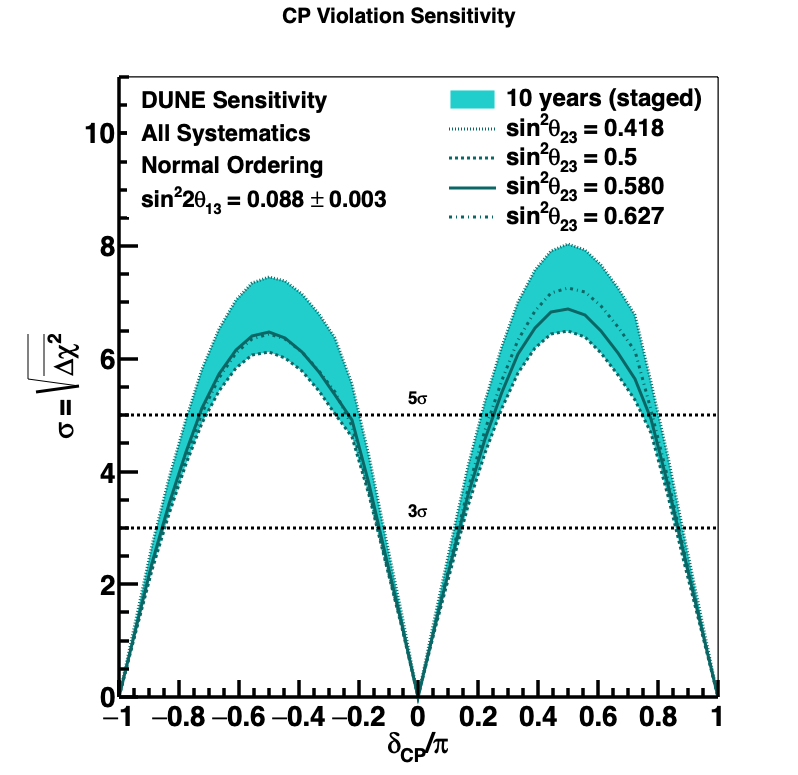}\\
    \includegraphics[width=0.98\linewidth, trim={0cm 0cm 0cm 2.3cm}, clip]{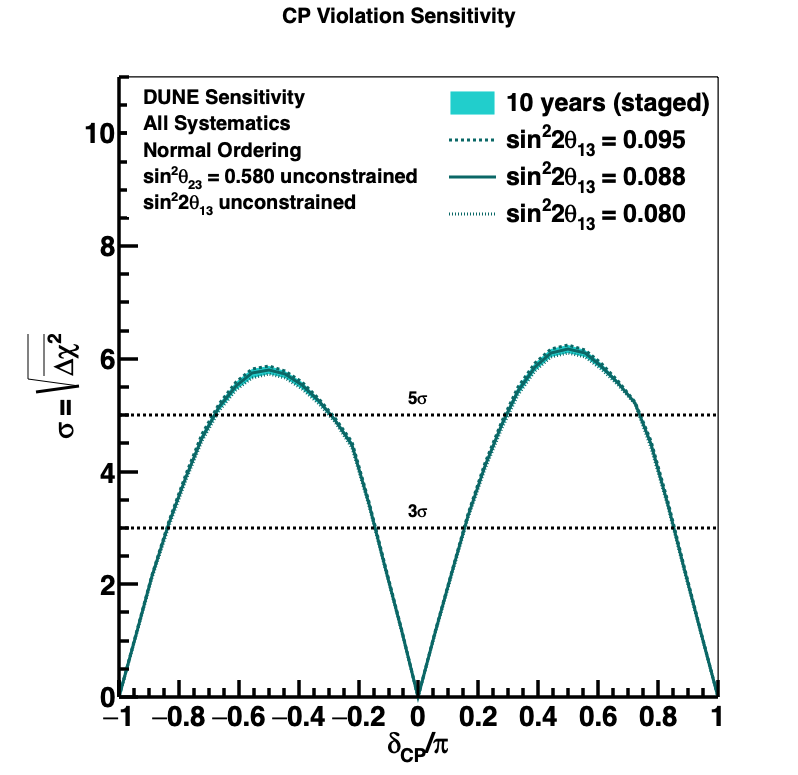}\\
    \includegraphics[width=0.98\linewidth, trim={0cm 0cm 0cm 2.3cm}, clip]{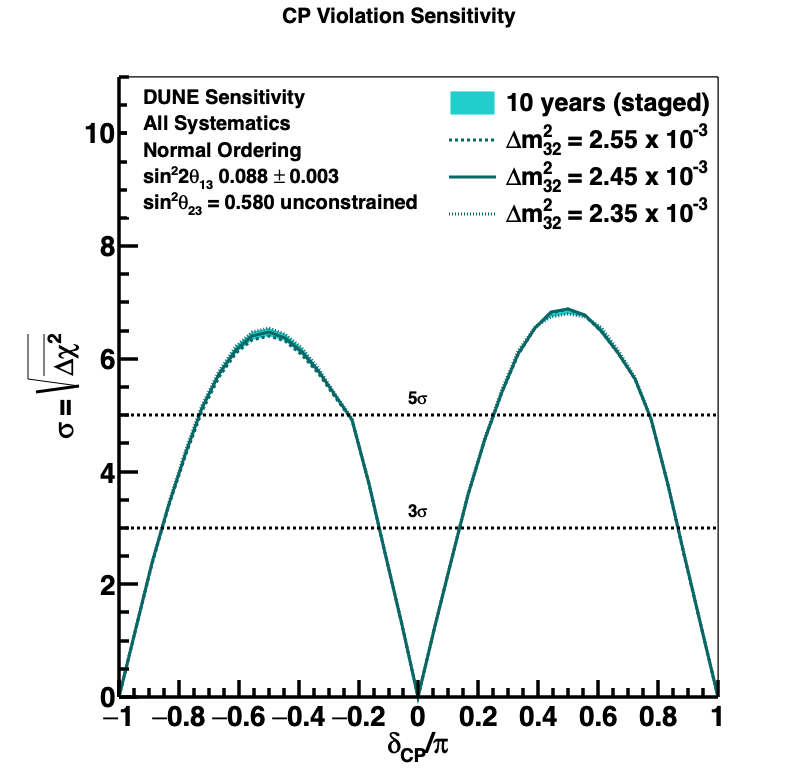}    
    \caption{Asimov sensitivity to CP violation, as a function of the true value of $\deltacp$, for ten years of exposure. Curves are shown for variations in the true values of $\theta_{23}$ (top), $\theta_{13}$ (middle) and $\Delta m^2_{32}$ (bottom), which correspond to their 3$\sigma$ \dword{nufit} range of values, as well as the \dword{nufit} central value, and maximal mixing.}
    \label{fig:cpv_oa_var}
\end{figure}
Figure~\ref{fig:cpv_oa_var} shows the \dword{dune} Asimov sensitivity to \dword{cpv} in \dword{no} when the true values of $\theta_{23}$, $\theta_{13}$, and $\Delta m^{2}_{32}$ vary within the 3$\sigma$ range allowed by \dword{nufit}. The largest effect is the variation in sensitivity with the true value of $\theta_{23}$, where degeneracy with $\deltacp$ and matter effects are significant. Values of $\theta_{23}$ in the lower octant lead to the best sensitivity to \dword{cpv}. The true values of $\theta_{13}$ and $\Delta m^2_{32}$ are highly constrained by global data and, within these constraints, do not have a dramatic impact on the sensitivity.
Note that in the Asimov cases shown in Figure~\ref{fig:cpv_oa_var}, the median sensitivity reaches 0 at \dword{cp}-conserving values of \deltacp (unlike the case with the throws as in Figure~\ref{fig:cpv_nominal}), but in regions far from \dword{cp}-conserving values, the Asimov sensitivity and the median sensitivity from the throws agree well.

\begin{figure}[htbp]
  \centering
  \includegraphics[width=0.98\linewidth, trim={0cm 0cm 0cm 2.3cm}, clip]{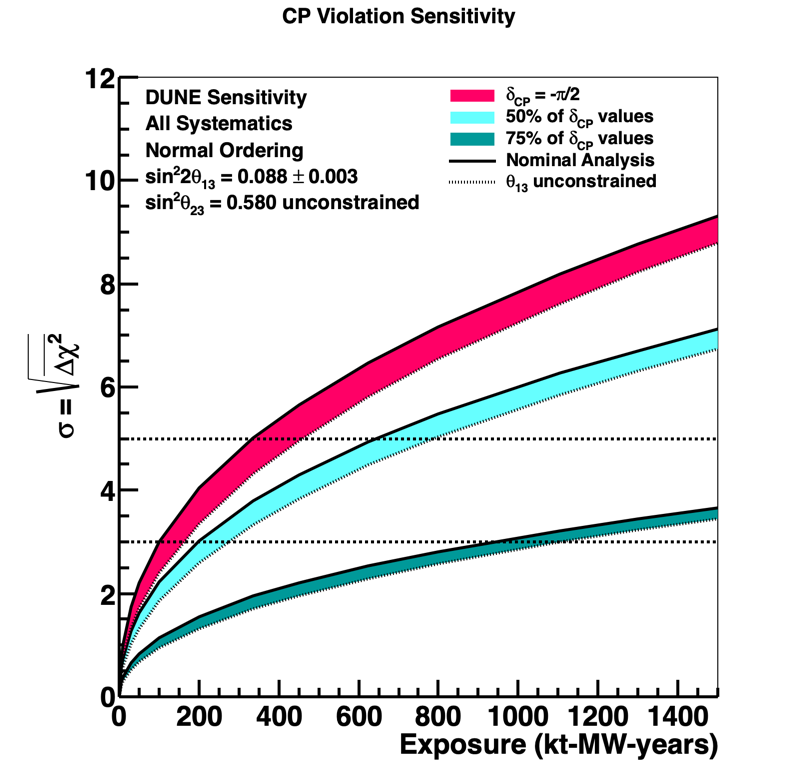}
  \includegraphics[width=0.98\linewidth, trim={0cm 0cm 0cm 2.3cm}, clip]{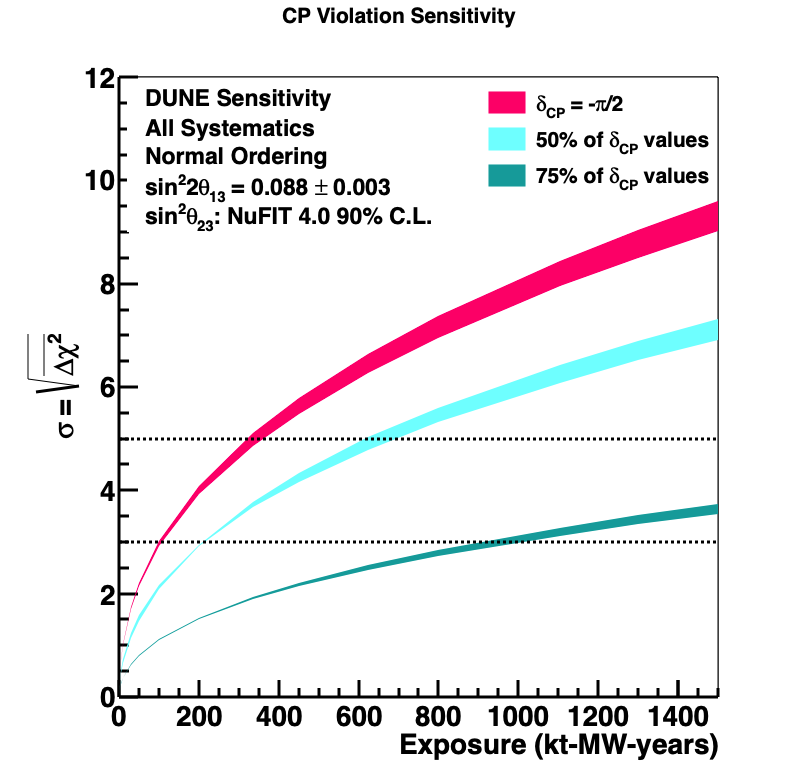}
  \caption[Significance of the DUNE determination of CP-violation as a function of exposure]{Significance of the DUNE determination of CP-violation ($\deltacp \neq [0,\pi]$) for the case when \deltacp=$-\pi/2$, and for 50\% and 75\% of possible true \deltacp values, as a function of exposure in kt-MW-years. Top: The width of the band shows the impact of applying an external constraint on $\theta_{13}$. Bottom: The width of the band shows the impact of varying the true value of \sinst{23} within the \dword{nufit} 90\% C.L. region.}
  \label{fig:cpv_exposure}
\end{figure}
Figure~\ref{fig:cpv_exposure} shows the result of Asimov studies investigating the significance
with which \dword{cpv} can be determined in \dword{no} for 75\% and 50\% of \deltacp values, and when $\deltacp=-\pi/2$, as a function of exposure in kt-MW-years, which can be converted to years using the staging scenario described in Section~\ref{sec:rate}. The width of the bands show the impact of applying an external constraint on $\theta_{13}$. CP violation can be observed with 5$\sigma$ significance after about seven years (336 kt-MW-years) if \deltacp = $-\pi/2$ and after about ten years (624 kt-MW-years) for 50\% of \deltacp values. CP violation can be observed with 3$\sigma$ significance for 75\% of \deltacp values after about 13 years of running. In the bottom plot of Figure~\ref{fig:cpv_exposure}, the width of the bands shows the impact of applying an external constraint on $\theta_{13}$, while in the bottom plot, the width of the bands is the result of varying the true value of \sinst{23} within the \dword{nufit} 90\% C.L. allowed region.

\begin{figure}[htbp]
  \centering
  \includegraphics[width=0.98\linewidth, trim={0cm 0cm 0cm 2.3cm}, clip]{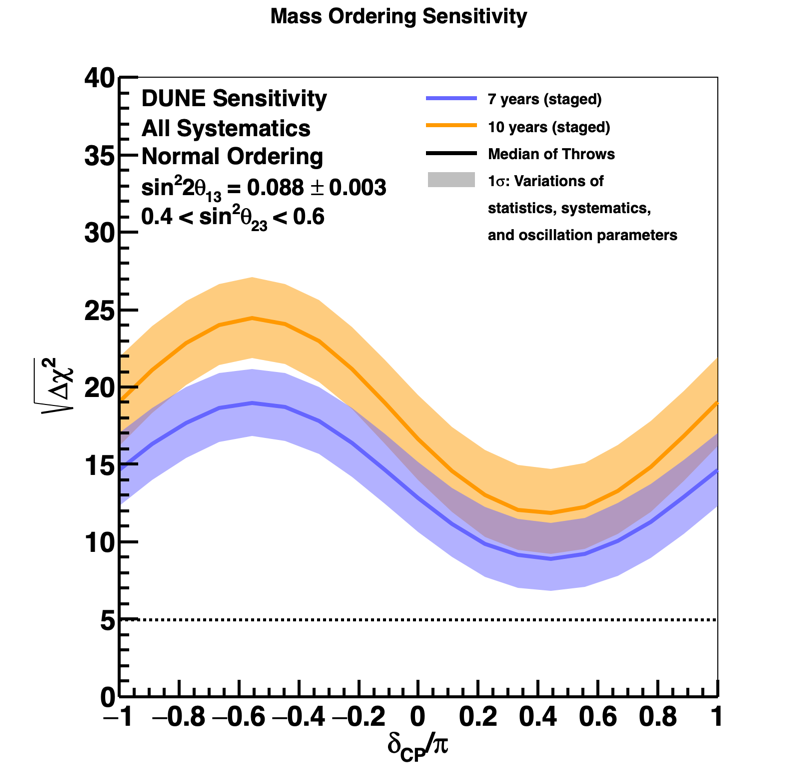}
  \includegraphics[width=0.98\linewidth, trim={0cm 0cm 0cm 2.3cm}, clip]{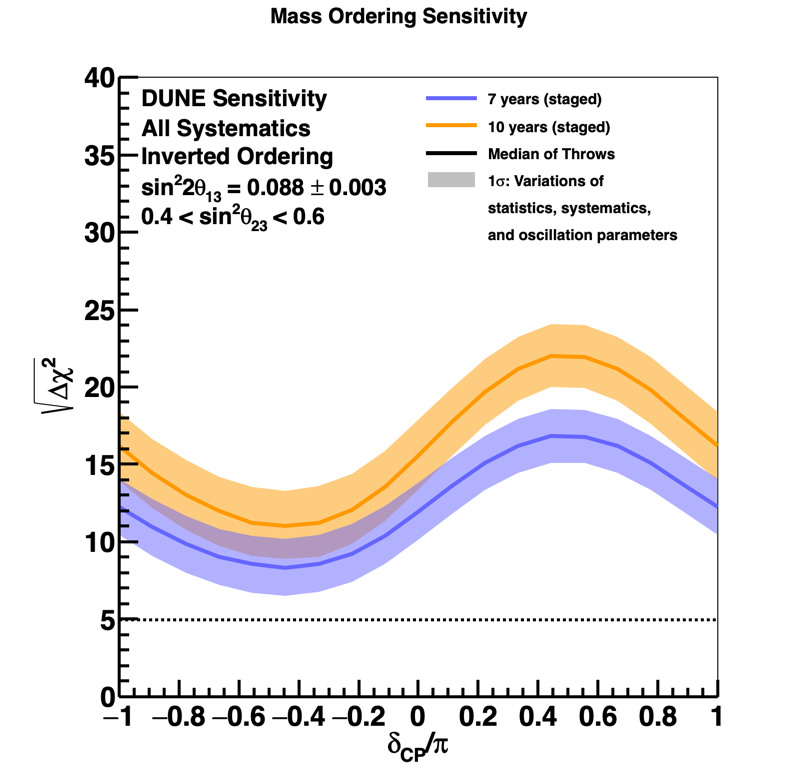}
  \caption[Significance of the DUNE neutrino mass ordering determination, as a function of \deltacp]{Significance of the DUNE determination of the neutrino mass ordering, as a function of the true value of \deltacp, for seven (blue) and ten (orange) years of exposure. The width of the transparent bands cover 68\% of fits in which random throws are used to simulate statistical variations and select true values of the oscillation and systematic uncertainty parameters, constrained by pre-fit uncertainties. The solid lines show the median sensitivity.}
  \label{fig:mh_nominal}
\end{figure}
Figure~\ref{fig:mh_nominal} shows the significance with which the neutrino mass ordering can be determined in both \dword{no} and \dword{io} as a function of the true value of \deltacp, for both seven and ten year exposures, including the effect of all other oscillation and systematic parameters using the toy throwing method described in Section~\ref{sec:methods}. The characteristic shape results from near degeneracy between matter and \dword{cpv} effects that occurs near $\deltacp=\pi/2$ ($-\deltacp=\pi/2$) for true normal (inverted) ordering. Studies have indicated that special attention must be paid to the statistical interpretation of neutrino mass ordering sensitivities~\cite{Ciuffoli:2013rza,Qian:2012zn,Blennow:2013oma} because the $\Delta\chi^2$ metric does not follow the expected chi-square function for one degree of freedom, so the interpretation of the $\sqrt{\Delta \chi^{2}}$ as the sensitivity is complicated. However, it is clear from Figure~\ref{fig:mh_nominal} that \dword{dune} is able to distinguish the mass ordering for both true \dword{no} and \dword{io}, and using the corrections from, for example, Ref.~\cite{Ciuffoli:2013rza}, DUNE would still achieve 5$\sigma$ significance for the central 68\% of all throws shown in Figure~\ref{fig:mh_nominal}. We note that for both seven and ten years (it was not checked for lower exposures), there were no parameter throws used in generating the plots ($\sim$300,000 each) for which the incorrect mass ordering was preferred.

\begin{figure}[htbp]
    \centering
    \includegraphics[width=0.98\linewidth, trim={0cm 0cm 0cm 2.3cm}, clip]{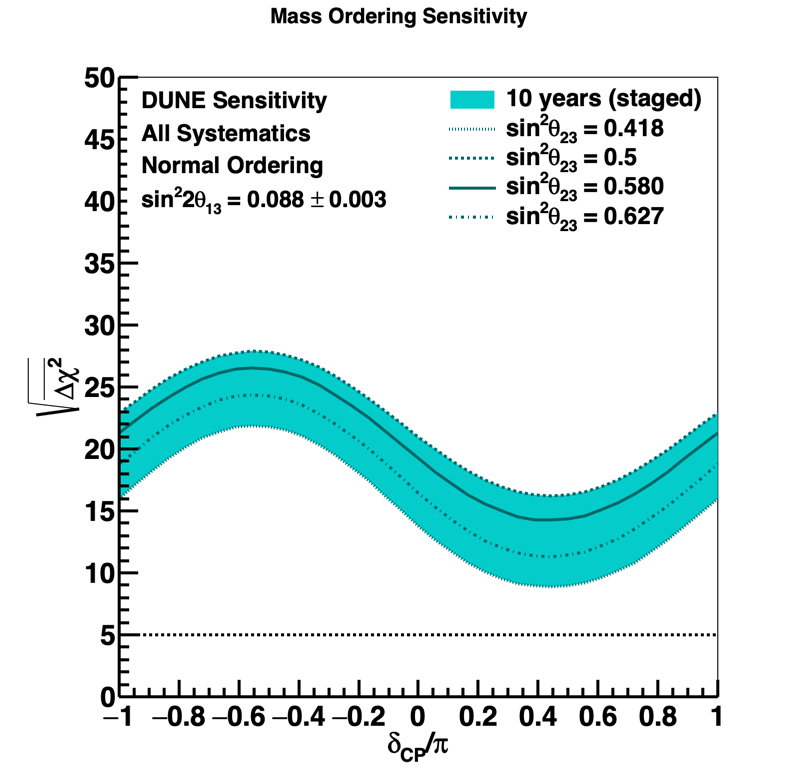}
    \includegraphics[width=0.98\linewidth, trim={0cm 0cm 0cm 2.3cm}, clip]{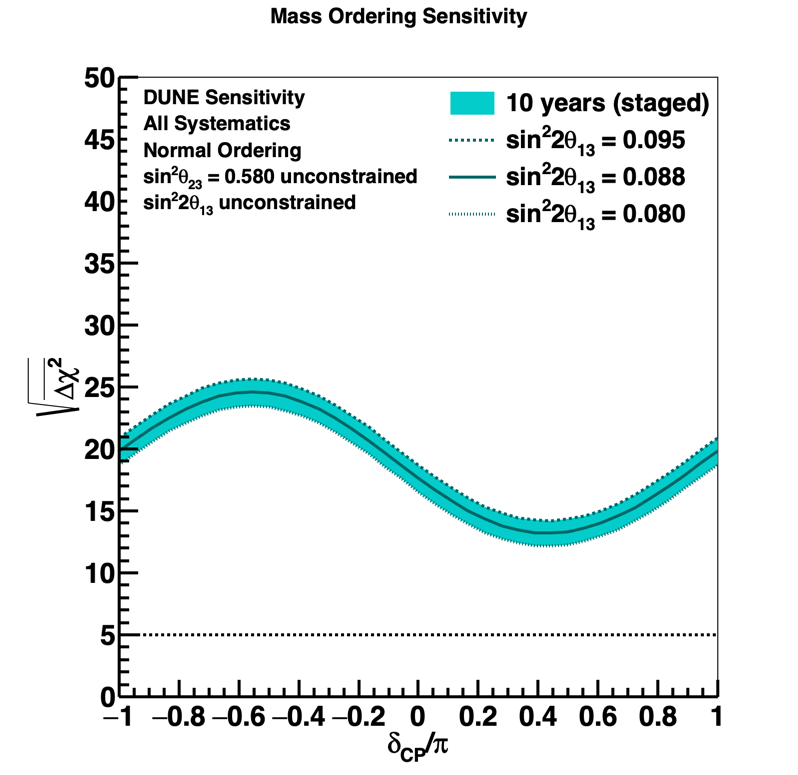}
    \includegraphics[width=0.98\linewidth, trim={0cm 0cm 0cm 2.3cm}, clip]{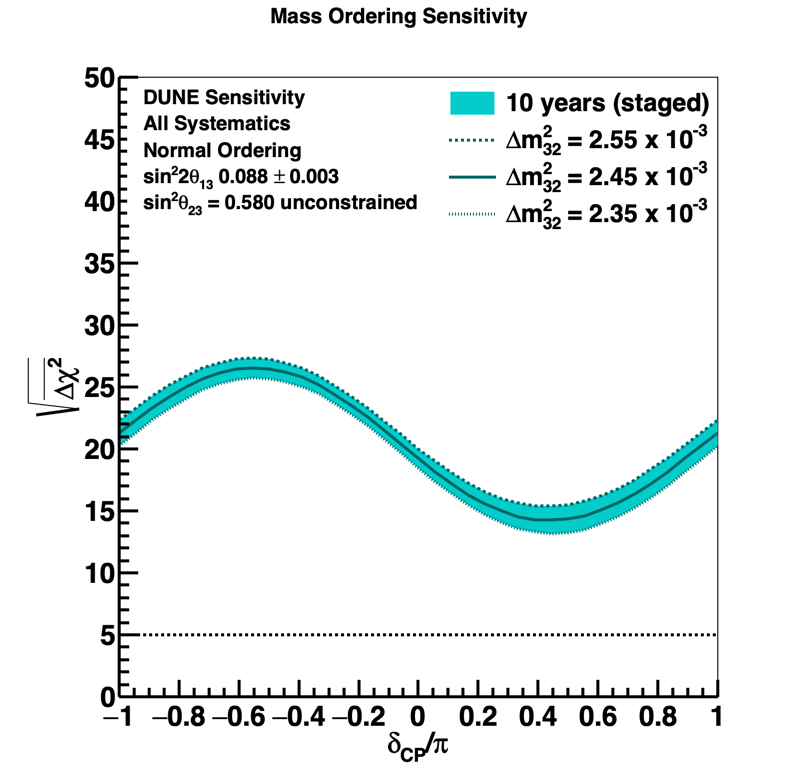}
    \caption{Asimov sensitivity to the neutrino mass ordering, as a function of the true value of $\deltacp$, for ten years of exposure. Curves are shown for variations in the true values of $\theta_{23}$ (top), $\theta_{13}$ (middle) and $\Delta m^2_{32}$ (bottom), which correspond to their 3$\sigma$ \dword{nufit} range of values, as well as the \dword{nufit} central value. and maximal mixing.}
    \label{fig:mh_oa_var}
\end{figure}
Figure~\ref{fig:mh_oa_var} shows the \dword{dune} Asimov sensitivity to the neutrino mass ordering when the true values of $\theta_{23}$, $\theta_{13}$, and $\Delta m^{2}_{32}$ vary within the 3$\sigma$ range allowed by \dword{nufit}. As for \dword{cpv} (in Figure~\ref{fig:cpv_oa_var}), the largest variation in sensitivity is with the true value of $\theta_{23}$, but in this case, the upper octant leads to the best sensitivity. Again, the true values of $\theta_{13}$ and $\Delta m^2_{32}$ do not have a dramatic impact on the sensitivity. The median Asimov sensitivity tracks the median throws shown in Figure~\ref{fig:mh_nominal} well for the reasonably high exposures tested --- this was not checked for exposures below seven years (336 kt-MW-years).

\begin{figure}[htbp]
    \centering
    \includegraphics[width=0.98\linewidth, trim={0cm 0cm 0cm 2.3cm}, clip]{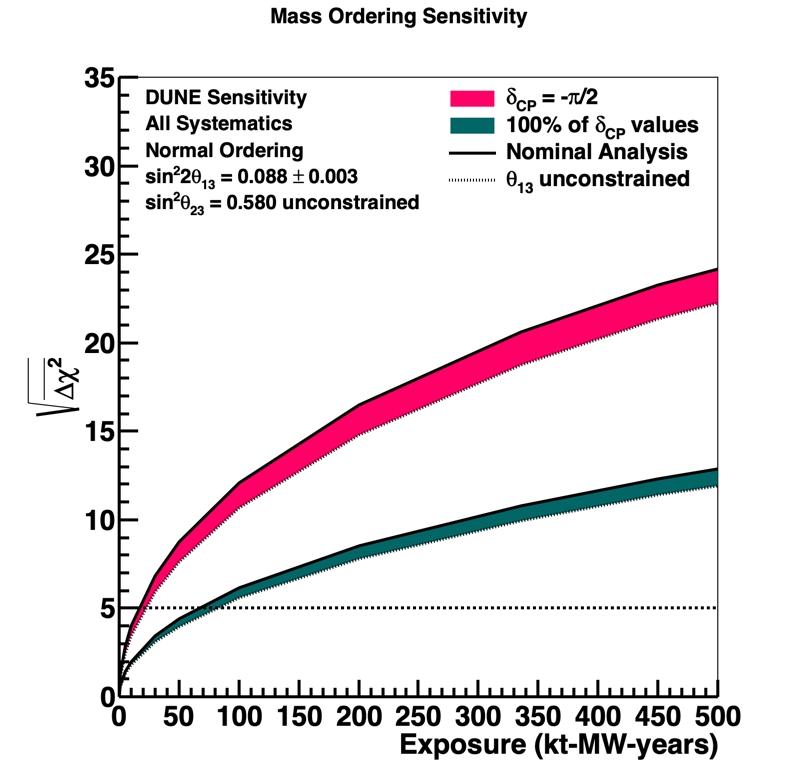}\\
    \includegraphics[width=0.98\linewidth, trim={0cm 0cm 0cm 2.3cm}, clip]{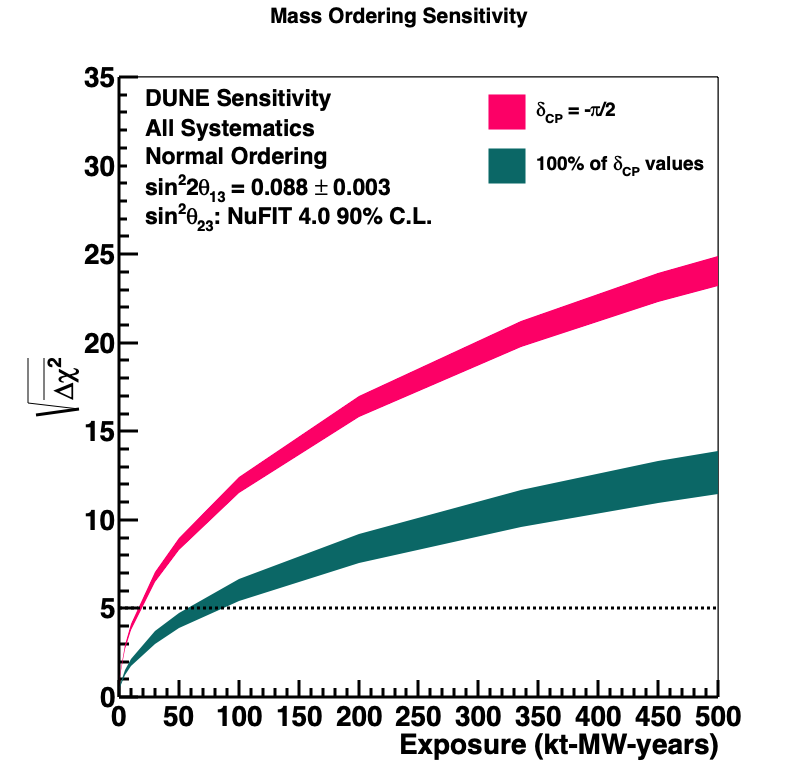} 
    \caption[Significance of the DUNE neutrino mass ordering determination as a function of exposure]{Significance of the DUNE determination of the neutrino mass ordering for the case when \deltacp=$-\pi/2$, and for 100\% of possible true \deltacp values, as a function of exposure in kt-MW-years. Top: The width of the band shows the impact of applying an external constraint on $\theta_{13}$. Bottom: The width of the band shows the impact of varying the true value of \sinst{23} within the \dword{nufit} 90\% C.L. region.}
    \label{fig:mh_exposure}
\end{figure}
Figure~\ref{fig:mh_exposure} shows the result of Asimov studies assessing the significance
with which the neutrino mass ordering can be determined for 100\% of \deltacp values, and when $\deltacp=-\pi/2$, as a function of exposure in kt-MW-years, for true \dword{no}. The width of the bands show the impact of applying an external constraint on $\theta_{13}$. The bottom plot shows the impact of varying the true value of \sinst{23} within the \dword{nufit} 90\% C.L. region. As DUNE will be able to establish the neutrino mass ordering at the 5$\sigma$ level for 100\% of \deltacp values after a relatively short period, these plots only extend to 500 kt-MW-years. 

\begin{figure}[htbp]
  \centering
  \includegraphics[width=0.98\linewidth]{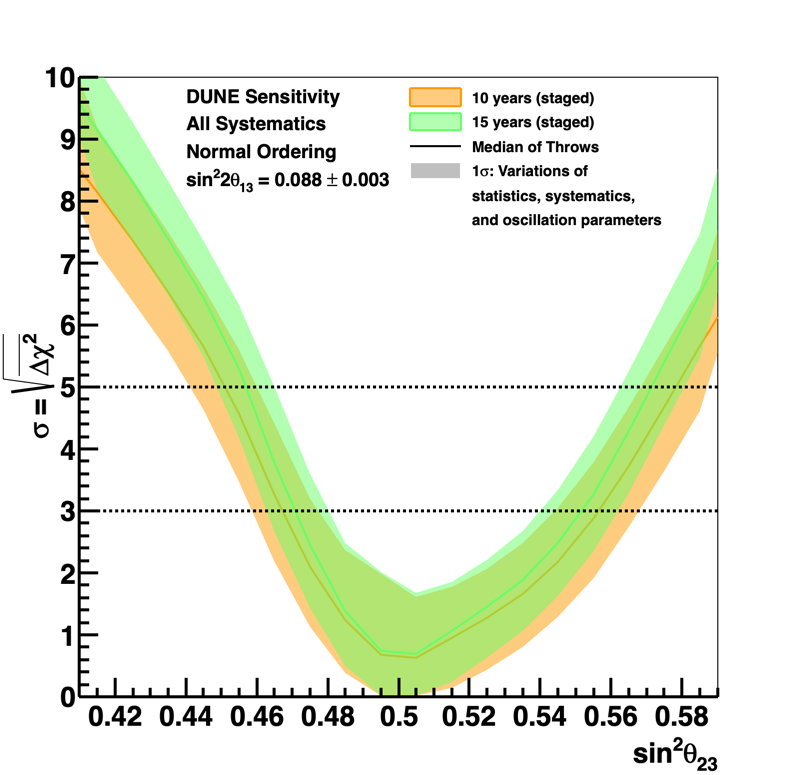}\\
  \includegraphics[width=0.98\linewidth]{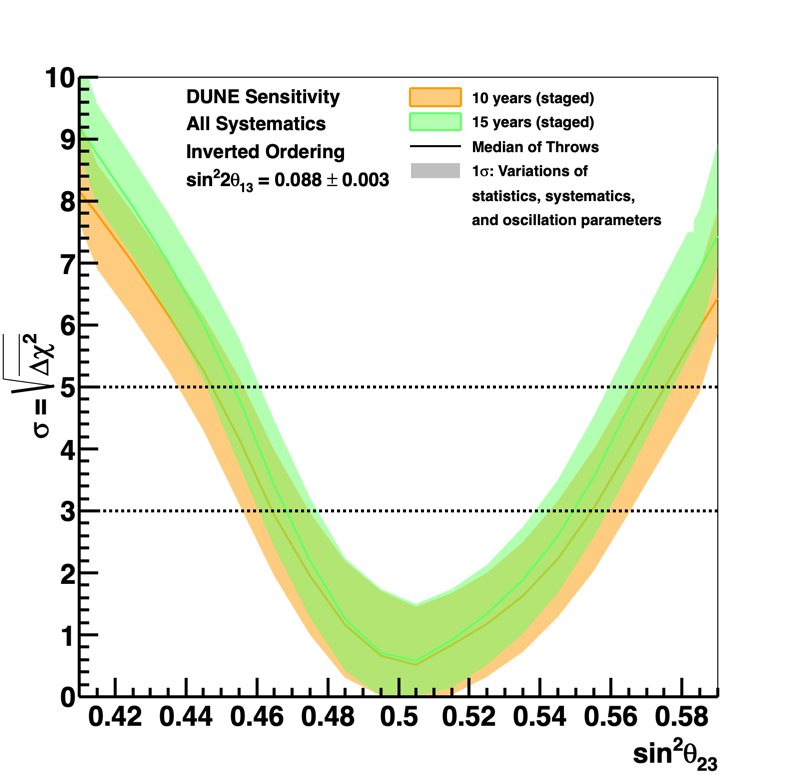}
  \caption[Sensitivity of determination of the $\theta_{23}$ octant as a function of \sinst{23} in both \dword{no} and \dword{io}]{Sensitivity to determination of the $\theta_{23}$ octant as a function of the true value of \sinst{23}, for ten (orange) and fifteen (green) years of exposure, for both normal (top) and inverted (bottom) ordering. The width of the transparent bands cover 68\% of fits in which random throws are used to simulate statistical variations and select true values of the oscillation and systematic uncertainty parameters, constrained by pre-fit uncertainties. The solid lines show the median sensitivity.}
    \label{fig:lbloctant}
\end{figure}
The measurement of $\nu_\mu \rightarrow \nu_\mu$ oscillations depends on $\sin ^2 2 \theta_{23}$, whereas the measurement of $\nu_\mu \rightarrow \nu_e$ oscillations depends on $\sin^2 \theta_{23}$.  A combination of both $\nu_e$ appearance and $\nu_\mu$ disappearance measurements can probe both maximal mixing and
the $\theta_{23}$ octant.  
Figure~\ref{fig:lbloctant} shows the sensitivity to determining the octant as a function of the true value of $\sinst{23}$, in both \dword{no} and \dword{io}. We note that the octant sensitivity strongly depends on the use of the external $\theta_{13}$ constraint.

In addition to the discovery potential for neutrino the mass ordering and \dword{cpv}, and sensitivity to the $\theta_{23}$ octant,  
\dword{dune} will improve the precision on key parameters that govern neutrino oscillations, including \deltacp, $\sin^22\theta_{13}$, \dm{31}, and $\sin^2\theta_{23}$.

\begin{figure}[htbp]
    \centering
    \includegraphics[width=0.98\linewidth]{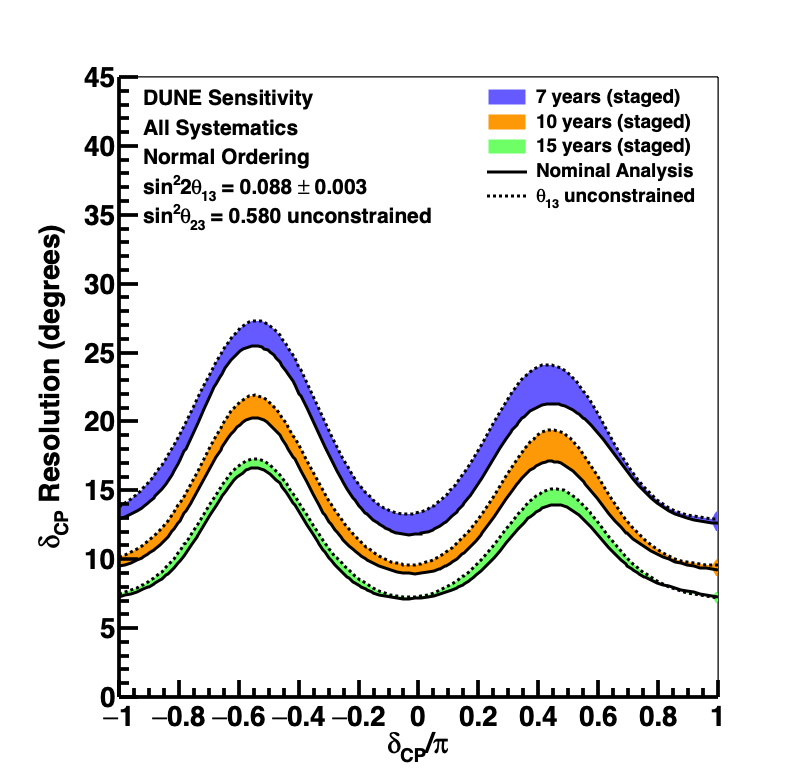}
    \caption[Resolution for the DUNE measurement of \deltacp as a function of \deltacp]
	    {Resolution in degrees for the DUNE measurement of \deltacp, as a function of the true value of \deltacp, for seven (blue), ten (orange), and fifteen (green) years of exposure. The width of the band shows the impact of applying an external constraint on $\theta_{13}$.}
    \label{fig:dcpresvdcp}
\end{figure}
Figure~\ref{fig:dcpresvdcp} shows the resolution, in degrees, of DUNE's measurement of \deltacp, as a function of the true value of \deltacp, for true \dword{no}. The resolution on a parameter is produced from the central 68\% of post-fit parameter values using many throws of the systematic and remaining oscillation parameters, and statistical throws. The resolution of this measurement is significantly better near CP-conserving values of \deltacp, compared to maximally CP-violating values. For fifteen years of exposure, resolutions between 5$^{\circ}$--15$^{\circ}$ are possible, depending on the true value of \deltacp. A smoothing algorithm has been applied to interpolate between values of \deltacp at which the full analysis has been performed.

\begin{figure}[htbp]
    \centering
    \includegraphics[width=0.98\linewidth]{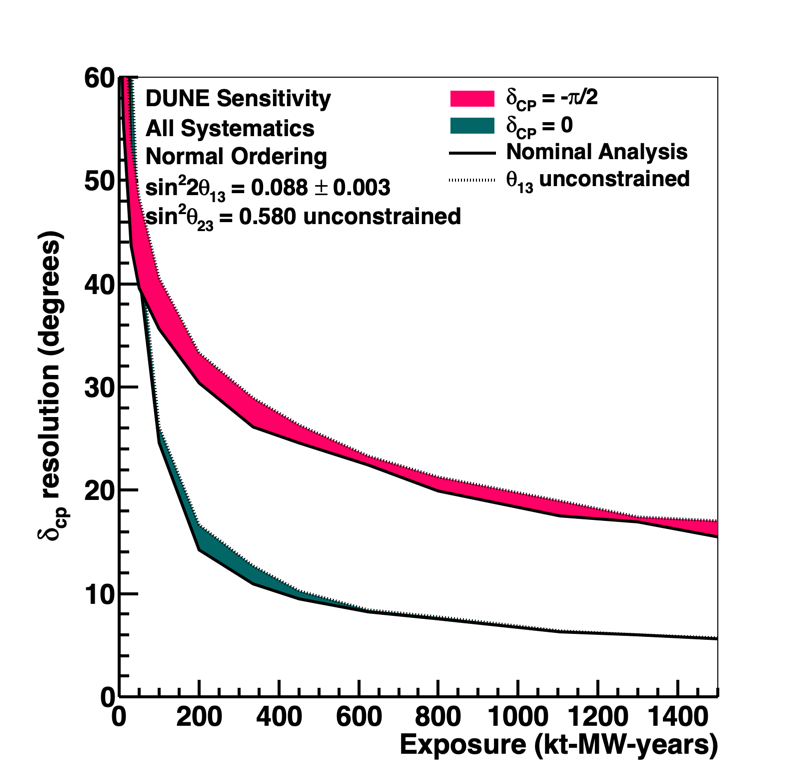}\\
    \includegraphics[width=0.98\linewidth]{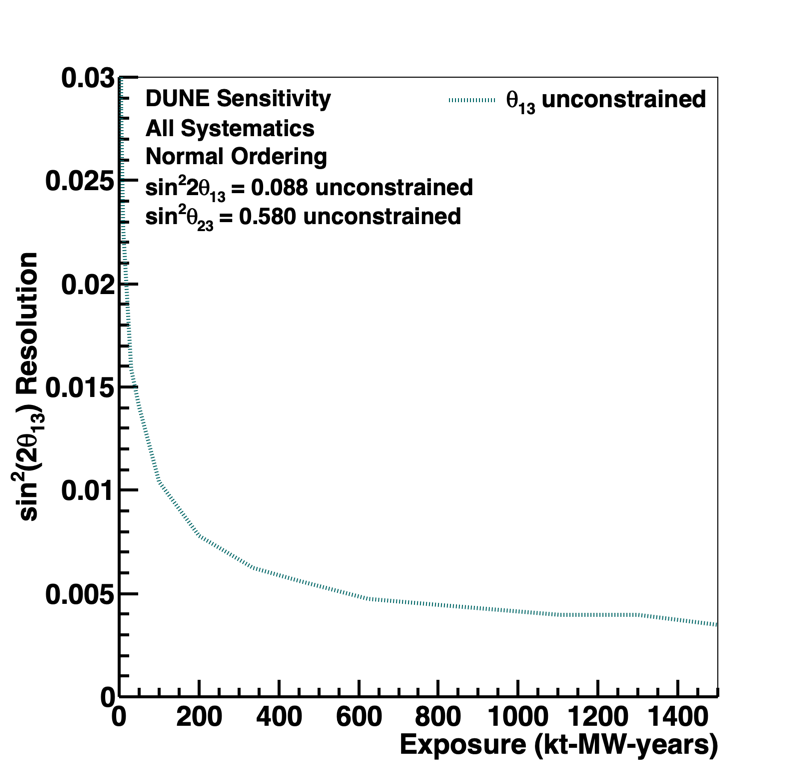} 
    \caption[Resolution of DUNE measurements of \deltacp and \sinstt{13}, as a function of exposure]{Resolution of DUNE measurements of \deltacp (top) and \sinstt{13} (bottom), as a function of exposure in kt-MW-years. As seen in Figure~\ref{fig:dcpresvdcp}, the \deltacp resolution has a significant dependence on the true value of \deltacp, so curves for $\deltacp=-\pi/2$ (red) and $\deltacp=0$ (green) are shown. For \deltacp, the width of the band shows the impact of applying an external constraint on $\theta_{13}$. No constraint is applied when calculating the \sinstt{13} resolution.}
    \label{fig:appres_exp}
\end{figure}
\begin{figure}[htbp]
    \centering
    \includegraphics[width=0.98\linewidth]{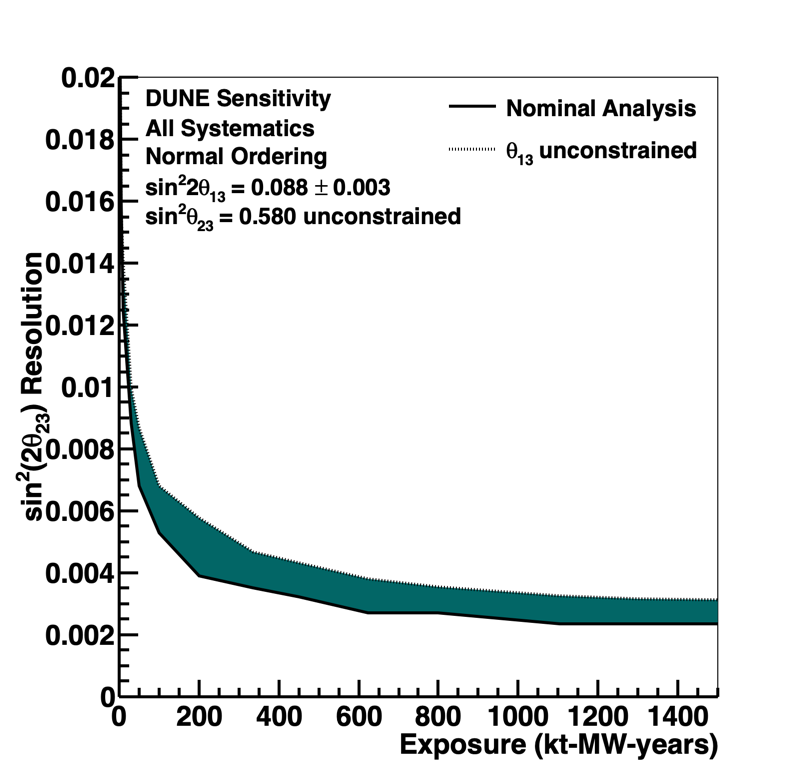}\\
    \includegraphics[width=0.98\linewidth]{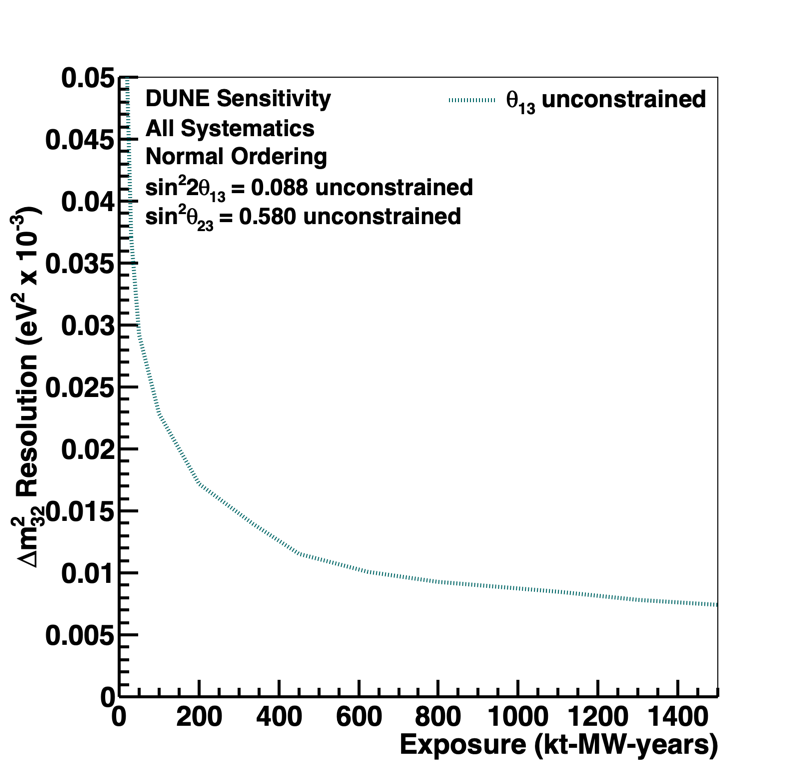} 
    \caption[Resolution of DUNE measurements of \sinstt{23} (top) and $\Delta m^{2}_{32}$, as a function of exposure]{Resolution of DUNE measurements of \sinstt{23} (top) and $\Delta m^{2}_{32}$ (bottom), as a function of exposure in kt-MW-years. The width of the band for the \sinstt{23} resolution shows the impact of applying an external constraint on $\theta_{13}$. For the $\Delta m^{2}_{32}$ resolution, an external constraint does not have a significant impact, so only the unconstrained curve is shown.}
    \label{fig:disres_exp}
\end{figure}
Figures \ref{fig:appres_exp} and \ref{fig:disres_exp} show the resolution of DUNE's measurements of \deltacp and \sinstt{13} and of \sinstt{23} and $\Delta m^{2}_{32}$, respectively, as a function of exposure in kt-MW-years. The resolution on a parameter is produced from the central 68\% of post-fit parameter values using many throws of the systematic other oscillation parameters, and statistical throws. As seen in Figure~\ref{fig:dcpresvdcp}, the \deltacp resolution varies significantly with the true value of \deltacp, but for favorable values, resolutions near five degrees are possible for large exposure. The DUNE measurement of \sinstt{13} approaches the precision of reactor experiments for high exposure, allowing a comparison between the two results, which is of interest as a test of the unitarity of the PMNS matrix. 

\begin{figure}[htbp]
  \centering
  \includegraphics[width=0.98\linewidth]{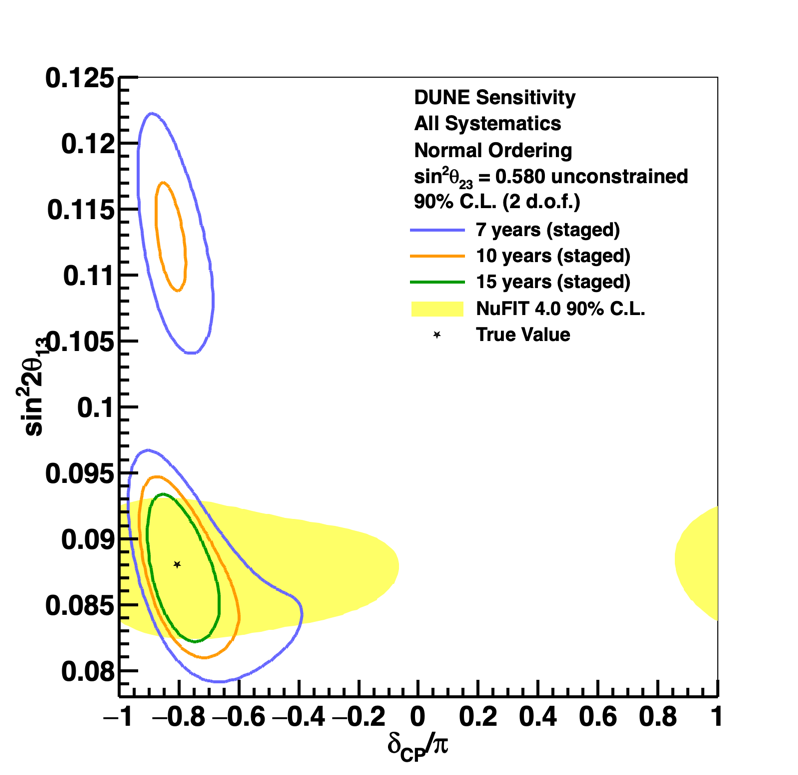}
  \includegraphics[width=0.98\linewidth]{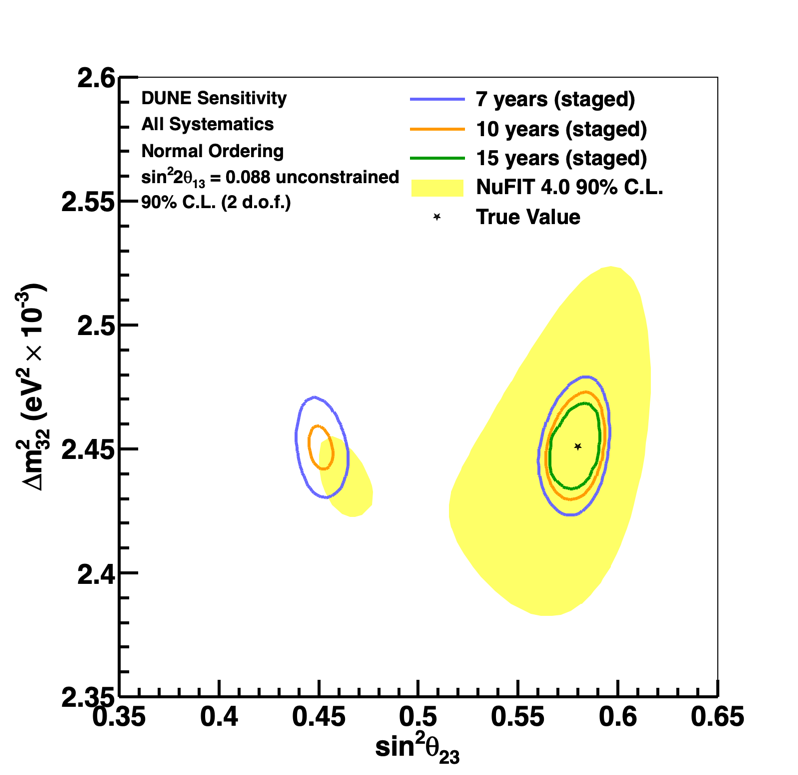}
  \caption[Two-dimensional 90\% constant $\Delta\chi^{2}$ confidence region in the \sinstt{13}--\deltacp and \sinst{23}--\dm{32} planes]{Two-dimensional 90\% constant $\Delta\chi^{2}$ confidence regions in the \sinstt{13}--\deltacp (top) and \sinst{23}--\dm{32} (botton) planes, for seven, ten, and fifteen years of exposure, with equal running in neutrino and antineutrino mode. The 90\% C.L. region for the \dword{nufit} global fit is shown in yellow for comparison. The true values of the oscillation parameters are assumed to be the central values of the \dword{nufit} global fit and the oscillation parameters governing long-baseline oscillation are unconstrained.}
    \label{fig:res_nopen_asimov0}
\end{figure}
\begin{figure}[htbp]
    \centering
    \includegraphics[width=0.98\linewidth]{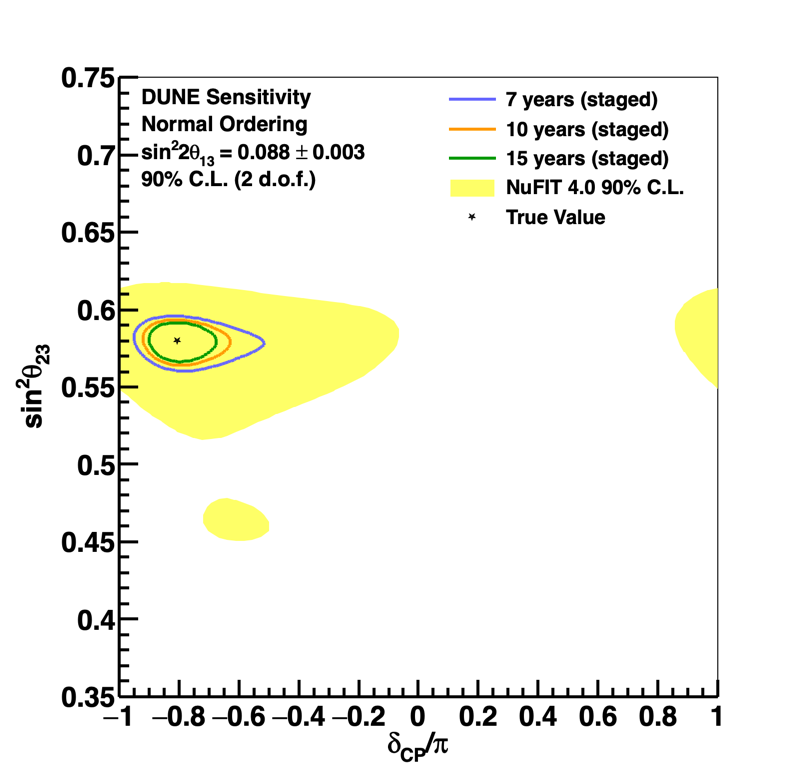}
    \caption[Two-dimensional 90\% constant $\Delta\chi^{2}$ confidence regions in the \sinst{23}--\deltacp plane]{Two-dimensional 90\% constant $\Delta\chi^{2}$ confidence regions in \sinst{23}--\deltacp plane, for seven, ten, and fifteen years of exposure, with equal running in neutrino and antineutrino mode. The 90\% C.L. region for the \dword{nufit} global fit is shown in yellow for comparison. The true values of the oscillation parameters are assumed to be the central values of the \dword{nufit} global fit and $\theta_{13}$ is constrained by \dword{nufit}.}
    \label{fig:res_th23vdcp}
\end{figure}
\begin{figure*}[htbp]
    \centering
    \includegraphics[width=0.49\linewidth]{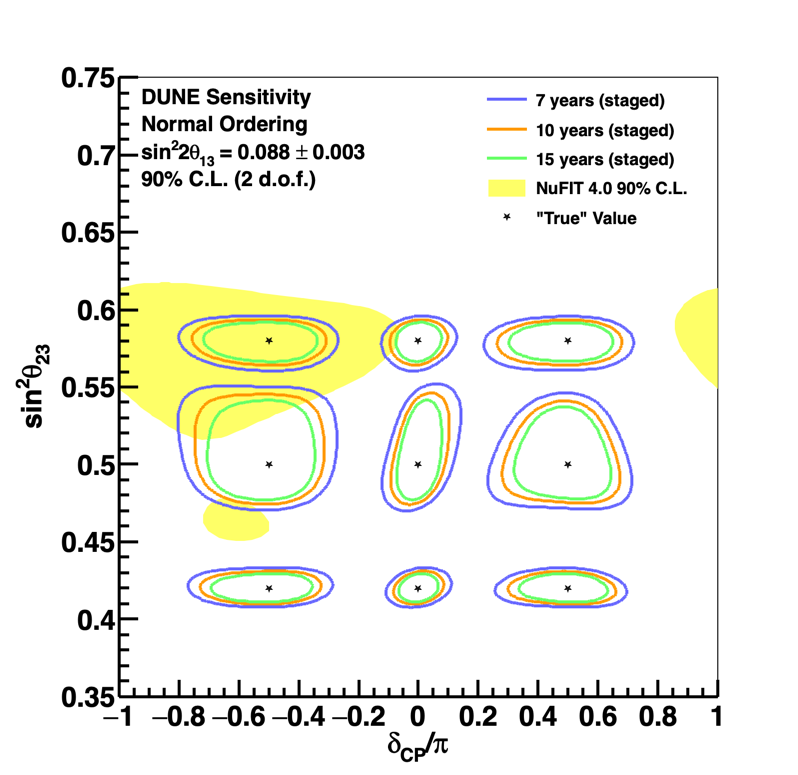}
    \includegraphics[width=0.49\linewidth]{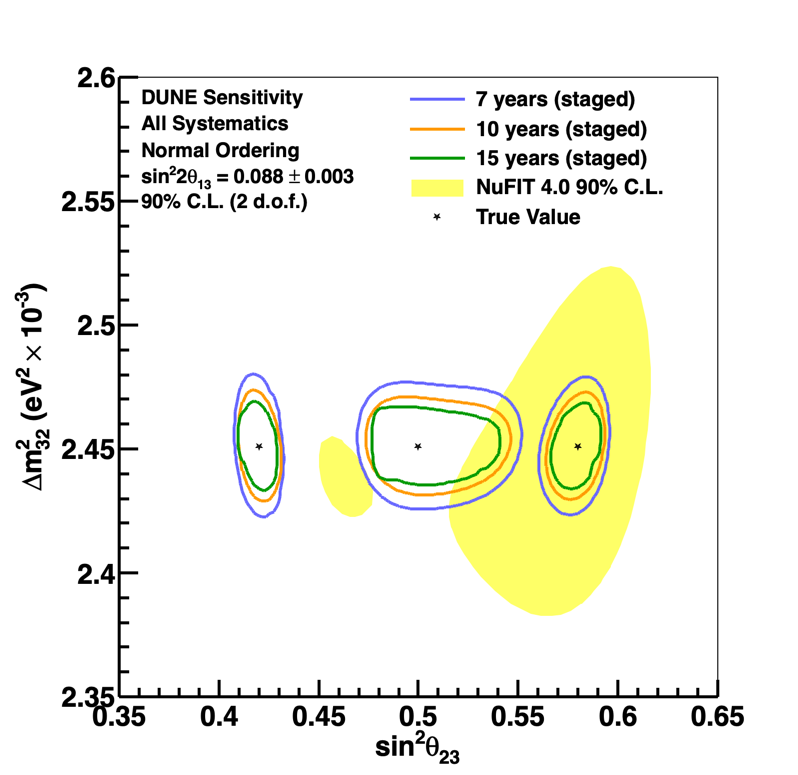}
    \caption[Two-dimensional 90\% constant $\Delta\chi^{2}$ confidence regions in the \sinst{23}--\deltacp and \sinst{23}--\dm{32} planes for different oscillation parameter values]{Two-dimensional 90\% constant $\Delta\chi^{2}$ confidence regions in the \sinst{23}--\deltacp (left) and \sinst{23}--\dm{32} (right) planes for different oscillation parameter values and seven, ten, and fifteen years of exposure, with equal running in neutrino and antineutrino mode. The 90\% C.L. region for the \dword{nufit} global fit is included in yellow for comparison. In all cases, an external constraint on the value of $\theta_{13}$ is applied. The true oscillation parameter values used are denoted by stars, and the \dword{nufit} best fit values are used as the true value of all those not explicitly shown. Test values of \sinst{23} = 0.42, 0.5, 0.58 were used for both top and bottom plots. In the top plot, test values of \deltacp = -$\pi/2$, 0, $\pi/2$ were used.}
    \label{fig:res_pen_various}
\end{figure*}

One of the primary physics goals for DUNE is the simultaneous measurement of all oscillation parameters governing long-baseline neutrino oscillation, without a need for external constraints. Figure~\ref{fig:res_nopen_asimov0} shows the 90\% constant $\Delta\chi^{2}$ allowed regions in the \sinstt{13}--\deltacp and \sinst{23}--\dm{32} planes for seven, ten, and fifteen years of running, when no external constraints are applied, compared to the current measurements from world data. An additional degenerate lobe visible at higher values of \sinstt{13} and in the wrong \sinst{23} octant is present in the seven and ten year exposures, but is resolved after long exposures. The time to resolve the degeneracy with \dword{dune} data alone depends on the true oscillation parameter values. For shorter exposures, the degeneracy observed in Figure~\ref{fig:res_nopen_asimov0} can be resolved by introducing an external constraint on the value of $\theta_{13}$. Figure~\ref{fig:res_th23vdcp} shows two-dimensional 90\% constant $\Delta\chi^{2}$ allowed regions in the \sinst{23}--\deltacp plane with an external constraint on $\theta_{13}$ applied. In this case, the degenerate octant solution has disappeared for all exposures shown.

Figure~\ref{fig:res_pen_various} explores the resolution sensitivity that is expected in the \sinst{23}--\deltacp and \sinst{23}--\dm{32} planes for various true oscillation parameter values, with an external constraint on $\theta_{13}$. The true oscillation parameter values used are denoted by stars, and the \dword{nufit} best fit values are used as the true value of all those not explicitly shown. Values of \sinst{23} = 0.42, 0.5, 0.58 were used in both planes, and additionally, values of \deltacp = -$\pi/2$, 0, $\pi/2$ were used in the \sinst{23}--\deltacp plane. It can be observed that the resolution in the value of \sinst{23} is worse at \sinst{23} = 0.5, and improves for values away from maximal in either octant. As was seen in Figure~\ref{fig:dcpresvdcp}, the resolution of \deltacp is smaller near the \dword{cp}-conserving value of \deltacp = 0, and increases towards the maximally \dword{cp}-violating values $\deltacp = \pm\pi/2$.

\begin{table}[htbp]
    \centering
    \begin{tabular}{lcc}
      \hline
 Physics Milestone & \multicolumn{2}{c}{Exposure} \\
 (\sinst{23} = 0.580) & Staged years & kt-MW-years \\
\hline\hline
 5$\sigma$ Mass Ordering & \multirow{2}{*}{1} & \multirow{2}{*}{16} \\
 \deltacp = -$\pi/2$ & & \\ \hline
 5$\sigma$ Mass Ordering & \multirow{2}{*}{2} & \multirow{2}{*}{66} \\
 100\% of \deltacp values & & \\ \hline
 3$\sigma$ CP Violation & \multirow{2}{*}{3} & \multirow{2}{*}{100} \\
 \deltacp = -$\pi/2$ & & \\ \hline
 3$\sigma$ CP Violation & \multirow{2}{*}{5} & \multirow{2}{*}{197} \\
 50\% of \deltacp values & & \\ \hline
 5$\sigma$ CP Violation & \multirow{2}{*}{7} & \multirow{2}{*}{334} \\
 \deltacp = -$\pi/2$ & & \\ \hline
 5$\sigma$ CP Violation & \multirow{2}{*}{10} & \multirow{2}{*}{646} \\
 50\% of \deltacp values & & \\ \hline
 3$\sigma$ CP Violation & \multirow{2}{*}{13} & \multirow{2}{*}{936} \\
 75\% of \deltacp values & & \\ \hline
 \deltacp Resolution of 10 degrees & \multirow{2}{*}{8} & \multirow{2}{*}{400} \\
 \deltacp = 0 & & \\ \hline
 \deltacp Resolution of 20 degrees & \multirow{2}{*}{12} & \multirow{2}{*}{806} \\
 \deltacp = -$\pi/2$ & & \\ \hline
 \sinstt{13} Resolution of 0.004 & 15 & 1079 \\ \hline
    \end{tabular}
    \caption[Projected DUNE oscillation physics milestones]{Exposure in years, assuming true normal ordering and equal running in neutrino and antineutrino mode, required to reach selected physics milestones in the nominal analysis, using the \dword{nufit} best-fit values for the oscillation parameters. The staging scenario described in Section~\ref{sec:rate} is assumed. Exposures are rounded to the nearest year.}
    \label{tab:milestones}
\end{table}
The exposures required to reach selected sensitivity milestones for the nominal analysis are summarized in Table~\ref{tab:milestones}. Note that the sensitivity to \dword{cpv} and for determining the neutrino mass ordering was shown to be dependent on the value of $\theta_{23}$ in Figures~\ref{fig:cpv_oa_var} and~\ref{fig:mh_oa_var}, so these milestones should be treated as approximate. $\deltacp = -\pi/2$ is taken as a reference value of maximal \dword{cpv} close to the current global best fit. Similarly, a resolution of 0.004 on \sinstt{13} is used as a reference as the current resolution obtained by reactor experiments.

%% file: sections/conclusion.tex
\section{Conclusion}
\label{sec:conclude}

The analyses presented here are based on full, end-to-end simulation, reconstruction, and event selection of \dword{fd} Monte Carlo and parameterized analysis of \dword{nd} Monte Carlo of the \dword{dune} experiment. Detailed uncertainties from flux, the neutrino interaction model, and detector effects have been included in the analysis. Sensitivity results are obtained using a sophisticated, custom fitting framework. These studies demonstrate that DUNE will be able to measure \deltacp to high precision, unequivocally determine the neutrino mass ordering, and make precise measurements of the parameters governing long-baseline neutrino oscillation.

We note that further improvements are expected once the full potential of the \dword{dune} \dword{nd} is included in the analysis. In addition to the samples used explicitly in this analysis, the \dword{lartpc} is expected to measure numerous exclusive final-state \dword{cc} channels, as well as \nue and \dword{nc} events. Additionally, neutrino-electron elastic scattering~\cite{dune_nue} and the low-$\nu$ technique~\cite{Quintas:1992yv,Yang:2000ju,Tzanov:2005kr,Adamson:2009ju,DeVan:2016rkm,Ren:2017xov} may be used to constrain the flux. Additional samples of events from other detectors in the \dword{dune} \dword{nd} complex are not explicitly included here, but there is an assumption that we will be able to control the uncertainties to the level used in the analysis, and it should be understood that that implicitly relies on having a highly capable \dword{nd}.

DUNE will be able to establish the neutrino mass ordering at the 5$\sigma$ level for 100\% of \deltacp values between two and three years. CP violation can be observed with 5$\sigma$ significance after $\sim$7 years if \deltacp = $-\pi/2$ and after $\sim$10 years for 50\% of \deltacp values. CP violation can be observed with 3$\sigma$ significance for 75\% of \deltacp values after $\sim$13 years of running. For 15 years of exposure, \deltacp resolution between five and fifteen degrees are possible, depending on the true value of \deltacp. The DUNE measurement of \sinstt{13} approaches the precision of reactor experiments for high exposure, allowing measurements that do not rely on an external \sinstt{13} constraint and facilitating a comparison between the DUNE and reactor \sinstt{13}  results, which is of interest as a potential signature for physics beyond the standard model. DUNE will have significant sensitivity to the $\theta_{23}$ octant for values of \sinst{23} less than about 0.47 and greater than about 0.55. We note that the results found are broadly consistent with those found in Ref.~\cite{Acciarri:2015uup}, using a much simpler analysis.

The measurements made by \dword{dune} will make significant contributions to completion of the standard three-flavor mixing picture, and provide invaluable inputs to theory work understanding whether there are new symmetries in the neutrino sector and the relationship between the generational structure of quarks and leptons. The observation of \dword{cpv} in neutrinos would be an important step in understanding the origin of the baryon asymmetry of the universe. The precise measurements of the three-flavor mixing parameters that \dword{dune} will provide may also yield inconsistencies that point us to physics beyond the standard three-flavor model.

%% file: sections/acknowledgements.tex
\begin{acknowledgements}

This document was prepared by the DUNE collaboration using the
resources of the Fermi National Accelerator Laboratory
(Fermilab), a U.S. Department of Energy, Office of Science,
HEP User Facility. Fermilab is managed by Fermi Research Alliance,
LLC (FRA), acting under Contract No. DE-AC02-07CH11359.
This work was supported by
CNPq, FAPERJ, FAPEG and FAPESP,              Brazil;
CFI, IPP and NSERC,                          Canada;
CERN;
M\v{S}MT,                                        Czech Republic;
ERDF, H2020-EU and MSCA,                     European Union;
CNRS/IN2P3 and CEA,                          France;
INFN,                                        Italy;
FCT,                                         Portugal;
NRF,                                         South Korea;
CAM, Fundaci\'{o}n ``La Caixa'' and MICINN,  Spain;
SERI and SNSF,                               Switzerland;
T\"UB\.ITAK,                                 Turkey;
The Royal Society and UKRI/STFC,             United Kingdom;
DOE and NSF,                                 United States of America.
This research used resources of the
National Energy Research Scientific Computing Center (NERSC),
a U.S. Department of Energy Office of Science User Facility
operated under Contract No. DE-AC02-05CH11231.
\end{acknowledgements}